%% file: book_layout.tex
\documentclass{cplslarge}%

\usepackage[authoryear]{natbib}
\usepackage{makeidx}
\makeindex

\usepackage{amsmath,amssymb}
\usepackage{bm}
\usepackage{graphicx}
\usepackage{textcomp}

\begin{document}

\frontmatter
  \maketitle
  \tableofcontents
  \listoffigures
  \listofcontributors
\mainmatter

\setcounter{part}{4}
\setcounter{chapter}{0}

  \part{Dynamical theories of jet formation:  Statistical and deterministic
    approaches} 

\chapter{Statistical Theories}

\section{Statistical models --- J. A. Krommes and J. B. Parker}
\label{KP}

\section{Direct statistical simulation of jets --- Tobias \& Marston}
\label{Marston}

\section{Equilibrium statistical mechanics of quasi-geostrophic models and
  zonal jets --- Bouchet \& Venaille}

\section{Zonostrophy and other quadratic invariants --- Nazarenko et~al.}

\chapter{Dynamical Second Order Closure Theories}

\section{Stochastic averaging, non-equilibrium statistical mechanics, and
  quasilinear approaches --- Bouchet et~al.}

\section{Stochastic structural stability theory --- Farrell \& Ioannou}
\label{FI}

\section{Zonostrophic instability --- Young \& Srinivasan}
\label{SY}

\input{zf_pf_includefile}
\input{modtext}

\input{jp_appendices}

\section{Emergence of non-zonal coherent structures --- Ioannou \& Bakas}

\contributor{john a. krommes}
{Plasma Physics Laboratory, MS 28\\
Princeton University\\
P.O. Box 451\\
Princeton, NJ  08543--0451\\
USA}

\contributor{jeffrey b. parker}
{Plasma Physics Laboratory, MS 29\\
Princeton University\\
P.O. Box 451\\
Princeton, NJ  08543--0451\\
USA}

\backmatter


\bibliography{./ZonalJetsReference}\bibliographystyle{cambridgeauthordate}
\label{refs}

\printindex

\end{document}

%% file: zf_pf_includefile.tex
\begingroup


\renewcommand{\a}{\alpha}
\renewcommand{\b}{\beta}
\newcommand{\de}{\delta}
\newcommand{\D}{\Delta}
\newcommand{\e}{\epsilon}
\newcommand{\ve}{\varepsilon}
\newcommand{\g}{\gamma}
\newcommand{\G}{\Gamma}
\renewcommand{\k}{\kappa}
\renewcommand{\l}{\lambda}
\renewcommand{\L}{\Lambda}
\newcommand{\m}{\mu}
\newcommand{\n}{\nu}
\newcommand{\p}{\phi}
\newcommand{\vp}{\varphi}
\renewcommand{\P}{\Phi}
\renewcommand{\r}{\rho}
\newcommand{\s}{\sigma}
\renewcommand{\t}{\tau}
\renewcommand{\th}{\theta}
\newcommand{\w}{\omega}
\newcommand{\W}{\Omega}
\newcommand{\z}{\zeta}

\newcommand{\la}{\langle}
\newcommand{\ra}{\rangle}

\renewcommand{\Re}{\operatorname{Re}}
\renewcommand{\Im}{\operatorname{Im}}
\newcommand{\sign}{\operatorname{sign}}

\renewcommand{\d}[2]{\frac{d #1}{d #2}}						
\newcommand{\dd}[2]{\frac{d^2 #1}{d #2^2}}					
\renewcommand{\v}[1]{\mathbf{#1}}				
\newcommand{\unit}[1]{{\v{\hat{#1}}}}			
\newcommand{\pd}[2]{\frac{\partial #1}{\partial #2}}		
\newcommand{\pdd}[2]{\frac{\partial^2 #1}{\partial #2^2}}	
\newcommand{\pddm}[3]{\frac{\partial^2 #1}{\partial #2 \partial #3}}	
\newcommand{\pddd}[2]{\frac{\partial^3 #1}{\partial #2^3}}	
\newcommand{\fd}[2]{\frac{\delta #1}{\delta #2}}					
\newcommand{\avg}[1]{\langle #1 \rangle}							
\newcommand{\bavg}[1]{\left\langle #1 \right\rangle}			

\newcommand{\comments}[1]{}									

\renewcommand{\O}{O}	
\newcommand{\defineas}{\equiv}

\newcommand{\wh}[1]{\widehat{#1}}
\providecommand{\ol}{}		
\renewcommand{\ol}[1]{\overline{#1}}

\newcommand{\vk}{\v{k}}
\newcommand{\ti}[1]{\widetilde{#1}}		

\newcommand{\azf}{\hat{\a}_{ZF}}
\newcommand{\LD}{L_d^{-2}}
\newcommand{\nablabarsq}{\ol{\nabla}^2}
\newcommand{\kbsq}{\ol{k}^2}
\newcommand{\xbar}{{\ol{x}}}
\newcommand{\ybar}{{\ol{y}}}
\newcommand{\qbsq}{\ol{q}^2}
\newcommand{\pbsq}{\ol{p}^2}
\newcommand{\hbpsq}{\ol{h}_+^2}
\newcommand{\hbmsq}{\ol{h}_-^2}

\newcommand{\eref}[1]{Eq.~\eqref{#1}}
\newcommand{\eqnref}[1]{Eq.~\eqref{eqn:#1}}					

\newcommand{\RB}{Rayleigh--B\'{e}nard\ }
\newcommand{\todo}[1]{\textbf{\emph{TODO:}#1}}
\renewcommand{\cite}{\citep}

\section{Zonal Flow as Pattern Formation --- J. B. Parker and J. A. Krommes}
\subsection{Introduction}
\label{Jeff1}
This section continues the use of statistical methods to investigate the physics of zonal jets.  Our interest in the problem of turbulent-driven zonal flows stems from wanting to understand their behavior in magnetized plasmas.  Plasmas possess their own host of complexities distinct from those of geophysics, including the mass differences of ions and electrons, kinetic effects such as wave--particle resonances, and electromagnetic effects.  It is a marvel that despite the immense disparities between laboratory plasmas and planetary atmospheres, similar physics in each conspire to organize regular flows out of turbulence.

In the challenging environments of plasmas devices, zonal flows have been detected \cite{fujisawa2009}.  Likewise, sophisticated gyrokinetic simulations, which are thought to retain all of the important physics of plasma microturbulence, exhibit the formation of zonal flows \cite{linhahmetal1998}.  Zonal flows have taken on special significance in magnetized plasmas because these flows are thought to suppress turbulent transport of heat \cite{diamonditohetal2005}, an advantageous feature when the ultimate goal is to keep hot the core plasma of fusion reactors.  As a result, much research into the physics of zonal flows in plasmas has been undertaken, and a great deal has been learned (see sections 3.2.1 and 3.2.2).

But several questions remain.  One problem of basic interest involves the length scale of the zonal flows (the width of the jets).  No one has yet found a heuristic estimate of the jet width in plasmas that enjoys as much success as the Rhines scale in geophysical contexts.  Another question is the detailed mechanism by which the flows suppress turbulence \cite{hatchjenkoetal2013}.  The flows can stabilize the linear modes responsible for driving the turbulence.  Additionally, plasma possess a complex array of feedbacks as well as velocity-space structure, any of which might be responsible for dissipation.  Strong interactions between turbulence and zonal flows can modify the energetics.  This is an active research area which requires strong collaboration between experiment, computation, and theory.

With that in mind, using the simplest paradigm systems we seek to develop a basic theory of zonal flows that can serve as the foundation for more complete plasma models.  To that end we begin from the Hasegawa--Mima equation, which describes electrostatic plasma turbulence in a uniform magnetic field in the presence of a background plasma density gradient \cite{hasegawamima1978,smolyakovdiamondetal2000a,krommeskim2000}.  This equation is equivalent to the quasigeostrophic barotropic vorticity equation in a certain limit.

We have found that from a statistical perspective, zonal flows constitute pattern formation amid a turbulent bath \cite{parkerkrommes2013a}.  Our account here emphasizes the role of a symmetry breaking and its consequences.  Some of the key insights to emerge are a mathematical prediction of a nonunique jet wavelength and a close linking of the phenomenon of jet merging with stability of the zonal flow--turbulence equilibrium.  We also study the symmetry-breaking zonostrophic instability in some detail and add some novel insights.

We are mindful that the Hasegawa--Mima equation is not a realistic or quantitatively accurate model of magnetized plasmas.  We are not primarily concerned with specific parameter dependencies, but rather we wish to establish general principles that will act as the building blocks in more elaborate theories and models.  The Hasegawa--Mima equation is a minimal model in that it contains the necessary physics for zonal flows to form but other complicating details are stripped away.

The zonal flows that are generated in these simple models can in some regimes be \emph{steady} in time, or at least nearing an idealization where that is true.  Such steady zonal flows may in fact occur in nature, such as the jets in Jupiter's atmosphere \cite{vasavadashowman2005}.  Steady jets may also be a robust feature of plasma turbulence in uniform magnetic geometry \cite{numataballetal2007}.  In the toroidal geometry relevant to fusion plasma, the zonal flows may fluctuate in time.  Nevertheless, for a tractable starting point we restrict ourselves to consideration of zonal flows that are steady or perhaps evolving much more slowly than the turbulence.  Although the Hasegawa--Mima equation in some regimes produces nonsteady jets, we select a parameter regime where the jets are nearly steady.  In any realistic situation there will always be some variation in time, and finding a true steady state requires a statistical perspective.

We have adopted the statistical approach to understanding turbulence.  While a brief review is given here, a broad introduction to statistical turbulence is given in Section 5.1.1. This approach complements other methods such as making detailed measurements of plasma fluctuations or performing direct numerical simulations (DNS), which accumulate reams of data so vast that it can be unclear how one should go about making sense of it all.  The aim of the statistical approach is to focus on the macroscopic quantities of interest, such as transport coefficients, energy spectra, and the like.  By working with averaged quantities from the outset, one can circumvent the rapid spatiotemporal fluctuations and potentially see a clearer view of the physics.  Of course, there is no free lunch.  As a consequence of averaging a nonlinear equation, one is generally left with the average of an unknown quantity: a closure problem.  Various statistical closures, perhaps motivated by physical considerations, provide different approximations for the unknown terms.  A major difficulty with this approach is that the closures are essentially uncontrolled approximations; the nonlinearity inherent to turbulence makes it hard to know exactly what is lost.  The closure might obliterate some highly coherent or correlated phenomena.  Nevertheless, these difficulties do not invalidate the statistical approach, from which much has been learned \cite{frisch1995,krommes2002,kraichnan1959b,kraichnan1964b}.  Historically, the majority of theoretical studies into turbulence that follow this approach assume \emph{homogeneous} statistics, where the statistics of turbulent quantities do not depend on position.  Consequently, most of the theoretical machinery that has been developed also applies only to homogeneous statistics, with comparatively little devoted to \emph{inhomogeneous} statistics.

In the presence of steady zonal flows, one is inevitably led to the conclusion that a proper statistical description must allow for inhomogeneous turbulence.  The turbulence cannot be statistically homogeneous; a location within the peak of a jet differs physically from a location at the node of a jet.  Inhomogeneous turbulence often arises on large spatial scales due to the presence of boundaries, topography, inhomogeneous driving forces, or other external factors.  These may be present even though it is often assumed that turbulence homogenizes on small scales.  But external influence is not the only way that inhomogeneities might develop.  Even if topography and boundaries are removed and the problem is contrived such that the governing equations of motion are independent of position, that translational symmetry may be broken spontaneously.  Zonal flows and inhomogeneous turbulence can result from such a scenario.  In this subtler development of inhomogeneous turbulence where the underlying physics is translationally invariant, a satisfactory statistical description must still allow for inhomogeneity.

As with several other sections in this chapter, we employ the second-order cumulant (CE2) framework \cite{marstonconoveretal2008,tobiasdagonetal2011,tobiasmarston2013,farrellioannou2003,farrellioannou2007,farrellioannou2009,bakasioannou2011,bakasioannou2013b,constantinoufarrelletal2013,parkerkrommes2013a} (see also sections 5.1.1, 5.1.2, 5.2.2, 5.2.3, and 5.2.4).  The CE2 formalism is the simplest possible setting in which to study inhomogeneous turbulence with statistical equations.  One way of deriving CE2 is through the \emph{quasilinear} (QL) approximation \cite{srinivasanyoung2012}.

We remark that the CE2 framework is equivalent to the Wigner-Moyal formalism.  The Wigner--Moyal formalism has been used in studies of wave physics in inhomogeneous media \cite{mcdonaldkaufman1985,halllisaketal2002}.  The Wigner distribution function, assuming an appropriate average is used in its definition, is closely related to the CE2 correlation function: they are both the two-point, one-time, second-order correlation of fluctuations.  The Wigner--Moyal equation, which describes the evolution of the distribution function, is the analog of the CE2 equation.  

In this article we use the 2D Charney--Hasegawa--Mima equation (CHME), written in a form to also encompass the modified Hasegawa--Mima equation (mHME).  The fundamental equation is\footnote{To obtain the coordinates conventionally used in the plasma literature, let $\{x,y\} \to \{-y,x\}$.}
	\begin{equation}
		\partial_t w + \v{v} \cdot \nabla w + \b \partial_x \psi = \ti{f} + D
		\label{jp:vorticityeqngeneral}
	\end{equation}
where $w$ is potential vorticity, $\v{v} = \unit{z} \times \nabla \psi$ is the horizontal velocity, $\psi$ is the stream function such that $w = \nabla^2 \psi - \azf \LD \psi$.  The standard CHME is obtained with $\azf=1$.  The mHME involves setting $\azf=0$ only for modes with $k_x=0$, i.e., zonal flow modes, and $\azf=1$ for all other modes.  This is done to more accurately model the zonal flow response in plasmas.  Additionally, in the plasma context, $L_d$ is the sound gyroradius $\r_s$.  $\ti{f}$ is the external forcing and can be thought of as a stirring, e.g., some idealization of excitation by buoyant convection or thermal gradient instability.  $D$ is dissipation and might represent frictional or viscous damping, or more generally any net transfer to external degrees of freedom.  Since the system is both driven and damped, it reaches an equilibrium where energy injection is balanced by energy dissipation.

To illustrate the QL approximation, we temporarily set $\ti{f}$ and $D$ to zero.  Conceptually, one decomposes the flow field into a zonally symmetric part (the zonal flow) and the residual (the eddies or tubulence).   Let $w = \ol{w} + w'$, where the overbar represents a zonal average, or average over $x$.  Equation \eqref{jp:vorticityeqngeneral} can be decomposed as
	\begin{subequations}
	\begin{align}
		\partial_t \ol{w} &+ \ol{ \v{v}'  \cdot \nabla w' } = 0, \\
		\partial_t w' &+ \ol{\v{v}} \cdot \nabla w' + \v{v}' \cdot \nabla \ol{w} \notag \\
		&+ \v{v}'  \cdot \nabla w' - \ol{ \v{v}'  \cdot \nabla w' } + \b \partial_x \psi' = 0. \label{jp:eddyeqngeneral}
	\end{align}
	\end{subequations}
No approximation has been made thus far.  At this point, one can make certain approximations that treat the eddies and the zonal flows differently.  The QL approximation proceeds by dropping, within the eddy equation \eqref{jp:eddyeqngeneral}, the terms quadratic in the eddy quantity (the advective nonlinearity).  The QL system is
	\begin{subequations}
		\label{jp:qlsystemgeneral}
	\begin{gather}
		\partial_t \ol{w} + \ol{ \v{v}'  \cdot \nabla w' } = 0, \\
		\partial_t w' + \ol{\v{v}} \cdot \nabla w' + \v{v}' \cdot \nabla \ol{w} + \b \partial_x \psi' = 0. \label{jp:qleddyeqngeneral}
	\end{gather}
	\end{subequations}
An alternative way of thinking about the QL approximation is in Fourier space.  All triad interactions between three Fourier modes are neglected except for those triads that contain one zonally symmetric mode ($k_x=0$).  The QL system respects the nonlinear conservation of energy and enstrophy.  It should be noted, however, that the QL approximation destroys exact material conservation of potential vorticity.

One of the uses of the QL system is that, subject only to an ergodic assumption, a statistical description can be obtained \emph{without} a closure problem.  Thus can one obtain CE2.  In the CE2 framework, the dynamical variables are the zonal-mean field $\ol{w}$ and the two-point covariance $W = \langle w'(\v{x}_1) w'(\v{x}_2) \rangle$.  The angle brackets may refer to a zonal average, ensemble average, or some other appropriate operation.  The derivation of CE2 from QL ensures that CE2 is statistically realizable with well-behaved statistics.

If the flow is predominantly zonal, then the QL approximation may be valid, at least for eddies at large scales \cite{bouchetnardinietal2013}.  However, we do not wish to restrict ourselves to discussion of a particular regime where zonal flow dominates.  In fact, most of our work focuses on the opposite limit where the zonal flow is weak relative to the turbulent flow.  We argue that even though the QL system is not rigorously valid as an approximation, it is useful as a model which contains some of the same behavior as the true system.  In particular, numerical evidence shows that the same symmetry breaking occurs in the QL system as in the original system.  The QL model is simpler and more tractable, however, and so provides a window into understanding the physics.

Although there are quantitative difference between CE2 and the original nonlinear dynamical system, CE2 does have something useful to offer about the physics of zonal flows.  CE2 provides a tractable problem with which to gain fundamental insight into the behavior of zonal flows and their interaction with turbulence.

\subsubsection{Analogy between zonal flows and \RB convection rolls}
\label{jp:sec:zfrbanalogy}
The notion of spontaneous symmetry breaking with respect to zonal flows has been discussed before \cite{farrellioannou2007,srinivasanyoung2012}.  This section will expand on that in discussing the mechanics of the symmetry breaking, as well as specific consequences it has for the physics of zonal flows \cite{parkerkrommes2013a,parkerkrommes2013b}.

An important aspect of zonostrophic instability is that it involves a spontaneous symmetry breaking.  A spontaneous symmetry breaking occurs when a situation's governing physics are invariant under a symmetry transformation but a physical realization is not invariant under the same transformation.  A simple example would be a ball moving in a symmetric double-well potential, as in Figure \ref{jp:fig:discretesymmetrybreaking}.  The equations of motion of the ball are invariant to reflection about the center line.  But with friction the ball must eventually end up in one of the wells, a state which breaks the symmetry.

Another well-known example of spontaneous symmetry breaking is the formation of convection rolls in \RB convection \cite{busse1978}.  A box of fluid, taken to be infinite in both horizontal directions and finite in vertical extent, is heated from below.  At weak heating, the heat is transferred to the cooler top surface solely by conduction, and the fluid is motionless.  But at sufficiently high heating, buoyancy forces overcome the inherent dissipation and the conduction state becomes unstable to the formation of convection rolls, as shown schematically in Figure \ref{jp:fig:schematic_convection_rolls}.  The convection rolls are spatially periodic but steady in time.  This transition to convection is analogous to the generation of zonal flows out of homogeneous turbulence.  Like the conduction state, homogeneous turbulence is (statistically) uniform in space.  And as a drive parameter such as the strength of the forcing is varied, that uniform state becomes unstable to the formation of a periodic structure.  Born out of turbulence are spatially periodic, steady-in-time zonal flows, which are analogous to the convection rolls (see Figure \ref{jp:fig:schematic_zonal_flows}).  More than merely descriptive, this analogy will be made mathematically precise in section \ref{jp:sec:bifurcation}.

	\begin{figure}
		\figurebox{3in}{}{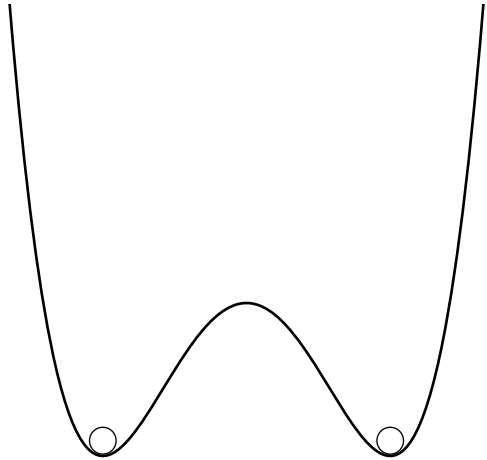}
		\caption{Discrete spontaneous symmetry breaking occurs when a ball moving in a symmetric double-well potential must, due to friction, end up in one of the wells.}
		\label{jp:fig:discretesymmetrybreaking}
	\end{figure}
	
	\begin{figure}
		\figurebox{3.1in}{}{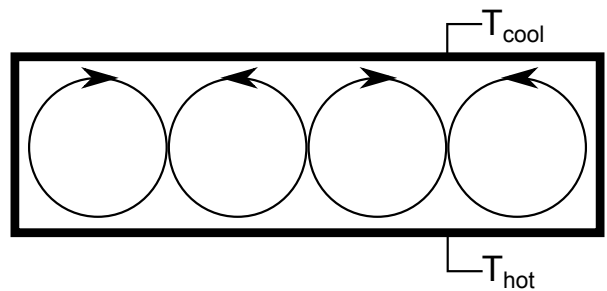}
		\caption{Convection rolls in \RB convection break the horizontal translational symmetry.}
		\label{jp:fig:schematic_convection_rolls}
	\end{figure}
	
	\begin{figure}
		\figurebox{3.1in}{}{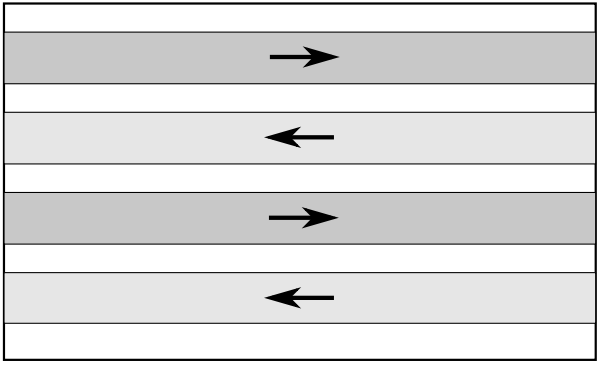}
		\caption{Zonal flows on a beta plane break the north-south (statistical) translational symmetry.}
		\label{jp:fig:schematic_zonal_flows}
	\end{figure}

\subsubsection{Outline}
This rest of this section is separated into two main parts.  The first part reexamines zonostrophic instability with the goal of improving physical understanding of the generation of zonal flows.  We show that zonostrophic instability contains the 4-mode modulational instability as a special case.  In \ref{jp:app:dispscale}, we provide a physical picture of the instability in the limit of long-wavelength zonal flows.

The second part studies the equilibrium between turbulence and zonal flows.  The symmetry breaking and the bifurcation to zonal flows is studied in some detail.  In addition, one method of numerical solution to the CE2 equations is offered.

\subsection{Zonal Flow Generation through Instability of Homogeneous Turbulence}
\label{jp:sec:zi}
\citet{srinivasanyoung2012} gave a detailed and insightful treatment of the so-called zonostrophic instability.  In zonostrophic instability, a homogeneous turbulent background is unstable to coherent zonal flow perturbations.  We give a brief overview before studying specific cases.  Srinivasan and Young began with a convenient, real-space form of the CE2 statistical equations.  We generalize their work to allow for finite $L_d$ and to unify the modified Hasegawa--Mima equation and the (equivalent) barotropic vorticity equation.  The appropriate CE2 equations, which can be derived from the QL equations \eqref{jp:qlsystemgeneral}, are
	\begin{subequations}
		\label{jp:CE2}
	\begin{align}
		\partial_t W &+ (U_+ - U_-) \partial_x W - \bigl(\ol{U}_+'' - \ol{U}_-''\bigr) \left( \nablabarsq + \frac{1}{4} \partial_\ybar^2 \right) \partial_x \Psi \notag \\
			& - \bigl[2\b - \bigl(\ol{U}_+'' + \ol{U}_-''\bigr)\bigr] \partial_\ybar \partial_x \partial_y \Psi = F(x,y) - 2\m W, \label{jp:CE2W} \\
		\partial_t \ol{I} U &+ \partial_\ybar \partial_x \partial_y \Psi(0,0 \mid \ybar,t) = -\m U, \label{jp:CE2U}
	\end{align}
	\end{subequations}
where $W(x,y \mid \ybar,t)$ is the 2-point covariance of vorticity, $x$ and $y$ are difference coodinates and $\ybar$ is the average coordinate, $\nablabarsq = \nabla^2 - \LD$, $U(\ybar, t)$ is the zonal-mean zonal velocity, $\m$ is the scale-independent friction, $U_\pm = U\bigl(\ybar \pm \frac12 y, t\bigr)$, $\ol{U}''_\pm = U''_\pm - \azf \LD U_\pm$, $\ol{I} = 1 - \azf \LD \partial_\ybar^{-2}$, and $\Psi$ is the covariance of stream function and is given by
	\begin{gather}
		W(x,y\mid \ybar,t) =  \hat{L} \Psi(x,y \mid \ybar, t), \label{jp:WfromPsi} \\
		\hat{L} = \left(\nablabarsq + \partial_y \partial_\ybar + \frac14 \partial_\ybar^2 \right)  \left(\nablabarsq - \partial_y \partial_\ybar + \frac14 \partial_\ybar^2 \right).
	\end{gather}
For simplicity the viscosity has been taken to be zero, though that is not necessary \cite{parkerkrommes2013a,parkerkrommes2013b}.  The external forcing $\ti{f}$ has been taken to be random white noise, and $F(x,y)$ is its covariance.

One equilibrium of \eref{jp:CE2} has no mean zonal flow, $U=0$, and corresponds to homogeneous turbulence.  The covariance is independent of $\ybar$ and takes the simple form
	\begin{equation}
		W_H(x,y) = \frac{F(x,y)}{2\m}.
		\label{jp:WH}
	\end{equation}
This is \emph{always} a steady-state solution of \eref{jp:CE2}.  But that does not mean it will naturally occur; it may be unstable.  Particularly of interest is the stability to perturbations with a mean-field component, i.e., zonal flow.  These perturbations are assumed to have $e^{\l t} e^{i q \ybar}$ dependence such that $\l$ is the eigenvalue and $q$ is the wavenumber of the zonal flow.  The dispersion relation is \cite{srinivasanyoung2012}
	\begin{equation}
		\frac{\qbsq}{q^2} (\l + \m) = q\L_- - q\L_+,
		\label{jp:dispersionrelation}
	\end{equation}
where $\qbsq = q^2 + \azf \LD$,
	\begin{equation}
		\L_\pm = \int \frac{dk_x dk_y}{(2\pi)^2} \frac{k_x^2 k_y (1 - \qbsq / \ol{h}_\pm^2 ) W_H(k_x, k_y \pm \frac12 q )} {(\l + 2\m) \ol{h}_+^2 \ol{h}_-^2 + 2i \b q k_x k_y},
		\label{jp:lambda_pm}
	\end{equation}
$\ol{h}_\pm^2 = k_x^2 + \bigl(k_y \pm \tfrac12 q\bigr)^2 + \LD$, $k_x$ and $k_y$ are the Fourier conjugate variables of $x$ and $y$, and our Fourier transform convention is
	\begin{equation}
		f(k) = \int dx\, e^{-ikx} f(x).
	\end{equation}
Although this convention uses the same symbol $f$ for the real space and Fourier space functions, we always specify the argument of the function to make clear whether the real or Fourier domain is being used.  In \eref{jp:dispersionrelation}, the left-hand-side (LHS) is the zonal flow intrinsic response and the right-hand-side (RHS) is the Reynolds stress forcing term.  If the perturbations are unstable, corresponding to a solution with $\Re \l > 0$, then zonal flows grow in the so-called zonostrophic instability.  One can also let $\m$ and $F$ be zero; in that case, with $U=0$ any homogeneous $W_H$ is a steady state and has the dispersion relation above.

\subsubsection{Isotropic Background Spectrum with Finite Deformation Radius}
\label{jp:sec:isotropic}
We now specialize the dispersion relation to an isotropic background and examine various limits, allowing for finite deformation radius.  Although a purely isotropic spectrum is unlikely to obtain in practice when the beta effect is present, such an investigation helps to isolate the physical consequences of various effects.  

In the context of an infinite deformation radius $L_d$, the effect of an isotropic background spectrum has been studied before \cite{srinivasanyoung2012,bakasioannou2013}.  Those studies concluded that for an isotropic background, $\b \neq 0$ is required for instability.  Additionally, they found that for an isotropic background, the eddies acted on long-wavelength zonal flows as a negative hyperviscosity instead of negative viscosity.  That is, the eddy forcing on the RHS of \eqref{jp:dispersionrelation} behaves as $q^4$ rather than $q^2$ at small $q$.  In this section, we study how these results change when finite deformation length $L_d$ is allowed.

The dispersion relation for a homogeneous, isotropic background spectrum can be written as
	\begin{equation}
		\frac{\qbsq}{q^2}(\l + \m ) = \frac{1}{\b} \int_0^\infty \frac{dk}{2\pi} k^2 \left(1 - \frac{\qbsq}{\kbsq} \right) W_H(k) S(\chi, n, m)
		\label{jp:disprelation_isotropic}
	\end{equation}
where
	\begin{gather}
		S(\chi, n, m) = \int_0^{2\pi} \frac{d\p}{2\pi} K, \\
		K = \frac{ (n - 2\cos\p) \sin^2 \p}{\chi (1 - 2n\cos\p + n^2 + m) + i (n - 2\cos\p) \sin \p}, \\
		\chi = \frac{(\l+2\m)\kbsq}{\b q}, \\
		n = \frac{q}{k}, \\
		m = (kL_d)^{-2}.
	\end{gather}
We now examine the limit of large $\chi$, which could correspond to either small $\b$ or small $q$.  Asymptotic expansion of $S$ for large $\chi$ reveals interesting behavior that can differ for finite vs.\ infinite $L_d$.

For infinite $L_d$ (i.e., $m=0$), $S$ behaves as\footnote{Validity of this formula requires that $1-n^2$ is not too small.} \cite{srinivasanyoung2012}
	\begin{equation}
		S(\chi, n, 0) =
			\begin{cases} \displaystyle \frac{n}{\chi^3} \frac{3}{8(1-n^2)} + \O(\chi^{-5}), 	&	n^2 < 1, \\[0.5cm]
				\displaystyle \frac{1}{\chi} \frac{n^2-1}{2n^3} + \O(\chi^{-3}),		&	n^2 > 1.
			\end{cases}
	\end{equation}
For small $q$, we recover $S \sim q^4$.  Additionally, we can consider the case of finite $q$ but small $\b$.  For $n^2 < 1$, the RHS of \eref{jp:disprelation_isotropic} goes as $\b^2$, which vanishes at $\b=0$.  This result was also found by \citet{srinivasanyoung2012}.  Therefore, at $\b=0$ any thin ring of an isotropic spectrum with $k > q$ has no net effect on the zonal flow.  On the other hand, for $n^2 > 1$ the $\b$ dependence in the RHS of \eref{jp:disprelation_isotropic} vanishes.  Thus, at $\b=0$ a thin ring with $k < q$ has a net damping effect on the zonal flow.

For finite $L_d$, $S$ behaves as\footnote{Validity requires that $m \neq 0$, because for $m=0$ and $n^2 < 1$, the lowest order result vanishes.}
	\begin{align}
		S(\chi,n,m&) = (4 n^3 \chi)^{-1} \left[  -n^2(-1+m) + (1+m) \Big( -1 - m  \right. \notag  \\
						& \left. + \sqrt{[(-1+n)^2 + m][(1+n)^2 + m]} \Big)\right] +\O\left( \chi_1^{-3} \right). \label{jp:S_finiteLD}
	\end{align}
For small $\b$, the $\b$ dependence cancels out of the RHS of \eref{jp:disprelation_isotropic}.  Hence, zonostrophic instability is possible even with $\b=0$.  For concreteness, one might take $m=1$, for which $S$ simplifies to 
	\begin{equation}
		S(\chi,n,1) = \frac{1}{n^3 \chi} \left( -1 + \sqrt{1 + \frac{n^4}{4}} \right) +\O\left( \chi^{-3} \right).
	\end{equation}
Additionally, the small $q$ limit of \eref{jp:S_finiteLD} is
	\begin{equation}
		S(\chi,n,m) = \frac{n}{\chi} \frac{m}{2(1+m)^2} + \cdots.
	\end{equation}
Thus, for an isotropic spectrum and finite $L_d$, $S$ goes as $q^2$ at small $q$, rather than like $q^4$ as in the case of infinite $L_d$.

These issues will be reexamined from another light in \ref{jp:app:dispscale}, where we give some physical understanding of the transfer of energy to long wavelengths. 

\subsubsection{Instability of a Primary Wave to a Secondary Wave}
\label{jp:sec:parametricinst}
Zonostrophic instability can be understood in a very general way as the instability of some turbulent background spectrum to a (zonally symmetric) coherent mode.  As a special case, one can consider the background spectrum to consist of only a single mode.  We show that in this case the dispersion relation of zonostrophic instability reduces exactly to that of the 4-mode modulational instability (sometimes called parametric instability).  This correspondence was first noted by \citet{carnevalemartin1982} but was not understood in the context of the generation of zonal flows.

The stability of a single, primary wave $\v{p}$ to perturbations is a problem that has received attention in the past \cite{lorenz1972,gill1974,krommes2006,connaughtonnadigaetal2010,gallagherhnatetal2012}.  These calculations have used the fluctuating dynamical equations such as \eref{jp:vorticityeqngeneral} and not a statistically averaged system.  Generally one considers the unforced, undamped case, for which a single wave is an exact solution of the nonlinear dynamical equations.  Conceptually similar is the so-called secondary instability, where a growing, primary eigenmode gives rise to a secondary mode \cite{rogersdorlandetal2000,plunk2007,pueschelgorleretal2013}.  When the secondary mode grows much faster, the primary mode is treated as a stationary background.  These secondary instabilities can be more complicated, where, for example, the toroidal geometry of magnetically confined plasmas results in the growing primary eigenmode having nontrivial spatial dependence.

To calculate the stability of the primary wave using \eref{jp:vorticityeqngeneral}, in general one needs to retain an infinite number of coupled, perturbing modes.  However, typically one truncates the system, for example retaining a secondary mode $\v{q}$ and the sideband pair $\v{p} \pm \v{q}$.  Within this 4-mode approximation and the further assumption that the primary has $p_y=0$ such as a pure Rossby or drift wave and the secondary has $q_x=0$, the dispersion relation for 4-mode modulational instability is given by \cite{connaughtonnadigaetal2010}
	\begin{equation}
		\l'^3 = \l' s^4 \left( \frac{ 2M^2 (1-s^2)(1+s^2+f)(1+f)^2 - (s^2+f)}{ (1+f)^2 (1+s^2+f)^2 (s^2+f) } \right),
		\label{jp:disprelation_connaughton}
	\end{equation}
where $\l' = p \l / \b$, $s=q/p$, $f = p^{-2} \LD$, $M = \psi_0 p^3 / \b$, and $\psi_0$ is the amplitude of the background stream function.

Some studies investigated this phenomenon by using a form of CE2 where the inhomogeneity is assumed to vary slowly in space compared to the turbulence \cite{dyachenkonazarenkoetal1992,maninyuetal1994,dubrullenazarenko1997,smolyakovdiamondetal2000b,wordsworth2009}.  With that assumption, the turbulence is described by a wave kinetic equation (see Section 5.1.1).  The wave kinetic equation can also be recovered from the CE2 equation \eref{jp:CE2W} by assuming $\partial_\ybar \ll \partial_y$, Taylor expanding, then Fourier transforming.\footnote{The proper dependent variable to use for the wave action (see section 5.1.1.4.2) is $\mathcal{N}(\v{k} \mid \ybar) = (1 - \azf \ol{k}^{-2} \LD ) W(\v{k} \mid \ybar)$.  For the CHME, this becomes $\mathcal{N} = (k^2 / \kbsq) W$.  With $\mathcal{N}$ as the dependent variable, the disparate-scale form of CE2 assumes wave-kinetic form.}  While those previous studies are limited to the regime of small $q$, the CE2 framework makes no assumption about the length scale of the inhomogeneity.  Moreover, the previous studies did not draw a direct connection between the results from the statistical calculation and from the 4-mode calculation.\footnote{One reason a connection may not have been made is that the small-$q$ results in \citet{maninyuetal1994} and \citet{smolyakovdiamondetal2000b} based on the wave kinetic equation are incomplete.  In their dissipationless ($\m=0$) formulation, they neglect the term $2i \b q k_x k_y$ compared to $\l$ in the denominator of \eref{jp:lambda_pm}.  But this is invalid if $\l \sim q^2$ because the neglected term is larger than the retained term.  For example, when specialized to a single primary mode, both papers state that for the CHME, instability occurs when $p_x^2 + \LD - 3p_y^2 > 0$, and that $\l \sim q^2$.  One can obtain this result from the small $q$ limit of \eref{jp:primarymode_dispersionrelation} if the $\b$ term is unjustifiably neglected.  Careful analysis shows this result also obtains, correctly, in the $\psi_0 \to \infty$ limit.  But contrary to statements made by \citet{connaughtonnadigaetal2010}, the wave-kinetic formalism is \emph{not} restricted to that large-amplitude regime.  If the $\b$ term is retained, the full answer at small $q$ can be recovered from the wave-kinetic formalism.}

This dispersion relation \eref{jp:disprelation_connaughton} can be recovered from CE2 and the zonostrophic instability dispersion relation \eref{jp:dispersionrelation}.  To precisely compare, one must carefully select the background spectrum $W_H$ to correspond to a wave of stream function $\psi_0$.  If the initial background amplitude of mode $\v{p}$ is $\psi_0$, then we write
	\begin{equation}
		\psi(x,y) = \psi_0 \left( e^{i \v{p} \cdot \v{x} - i\w t} + e^{-i\v{p}\cdot\v{x} + i\w t} \right)
	\end{equation}
\ref{jp:app:wavecorr} shows that this corresponds to a one-time, two-point covariance of streamfunction
	\begin{equation}
		\Psi_H(k_x,k_y) = (2\pi)^2 \psi_0^2 \bigl[ \de(\v{k} - \v{p}) + \de(\v{k} + \v{p}) \bigr].
	\end{equation}
From \eref{jp:WfromPsi}, the corresponding covariance of vorticity is given by $W_H(k_x,k_y) = \ol{k}^4 \Psi_H(k_x,k_y)$, and thus, because of the delta functions,
	\begin{equation}
		W_H(k_x,k_y) = (2\pi)^2 A \bigl[ \de(\v{k} - \v{p}) + \de(\v{k} + \v{p}) \bigr] \label{WH_from_psi0}
	\end{equation}
where we have defined $A = \psi_0^2 \bigl(p^2 + \LD \bigr)^2$.  There are two ways of achieving this background spectrum.  First, we could choose the external forcing to be $F(\v{k}) = 2\m W_H$.  Since we want the dissipation term $\m$ to disappear in the final expression, $\m$ can be chosen to be vanishingly small, in particular smaller than the eigenvalue $\l$.  Alternatively, as previously mentioned we could take the external forcing and the dissipation to be zero, in which case any arbitrary homogeneous spectrum trivially satisfies the CE2 equations.  This latter point of view is closer to the traditional primary wave stability calculations.

Substituting \eref{WH_from_psi0} into \eref{jp:dispersionrelation}, we find
	\begin{align}
		\frac{\qbsq}{q^2} \l = & 2q A p_x^2 \left( 1 - \frac{\qbsq}{\pbsq} \right) \left( \frac{p_y + \frac12 q}{\l \pbsq_{+} \pbsq + 2i\b q p_x(p_y + \frac12 q)} \right. \notag \\
			& - \left. \frac{p_y - \frac12 q}{\l \pbsq_{-} \pbsq + 2i\b q p_x(p_y - \frac12 q)} \right) \label{jp:primarymode_dispersionrelation}
	\end{align}
where dissipation has been neglected, $p_{\pm}^2 = p_x^2 + (p_y \pm q)^2$, and $\pbsq_{\pm} = p_{\pm}^2 + \LD$.

When specialized to the case of a primary wave with $p_y=0$, the dispersion relation becomes
	\begin{equation}
		\frac{\qbsq}{q^2} \l = 2q A p_x^2 \left( 1 - \frac{\qbsq}{\pbsq} \right) \frac{q}{2} \frac{2 \l \pbsq_{+} \pbsq}{\l^2 \ol{p}_{+}^4 \ol{p}^4 + \b^2 q^4 p^2}.
	\end{equation}
Now, taking $\azf=1$ and introducing the same normalizations as used in \eref{jp:disprelation_connaughton}, we obtain
	\begin{equation}
		\frac{s^2 + f}{s^2} \l' = \frac{ 2 A s^2 \l' (1-s^2) (1+s^2+f)}{ (\b/p)^2 [ \l'^2 (1 + s^2 + f)^2(1 + f)^2 + s^4]}.
	\end{equation}
Letting $A' = p^2 A / \b^2$, after some simplification we find
	\begin{equation}
		\l'^3 = \l' s^4 \left( \frac{ 2A' (1-s^2) (1+s^2+f) - (s^2+f) }{ (1+f)^2 (1+s^2+f)^2 (s^2+f) } \right).
		\label{secondary_disprelation_zonal}
	\end{equation}
Since $A' = p^6 \psi_0^2 (1+f)^2 / \b^2 = M^2 (1+f)^2$, this exactly matches the dispersion relation given by \citet{connaughtonnadigaetal2010} in \eref{jp:disprelation_connaughton}.
	
In the above calculation, we have shown that from CE2 we recover the 4-wave modulational instability in the special case of a primary wave with $p_y=0$ and a secondary wave with $q_x=0$.  In \ref{jp:app:arbwave} we generalize this to show that CE2 recovers the 4-wave modulational instability for an arbitrary primary wave and an arbitrary secondary wave.

It may be at first surprising that the two dispersion relations agree exactly, but retrospectively it makes sense.  The 4-wave modulational instability contains the primary wave $\v{p}$ and the perturbations at wave vectors $\v{q}$ and $\v{p} \pm \v{q}$.  From \eref{jp:W_manywaves} in \ref{jp:app:wavecorr} for the correlation between the primary mode $\v{k} = p \unit{x}$ and sidebands $\v{k'} = p \unit{x} \pm q \unit{y}$, we see that the spatial dependence of the correlation goes as $\cos( px \pm \tfrac12 qy \pm q \ybar)$.  Upon examining the CE2 calculations, we see that the retained modes are the zonal flow $\de U e^{\pm i q \ybar}$ (which corresponds to mode $\pm \v{q}$) and the perturbations to the spectrum $\de W(k_x,k_y) e^{\pm i q \ybar}$.  The perturbation $\de W(k_x,k_y)$ is proportional to $W_H(k_x, k_y \pm \frac12 q)$, which is nonzero at $k_x = p$ and $k_y = \pm \tfrac12 q$ for the given primary mode.  Therefore the perturbations kept within CE2 are precisely the corresponding modes kept in the 4-mode truncation.  The CE2 instability calculation neglects higher harmonics of $\v{q}$ such as $e^{2iq\ybar}$ at the linear level.  These higher harmonics are precisely what is neglected by truncation to 4 modes instead of retaining higher sidebands.

In summary, the instability of a single primary mode can be thought of as a special case of the instability of an arbitrary background spectrum.  In the fluctuating dynamical equations it is difficult to represent a turbulent spectrum as an exact solution and hence calculate its stability.  For this purpose, a statistical formulation such as CE2 is advantageous, since a homogeneous turbulent background can be represented more easily, particularly as a time-independent spectrum.

When the homogeneous state is unstable, it is not obvious \emph{a priori} what happens to the zonal flows.  Do the zonal flows grow and saturate into a steady state?  Or do they fluctuate turbulently, unable to persist in a steady state?  Even though the answer is not obvious, within the Hasegawa--Mima or barotropic vorticity equation framework numerical simulations sometimes find steady zonal flows.  But in other situations there may be fluctuating zonal flows.  Complicated nonlinear physics determine what happens and it is difficult to determine what actually occurs without simulations.

Another advantage of CE2 in particular is that it is possible to calculate not only the instability of a turbulent background, but also how the instability saturates.  Analytic solutions are even possible in some regimes.  This is undertaken in section \ref{jp:sec:bifurcation}.

\subsection{Pattern Formation}
In this first half of this article, we discussed the tendency for zonal flows to grow if they are not already present.  Now, we consider what happens to such zonal flows beyond the initial stages of the instability.  As the zonal flows grow larger, they reach an amplitude where they significantly modify the turbulence.  Eventually, some kind of equilibrium between the turbulence and the zonal flows is reached.  It is this saturated state that is of main interest in understanding the observable turbulence.

Compared to zonal flow generation, the problem of zonal flow saturation has received much less attention in the theoretical literature.  Many of the works that have considered it typically make an assumption of scale separation where the scale or wavelength of the zonal flows is much larger than the scale of the turbulence \cite{diamondrosenbluthetal1998,connaughtonnazarenkoetal2011}.  Zonal flows in plasmas, however, are often observed to be of comparable scale to the turbulence \cite{guptafoncketal2006,fujisawaitohetal2004}.  Another line of inquiry is based on potential vorticity mixing (see Section 4.2).

The CE2 equations provide a well-posed nonlinear system whereby the saturation of zonal flows can be investigated.  As discussed at length previously, CE2 can describe the generation of zonal flows through zonostrophic instability.  But it can also describe the statistically steady inhomogeneous turbulence that results.

Numerical simulations of CE2 have also been performed \cite{farrellioannou2003,farrellioannou2007,farrellioannou2009,bakasioannou2013b,constantinoufarrelletal2013,marstonconoveretal2008,tobiasdagonetal2011,tobiasmarston2013} (see also sections 5.1.2, 5.2.2, and 5.2.4).  Simulations of statistical equations, which evolve in time the covariances of the fluctuating fields, are distinct from conventional DNS, which evolve the amplitudes.  The CE2 simulations have explored zonal flow physics in interesting ways distinct from DNS.

In addition, and especially relevant for this section, the CE2 simulations have yielded important information that inform our analytic calculations.  First, the simulations confirm that CE2, like both the quasilinear and original systems, exhibit zonal flows that can reach a steady state.  Second, within the CE2 simulations one also sees the phenomenon of merging jets, which is ubiquitous in DNS but has yet to be fully understood.  Third, \citet{farrellioannou2007} have found that CE2 can exhibit nonunique solutions, where the number of jets in the steady state depends on initial conditions.  They also discovered that zonal flows emerge from homogeneous turbulence in a bifurcation triggered by zonostrophic instability, and furthermore that the bifurcation is supercritical.  Not only do these features guide our calculations, they also demonstrate that CE2 possesses at least some of the essential physics of zonal flows as well as interesting and relevant nonlinear behavior.  CE2 is therefore a system worthy of further understanding.

In this second half of the section, we show that zonal flows can be understood as pattern formation.

\subsubsection{Bifurcation of homogeneous turbulence}
\label{jp:sec:bifurcation}
Zonostrophic instability, which was discussed previously in Section \ref{jp:sec:zi}, provides the starting point for our theoretical considerations.  In some parameter regime, the coherent perturbations are stable and homogeneous turbulence is stable.  But as a control parameter $\r$, such as the friction $\m$ or the forcing strength $F$, is adjusted, the homogeneous state becomes unstable \cite{farrellioannou2007,srinivasanyoung2012}.  On either side of this instability threshold, or bifurcation point, the behavior of the system must be qualitatively different.  Numerical simulations show that beyond the threshold the result is steady zonal flows and inhomogeneous turbulence.

In mathematical terms, the statistical CE2 equations \eref{jp:CE2} possess translational symmetry $\ybar \to \ybar + \de \ybar$.  In other words, if $\{W(x,y \mid \ybar,t), U(\ybar,t) \}$ is a solution, then $\{W(x,y\mid \ybar + \de \ybar,t ), U(\ybar + \de \ybar, t) \}$ is too.  Quite separately, when the system is zonostrophically stable, the homogeneous \emph{solution}, \eref{jp:WH}, manifests this symmetry by being itself invariant to that transformation.  When the control parameter crosses the instability threshold, the system suddenly develops dependence on $\ybar$ as well as a mean field.  The new solution is not invariant under translation.

In order to fully understand the behavior of the system, analytic solutions would be beneficial in addition to numerical solutions.  But the complexity of the nonlinear CE2 equations means that finding a solution analytically is a formidable task and does not appear feasible in general.  One way to proceed is by considering a regime where additional approximations can be made.  Our approach is to investigate near the bifurcation point.  The distance from the bifurcation point serves as a small parameter and facilitates further progress.

The bifurcation analysis follows a standard procedure and involves a multiscale perturbation expansion about the threshold \cite{crosshohenberg1993,hoyle2006,crossgreenside2009}.  Since the bifurcation is supercritical, only the lowest-order terms in the bifurcation analysis are needed to provide saturation of the instability.  The instability is known in the pattern formation literature as a Type I$_{\rm s}$ instability.  This type of bifurcation generically consists of a symmetry-breaking instability, a spatially periodic but temporally nonoscillatory marginal eigenvector, and a supercritical transition.

We review the basic procedure of the perturbation expansion in an abstract notation.  The full details are reported elsewhere \cite{parkerkrommes2013b}.  Consider a system with quadratic nonlinearity.  Let $\phi$ be an abstract vector, $\Lambda$ be a linear operator, $N$ be a symmetric, bilinear operator, and $F$ be external forcing.  Let $\epsilon = (\rho - \rho_c) / \rho_c$ be the normalized bifurcation parameter.  Any of $\Lambda$, $N$, and $F$ may depend explicitly on $\e$.  The basic equation is assumed to be given as
	\begin{equation}
		0 = \Lambda \phi + N(\phi,\phi) + F.
	\end{equation}
Given a nonzero equilibrium $\phi_e$, we change variables by letting $\phi = \phi_e + u$ to obtain
	\begin{equation}
		0 = L u + N(u,u),
		\label{glderivation:abstractequation}
	\end{equation}
where $Lu$ = $\Lambda u + 2N(\phi_e,u)$.  In the context of the CE2 calculation, $\phi_e = \{W_H,0\}$ and $u=\{W-W_H, U\}$.  By assumption, the linearization $L$ around the equilibrium $\phi_e$ is stable for $\e<0$, neutrally stable at $\e=0$, and unstable for $\e>0$.

The perturbation procedure employs slowly varying space and time scales in a multiple-scale expansion.  We introduce the slow scales $Y=\e^{1/2} \ybar$ and $T = \e t$, then let $\partial_\ybar \to \partial_\ybar + \e^{1/2} \partial_Y$ and $\partial_t \to \partial_t + \e \partial_T$.  Using these, we expand the operators $L = L_0 + \e^{1/2} L_1 + \e L_2 + \e^{3/2} L_3 + \cdots$ and $N = N_0 + \e^{1/2} N_1 + \cdots$, and we expand the state vector $u = \e^{1/2} u_1 + \e u_2 + \cdots$\,.  Collecting terms of the same order, we obtain the equations at $\O\bigl(\e^{1/2}\bigr)$, $\O(\e)$, and $\O\bigl(\e^{3/2}\bigr)$:
	\begin{align}
		\O\bigl(\e^{1/2}\bigr): \qquad 0 &= L_0 u_1, \label{jp:eps12} \\
		\O(\e): \qquad 0 &= L_0 u_2 + L_1 u_1 + N_0(u_1,u_1), \label{jp:eps1} \\
		\O\bigl(\e^{3/2}\bigr): \qquad0 &= L_0 u_3 + L_1 u_2 + L_2 u_1 \notag \\
				& \qquad + 2 N_0(u_1,u_2) + N_1(u_1,u_1). \label{jp:eps32}
	\end{align}

At $\O\bigl(\e^{1/2}\bigr)$, \eref{jp:eps12} states that $u_1$ is a null eigenvector of $L_0$, meaning it has a zero eigenvalue.  The reality condition on $u$ restricts the form of $u_1$ to be
	\begin{equation}
		u_1 = A(Y,T) r + A(Y,T)^* r^*,
		\label{glderivation:u1form}
	\end{equation}
where $r \sim e^{i q_c \ybar}$ and its complex conjugate $r^*$ are the right null eigenvectors, and $A$ is the to-be-determined amplitude.  These eigenvectors are periodic in $\ybar$ with critical wave number $q_c$, which is the first wave number to go unstable as $\e$ crosses zero.  Once an inner product $(\cdot\, ,\cdot)$ is defined, then associated with the right null eigenvector $r$ is a left null eigenvector $l$ of $L_0$, which satisfies $( l, L_0 u ) = 0$ for any $u$.  The $\ybar$ dependence of $l$ is also $e^{i q_c \ybar}$.  As is common in perturbative procedures, the amplitude $A$ will be determined by nonlinearities occurring at higher order.

At $\O(\e)$, we first note that $L_1 u_1=0$ automatically.  This is because $q_c$ is marginally stable at the instability threshold: given a dispersion relation $\lambda(q,\e)$ as a function of wave number $q$ and control parameter $\e$, then both $\lambda(q_c,0)=0$ and $\partial \lambda / \partial q (q_c,0)=0$.  The former equality yields $L_0 u_1 = 0$, while the latter equality yields the condition $L_1 u_1=0$.  In order to ensure that a solution for $u_2$ exists, a solvability condition obtained by taking the inner product with the left null eigenvector must be satisfied.  Applying this to \eref{jp:eps1}, the solvability condition is 
	\begin{equation}
		\bigl(l, N_0(u_1,u_1)\bigr) = 0.
	\end{equation}
This solvability condition is automatically satisfied because the quadratic nonlinearity implies $N_0(u_1,u_1) \sim 1$ or $e^{\pm 2 i q_c \ybar}$, while $l \sim e^{i q_c \ybar}$, so the inner product $\bigl(l, N_0(u_1,u_1)\bigr)$ vanishes.  Therefore, given that a solution exists, one  may write $u_2$ as a linear combination of homogeneous and particular solutions:
	\begin{equation}
		u_2 = u_{2h} + u_{2p},
	\end{equation}
	where
	\begin{gather}
		u_{2h} = A_2(Y,T) r  + A_2(Y,T)^* r^*, \\
		L_0 u_{2p} = -N_0(u_1,u_1). \label{glderivation:u2peqn}
	\end{gather}
 Since we have not yet determined $A$, we must proceed to higher order.  Another unknown parameter $A_2$ has been introduced, but we will not need it in order to solve for $A$.

At $\O\bigl(\e^{3/2}\bigr)$, note that $L_1 u_{2h} = 0$ for the same reason that $L_1 u_1=0$.  Upon writing the solvability condition for \eref{jp:eps32}, one finds that several terms vanish, leaving
	\begin{equation}
		0 = (l, L_2 u_1) + \bigl(l, 2 N_0(u_1,u_{2p})\bigr).
		\label{jp:abstractsolvabilitycondition}
	\end{equation}
This is the desired partial differential equation that determines the amplitude $A$.  It turns out that one never explicitly needs $L_1$ or $N_1$ in order to obtain this equation.

We quote the results of the full analysis.  After returning to the unscaled variables, the amplitude equation for $A$ is
	\begin{equation}
		c_0 \partial_t A(\ybar, t) = \e c_1 A + c_2 \partial_\ybar^2 A - c_3 |A|^2 A,
		\label{jp:glequation}
	\end{equation}
where the $c_i$ are order unity, real constants.  All of the $c_i$ should be positive (negative $c_3/c_0$ corresponds to subcritical rather than supercritical instability).

Actually, one could have determined the form of \eref{jp:glequation} without going through the actual calculation \cite{crossgreenside2009}.  The symmetries inherent in the original equation constrain the forms of possible terms.  For instance, translational symmetry in $\ybar$ requires the amplitude equation to be invariant to phase shifts of $A$, so that the lowest-order nonlinear term is uniquely determined to be $|A|^2 A$.

Furthermore, the behavior of \eref{jp:glequation} is universal in the sense that, as long as all of the $c_i > 0$, the qualitative behavior does not depend on the value of any of the $c_i$.  This can be seen because a simple rescaling of $A$, $\ybar$, and $t$ eliminates $c_i$ dependence from the equation.

With \eref{jp:glequation}, the analogy between the zonal flows and the convection rolls in \RB convection is complete.  The transition to convection is governed by the same class of bifurcation and subject to the amplitude equation.  The similarities between zonal flows and convection rolls alluded to in section \ref{jp:sec:zfrbanalogy} are not merely descriptive, but mathematical as well.

The formulas for the $c_i$ are complicated but are written in terms of the external parameters and integrals over the spectrum of the forcing.  The formulas given in full by \citet{parkerkrommes2013b}.  The important point is that it is possible to find a complete solution to the nonlinear CE2 equations, at least in a certain regime.

There is an alternate method for obtaining $c_0$, $c_1$, and $c_2$, which govern the linear behavior of \eref{jp:glequation} for small $A$.  The dispersion relation \eref{jp:dispersionrelation} can be put into the form $D(\l,\e,q)=0$ and can be Taylor expanded about the threshold.  The conditions of the instability threshold require $D(0,0,q_c)=0$ and $\partial D/\partial q(0,0,q_c)=0$.  Upon expanding $D$ to lowest order about $(0,0,q_c)$, one finds
	\begin{equation}
		-\pd{D}{\l}(0,0,q_c)\, \l = \e \pd{D}{\e}(0,0,q_c) + \frac{1}{2} \pdd{D}{q}(0,0,q_c) (q-q_c)^2.
	\end{equation}
Up to a constant of proportionality, we identify $c_0 = -\partial D/\partial \l(0,0,q_c)$, $c_1 = \partial D/\partial \e(0,0,q_c)$, and $c_2 = -\frac12 \partial^2 D/ \partial q^2 (0,0,q_c)$.  This provides an independent check on the multiple-scale expansion calculation.  However, this approach does not give $c_3$; for that one needs the full bifurcation calculation which includes nonlinear terms.

Desired quantities of interest can be calculated analytically from \eref{jp:glequation}.  Linearizing about $A=0$ and substituting the form $A \sim e^{\l t} e^{ik\ybar}$, one calculates the growth rate to be
	\begin{equation}
		\l = \frac{ \e c_1 - c_2 k^2 }{c_0}.
		\label{jp:glderivation:analyticgrowthrate}
	\end{equation}
We recognize from \eref{glderivation:u1form} that $k$ is the wave number relative to $q_c$ so that $k=q-q_c$.  Steady state solutions including the nonlinear term also have the form $A=A_s(k) e^{ik\ybar}$, where
	\begin{equation}
		|A_s(k)| = \left( \frac{\e c_1 - c_2 k^2}{c_3} \right)^{1/2}.
		\label{jp:glderivation:zfamplitude}
	\end{equation}
The lowest-order correction to $\l$ is $\O(\e^2)$, while the lowest-order correction to $A_s$ is $\O(\e)$.  

Figure \ref{jp:fig:GL_coeffs} verifies that \eref{jp:glequation} provides an adequate description of CE2 near the instability threshold.  The analytical growth rate found from \eref{jp:glderivation:analyticgrowthrate} is compared with that from the exact dispersion relation \eref{jp:dispersionrelation}.  Similarly, the analytical zonal flow amplitude found from \eref{jp:glderivation:zfamplitude} is compared with that from solving the full CE2 system as in section \ref{jp:sec:numericalsoln}.  We identify the amplitude $A_s$ of the first harmonic $e^{iq_c\ybar}$ with the numerically determined coefficient $U_1$.  The results are in excellent agreement.

	\begin{figure}
		\figurebox{3.1in}{}{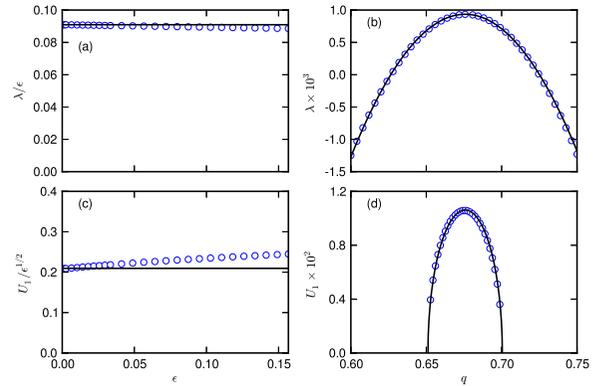}
		\caption{Comparison showing agreement between numerical solution (blue circles) and analytic solution (black line). (a)  Compensated growth rate $\l/\e$ as a function of $\e$ at $q=q_c$.  (b) Growth rate $\l$ as a function of $q$ at $\e=0.01$.  (c) Compensated zonal flow amplitude $U_1/e^{1/2}$ as a function of $\e$ at $q=q_c$.  (d) Zonal flow amplitude $U_1$ as a function of $q$ at $\e=0.0025$.  The zonal flow amplitude $U_1$ is the first Fourier component of the zonal flow velocity $U(\ybar).$  (Adapted from New J. Phys. \textcopyright 2014)}
		\label{jp:fig:GL_coeffs}
	\end{figure}
	
In addition to finding the steady states of the amplitude equation, one can ask whether those steady states are stable to small perturbations.  Linear stability analysis about the solution $A_s e^{ik\ybar}$ shows that it is unstable to the Eckhaus instability when $k^2 > \e c_1/ 3c_2$ \cite{crossgreenside2009}.

A stability diagram representing the various possibilities is shown in Figure \ref{jp:fig:gl_stability}.  The neutral curve (N) indicates marginal stability of the $A=0$ solution as a function of the wave number $k$ and control parameter $\e$.  The $A=0$ solution is unstable to those $k$ that are inside the neutral curve.  At a fixed $\e>0$, steady-state solutions with $A \neq 0$ exist at any of the $k$ inside the neutral curve.  The marginal stability of these $A \neq 0$ solutions is indicated by the Eckhaus curve (E).  Inside the E curve is a smaller band of wave numbers for which the steady-state solutions are stable.

If a solution with an unstable wavelength is slightly perturbed, it must evolve to reach a stable wavelength.  The plot of $\Re A(\ybar,t)$ in Figure \ref{jp:fig:merging_jets_amplitude}, with branches merging into wider branches, resembles similar plots of the zonal flow $U(y,t)$ in which jets merge.  In the amplitude equation \eqref{jp:glequation}, the merging occurs in the nonlinear stage of the Eckhaus instability.  At the instant of merging there is a topological defect known as a dislocation \cite{crossgreenside2009}.

	\begin{figure}
		\figurebox{3.1in}{}{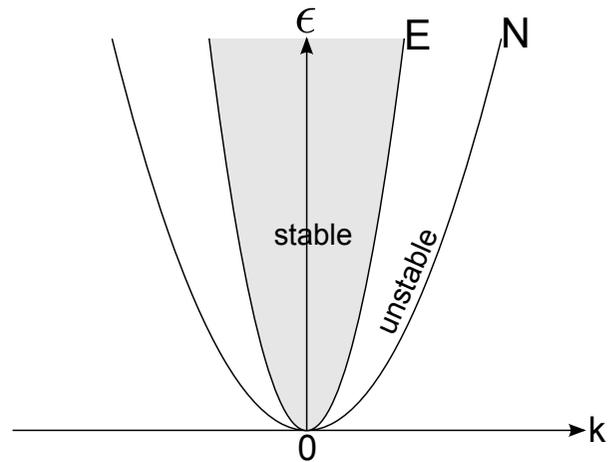}
		\caption{Stability diagram for the amplitude equation \eqref{jp:glequation}.  The labels `stable' and `unstable' refer to the nonzero-$A$ steady states.}
		\label{jp:fig:gl_stability}
	\end{figure}
	
	\begin{figure}
		\figurebox{}{}{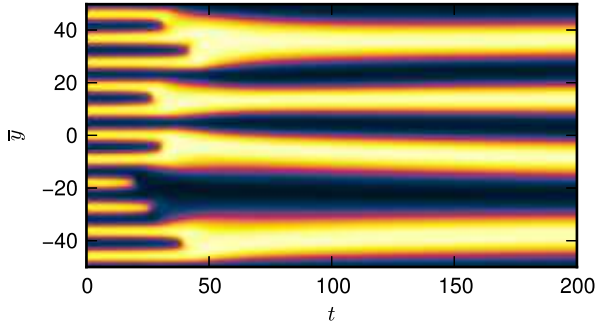}
		\caption{Merging behavior in the amplitude equation \eqref{jp:glequation} [$\Re A(\ybar,t)$ is shown].  (From New J. Phys. \textcopyright 2014)}
		\label{jp:fig:merging_jets_amplitude}
	\end{figure}
	
When the CE2 system is far from threshold, the amplitude equation ceases to be a quantitatively accurate description.  However, many of the basic behaviors just described about the amplitude equation hold also for steady solutions of the CE2 equations, as we now verify by numerical solution.

\subsection{Numerical Solution of CE2 and Stability Diagram}
\label{jp:sec:numericalsoln}
In general, CE2 must be numerically solved.  One approach is to evolve the CE2 equations in time until an equilibrium is reached.  Our approach differs in that we solve the steady-state limit directly, i.e., \eref{jp:CE2} with $\partial/\partial t = 0$.  We find steady-state solutions of zonal flows and turbulence using numerical techniques developed by Busse and Clever \cite{busse1967,cleverbusse1974,busseclever1979} for the \RB convection problem.  This method of solution uses a Galerkin expansion, where the dynamical variables are expanded in basis functions with unknown coefficients and substituted into the equations of motion.  The equations of motion are then projected onto the basis functions, yielding a set of nonlinear algebraic equations for the coefficients.  

The covariance of the turbulence and the zonal flow amplitude are expanded as
	\begin{gather}
		W(x,y \mid \ybar) = \sum_{m=-M}^M \sum_{n=-N}^N \sum_{p=-P}^P W_{mnp} e^{imax} e^{inby} e^{ipq\ybar}, \\
		U(\ybar) = \sum_{p=-P}^P U_p e^{ipq\ybar},
	\end{gather}
where $q$ is the fundamental wavenumber or $2\pi/q$ is the spatial periodicity of the zonal flows.  There is a range of $q$ that allows a solution.  We obtain a system of nonlinear algebraic equations for the coefficients $W_{mnp}$ and $U_p$ by substituting the Galerkin series into \eref{jp:CE2} and projecting onto the basis functions.  To demonstrate the projection for \eref{jp:CE2W}, let $\p_{mnp} = e^{imax} e^{inby} e^{ipq\ybar}$.  We project \eref{jp:CE2W} onto~$\p_{rst}$ by operating with 
	\begin{equation}
		\left(\frac{2\pi}{a} \frac{2\pi}{b} \frac{2\pi}{q}\right)^{-1}  \int_{-\pi/a}^{\pi/a} dx   \int_{-\pi/b}^{\pi/b} dy 	\int_{-\pi/q}^{\pi/q} d\ybar\,  \p_{rst}^*.
	\end{equation}
Projection of the first term, $(U_+ - U_-) \partial_x W$, yields $I_{rstp'mnp} U_{p'} W_{mnp}$, where repeated indices are summed over, $I_{rstp'mnp} = ima \de_{m,r} \de_{p'+p-t,0} (\s_+ - \s_-)$, $\s_{\pm} = \mathrm{sinc} ( \a_{\pm}\pi/b )$, and $\a_{\pm} = nb - sb \pm \frac12 p'q$.  The other terms of \eref{jp:CE2W}, as well as \eref{jp:CE2U}, are handled similarly.  In total, we generate as many equations as there are coefficients.

The system of nonlinear algebraic equations is solved with Newton's method.  One feature of Newton's method is that it requires a good initial guess.  We attain a suitable guess by using the bifurcation solution near threshold.  Then we adjust a parameter in small increments towards the desired value, a technique known as numerical continuation.  The solution at the previous value of the parameter can serve as the initial guess for the next value.

Once a steady-state solution is found, its stability can be assessed.  The general method, again following Busse and Clever, involves linearizing the equations of motion about the steady state.  Since the equilibrium is periodic in $\ybar$, the perturbation may be expressed as a Bloch state.  Then the perturbation is expanded in the same Fourier-Galerkin basis functions used to express the equilibrium.  The equilibrium is unstable if there are any eigenvalues with positive real part \cite{crossgreenside2009}.  Further details of the numerics regarding the equilibrium and stability may be found in \citet{parkerkrommes2013a,parkerkrommes2013b}.

In the same manner as for the amplitude equation, the results are organized into a stability diagram.  Figure \ref{jp:fig:CE2_stability_diagram} displays the stability diagram for the CE2 system with infinite deformation radius.  The control parameter on the $y$~axis is $\g = \ve^{1/4} \b^{1/2} \m^{-5/4}$, a fundamental parameter controlling the jet dynamics \cite{danilovgurarie2004,galperinsukorianskyetal2010,scottdritschel2012,tobiasmarston2013,bouchetnardinietal2013}.  This parameter is related to the zonostrophy parameter $R_\b$ by $\g = R_\b^5$.  Near the instability threshold, the stability diagram resembles that for the amplitude equation (see Figure \ref{jp:fig:gl_stability}), as it should.  The Eckhaus (E) instability forms the stability boundaries near the threshold in the sense that if one starts inside the stable region and increases or decreases $q$, the Eckhaus instability is the first instability triggered unstable.  Farther from threshold, at larger $\g$, other instabilities form the boundary (L$_1$ and R$_1$ in the diagram).  These other instabilities have not yet been studied in detail.

In Figure 1.2 of Section 5.2.2, Farrell and Ioannou show a similar stability diagram for the $\beta$ plane.  Their statistical approach, called S3T, is mathematically equivalent to CE2 although a different coordinate system and numerical method are used in practice in the computations.  Unlike our numerical method, which develops problems at larger values of $\g$, their method has no problem achieving values of $\g$ far from the critical value.  In their figure, as the strength of the forcing is increased (which corresponds to increasing $\g$) well beyond the critical value, the region of stability curves to the left toward small wavenumbers or larger jets.  Such is the behavior qualitatively expected in order to follow the Rhines scaling.

	\begin{figure}
		\figurebox{3.1in}{}{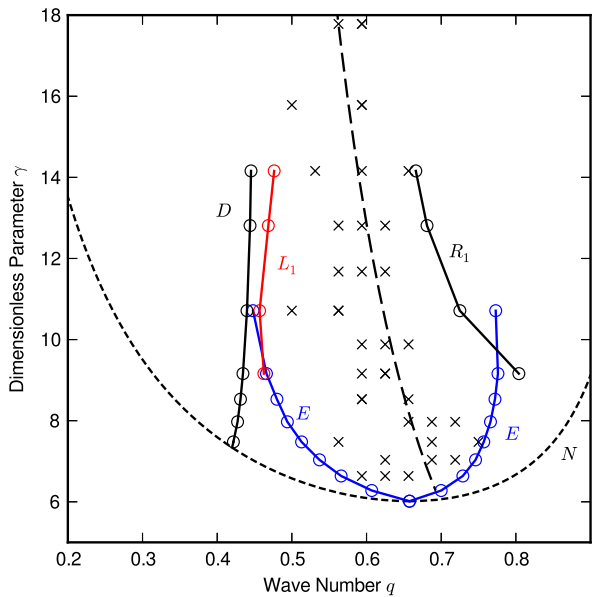}
		\caption{Stability diagram for the CE2 equations.  For $\g$ above the bottom of the neutral curve (N), the homogeneous turbulent state is zonostrophically unstable and  the result is inhomogeneous turbulence with zonal flows .  Ideal states are stable within the marginal stability curves E, L$_1$, R$_1$.  The stability curve is consistent with the dominant zonal flow wave number from independent QL simulations (crosses).  The stationary ideal states vanish to the left of curve $D$.  The black dashed line depicts the Rhines wave number. (Adapted from New J. Phys. \textcopyright 2014)}
		\label{jp:fig:CE2_stability_diagram}
	\end{figure}

\subsection{Summary}
In the first part of this article, we joined numerous other authors in offering a perspective on the generation of zonal flows.  We found a deep connection between the stability of a single wave and the zonostrophic instability of homogeneous turbulence.  In particular, the 4-wave modulational instability can be recovered exactly as a special case of zonostrophic instability within the CE2 formalism.  In addition to a single wave, we also examined the case when the background spectrum is isotropic.  When the deformation radius is finite, there are some notable differences in the physics of eddy forcing of zonal flows, especially for long-wavelength jets.

In the second part, we considered zonal flows beyond the initial stages of growth and asked how they saturated into a steady state.  We described zonal flows as pattern formation amid a bath of turbulence.  A deep understanding of the spontaneous symmetry breaking of statistical homogeneity attained through the CE2 framework reveals behaviors such as the existence of multiple solutions with different jet wavelengths and the phenomenon of jet merging to reach a stable wavelength.  These features have been observed in simulations.

The pattern formation view of zonal flows is quite general.  It possesses a far broader scope than the minimalistic 2D models considered here.  The behaviors predicted by the amplitude equation should be expected any time there is a spontaneous symmetry breaking with the appearance of steady zonal flows.  For example, in a generalization of the Hasegawa--Mima equation that includes a resistive instability, some of the expected features occur along with zonal flows \cite{numataballetal2007}.

We have been emphasizing the role of symmetry breaking.  But in reality, a $\b$ plane does not exist.  Moving to a more physical model such as the surface of a rotating sphere destroys the north--south translational symmetries associated with a $\b$ plane.  Do any of these results apply to zonal flows in spherical geometry?  Although this question should be studied in detail, we offer one possibility.  Due to the latitudinal variation of the Coriolis parameter, the turbulence is always inhomogeneous on the sphere.  A transition from homogeneous to inhomogeneous turbulence is not the right description, but perhaps some type of transition may still occur.  Besides for the development of inhomogeneity, another aspect of the bifurcation on a $\b$ plane is the spontaneous formation of a mean field, i.e., the zonal flow.  We suggest that this mean-field generation may survive for flow on a rotating sphere, and would be observable as a control parameter is varied.  The zonal flow still behaves as an order parameter in this more general type of scenario.



\endgroup

%% file: modtext.tex
%
%
%
%
%
%
%
%

\setcounter{subsection}{0}

\let\AppendixName\empty

\def\thesubsection{\AppendixName\thesection.\Alph{subsection}}

\let\subsectiono\subsection

\def\subsection{\def\AppendixName{Appendix }\subsectiono}

\let\subsubsectiono\subsubsection

\def\subsubsection{\let\AppendixName\empty\subsubsectiono}

\subsection{Zonostrophic instability and the physics of
  disparate-scale interactions} 
\label{jp:app:dispscale}

\begingroup 

\newcommand*\degrees{}
\newcommand*\tensor{}
\newcommand*\term{}
\newcommand*\vr{}

\catcode`\@=11 

\newif\ifjournal

\renewcommand*\boldmath{\protect\mathversion{bold}} 
\newcommand*\Unskip{\unskip~}	

\newcommand*\Appabbrev{Appendix} 
\newcommand*\Appsabbrev{Appendixes}
\newcommand*\Appo[1]{\Appabbrev~\ref{#1}} 
\newcommand*\App[1]{\Appabbrev~\seco{#1}}
\newcommand*\Apps[1]{\Appsabbrev~\seco{#1}}
\newcommand*\APP[1]{\Unskip\ref{#1}}
\newcommand*\Appendix{\App}	

\newcommand*\eq[1]{\label{#1}}	
\newcommand*\Eq[1]{Eq.~(\ref{#1})}
\newcommand*\Eqs[1]{Eqs.~(\ref{#1})}
\newcommand*\EQo[1]{(\ref{#1})} 
\newcommand*\EQ[1]{\Unskip\EQo{#1}}
\newcommand*\Equation[1]{Equation~(\ref{#1})}	
\newcommand*\Equations[1]{Equations~(\ref{#1})}

\newcommand*\Figo[1]{Fig.~\ref{Fg.#1}}		
\newcommand*\Fig[2][]{Fig.~\seco[#1]{Fg.#2}}
\newcommand*\Figs[2][]{Figs.~\seco[#1]{Fg.#2}}
\newcommand*\Figso[1]{Figs.~\ref{Fg.#1}}
\newcommand*\FIG[1]{\seco{Fg.#1}}
\newcommand*\FIGo[1]{\ref{Fg.#1}}
\newcommand*\Figure[1]{Figure~\seco{Fg.#1}}
\newcommand*\Figureo[1]{Figure~\ref{Fg.#1}}
\newcommand*\Figures[1]{Figures~\seco{Fg.#1}}

\newcommand*\Fnabbrev{footnote}
\newcommand*\Fn[1]{\Fnabbrev~\seco{#1}}
\newcommand*\Footnote[1]{Footnote~\seco{#1}}	

\newcommand*\Ref[1]{Ref.~\Onlinecite{#1}}
\newcommand*\Refs[1]{Refs.~\Onlinecite{#1}}
\newcommand*\REF[1]{\Unskip\Onlinecite{#1}}
\newcommand*\REFo[1]{{\let\ \relax\Onlinecite{#1}}} 

\newcommand*\Onlinecite{\onlinecite} 

%

\newcommand*\seco[2][]{\ref{#2}#1}	
\newcommand*\secos[1]{\ref{#1}}

\newcommand*\Chap[1]{Chap.~\seco{#1}}
\newcommand*\Seco[1]{Sec.~\ref{#1}}	
\newcommand*\Sec[1]{Sec.~\seco{#1}}
\newcommand*\Secs[1]{Secs.~\secos{#1}}
\newcommand*\Section[1]{Section~\seco{#1}}
\newcommand*\SECo[1]{\seco{#1}}
\newcommand*\SEC[1]{\Unskip\SECo{#1}}

\newcommand*\Table[1]{Table~\seco{Tb.#1}}
\newcommand*\Tableo[1]{Table~\ref{Tb.#1}}	
\newcommand*\Tables[1]{Tables~\seco{Tb.#1}}

\newcommand*\Ex[1]{Ex.~\ref{#1}}

\newcommand*\figpath{}

\newcommand*\Ifarrayflag{\ifarrayflag}
\newcommand*\Ifletterflag{\ifletterflag}
\newcommand*\Ifbeginningeq{\ifbeginningeq}

\newcommand*\BE{\arrayflagfalse\beginningeqfalse\begin{equation}} 

\newcommand*\BEA{\arrayflagtrue\beginningeqfalse\begin{eqnarray}} 
\newcommand*\BA{\BEA} 
\def\BAams#1\EAams{\begin{align}#1\end{align}} 
\newcommand*\BEAo{\BE\noand}
\newcommand*\BAo{\BEAo} 

\newcommand*\EEA{\end{eqnarray}}
\newcommand*\EA{\EEA} 
\newcommand*\EEAo{\EE}
\newcommand*\EAo{\EEAo} 


\newcommand*\BAL[1][]{\letterflagtrue\st@rtarray[#1]}
\newcommand*\EAL{\end{eqnarray}\EM}


\newcommand*\EE{%
\Ifbeginningeq
	\beginningeqfalse
	\BE
\else
	\Endanequation
	\beginningeqtrue
\fi
\Ifletterflag
	\EM
\fi
}

\newcommand*\Endanequation{%
\Ifarrayflag
	\end{eqnarray}%
\else
	\end{equation}%
\fi
}

\newcommand*\thebasicequation{\arabic{equation}}

\newcommand*\BM[1][]{\begin{subequations}%
	\gdef\theletters{a}%
	\def\NP##1{##1%
		\ifx\df@label\@empty\else\@xp\ltx@label\@xp{\df@label}\fi
		\let\df@label\@empty
		\def\theequation{\theparentequation\theletters}%
		\stepcounter{equation}%
		\protected@edef\@currentlabel{\theparentequation\alph{equation}}%
		\xdef\theletters{\theletters,\alph{equation}}%
		}%
	\def\NQ##1{\xdef\theletters{\alph{equation}}%
		\NP{##1}}%
	\l@belletters{#1}%
	}



\newcommand*\l@belletters[1]{\def\Jtemp{#1}\ifx\Jtemp\empty\else\eq{#1}\fi}

\newcommand*\EM{\end{subequations}\COMMENT
	\letterflagfalse
	}

\newcommand*\NN{\nonumber}

\newcommand*\BI{\begin{itemize}}
\newcommand*\EI{\end{itemize}}



\newcommand*\verticalbar{|}



 \let\barunder\b

\newcommand*\vhacek[1]{{\accent20 #1}}


\newcommand*\al{\alpha}
\newcommand*\be{\beta}
\newcommand*\ch{\chi}
\newcommand*\de{\delta}
\newcommand*\De{\Delta}
\newcommand*\e{\epsilon}
\newcommand*\eps{\varepsilon}
\newcommand*\g{\gamma}
\newcommand*\G{\Gamma}
\newcommand*\h{\eta }
\newcommand*\io{\iota}
\renewcommand*\k{\kappa}
\newcommand*\la{\lambda}
\newcommand*\La{\Lambda}
\newcommand*\n{\nu}
\newcommand*\p{\phi}
\newcommand*\ps{\psi}
\newcommand*\Ps{\Psi}
\newcommand*\ph{\varphi}
\renewcommand*\r{\rho}
\newcommand*\s{\sigma}
\renewcommand*\th{\theta}
\newcommand*\Th{\Theta}
\newcommand*\ups{\upsilon}
\newcommand*\U{\Upsilon}
\newcommand*\w{\omega}
\newcommand*\W{\Omega}
\newcommand*\X{\Xi}
\newcommand*\y{\relax}
\newcommand*\z{\zeta}


\newcommand*\shortGreek{
	\renewcommand*\a{\alpha}
	\renewcommand*\b{\beta}
	\renewcommand*\c{\chi}
	\renewcommand*\d{\delta}
	\newcommand*\D{\Delta}
	\renewcommand*\l{\lambda}
	\renewcommand*\L{\Lambda}
	\renewcommand*\t{\tau}
	\renewcommand*\u{\upsilon}
}

\newcommand*\rcvr@tmp[1]{\edef\next{\global\let\noexpand#1\csname#1@tmp}\next}

\newcommand*\longGreek{
	\rcvr@tmp{a}
	\rcvr@tmp{b}
	\rcvr@tmp{c}
	\rcvr@tmp{d}
	\rcvr@tmp{D}
	\rcvr@tmp{l}
	\rcvr@tmp{L}
	\rcvr@tmp{t}
	\rcvr@tmp{u}
}
	
\shortGreek


\newcommand*\BIGavg[1]{\left\langle#1\right\rangle}

\renewcommand*\({\left(}
\renewcommand*\){\right)}

\newcommand*\lp{\left(}
\newcommand*\rp{\right)} 


\renewcommand*\[{\left[}
\renewcommand*\]{\right]}

\newcommand*\lb{\left[}
\newcommand*\rb{\right]} 





\newcommand*\?{}


\newcommand*\Abar{{\Bar A}}	
\newcommand*\Ahat{{\Hat{A}}}	
\newcommand*\Alfven{Alfv\'en}	
\newcommand*\Ampere{Amp\`ere}
\newcommand*\Ansatze{Ans\"atze}
\newcommand*\Arg{\Mathop{Arg}}	
\newcommand*\Arnold{Arnold}	
\newcommand*\Apar{A_\parallel}	
\newcommand*\Aperp{A_\perp}	
\newcommand*\alhat{{\Hat{\alpha}}}	
\newcommand*\ahat{\unit{a}}	
\newcommand*\abar{{\Bar\alpha}}	
\newcommand*\abso[1]{\verticalbar#1\verticalbar} 

\newcommand*\abs[1]{\mathopen{\ifBIG\left\fi 
	\verticalbar}
	#1\mathclose{\ifBIG\global\BIGfalse\right\fi
	\verticalbar}} 
\newcommand*\Abs[1]{\left\verticalbar#1\right\verticalbar}

\newcommand*\adhoc{\Latin{ad~hoc}}
\newcommand*\adj{^{\dagger}}	
\newcommand*\aka{\Latin{a.k.a}}
\newcommand*\ala{\Latin{a~la}}
\newcommand*\alphahat{{\widehat{\alpha}}}	
\newcommand*\aposteriori{\Latin{a~posteriori}}
\newcommand*\Apriori{\Latin{A~priori}}
\newcommand*\apriori{\Latin{a~priori}}

\newcommand*\At[1]{{}_{\big\vert#1}} 
\newcommand*\at[1]{_{\vert#1}}

\renewcommand*\Bar[1]{{\overline{#1}}}	
\newcommand*\BarC[1]{\underline{#1}}	
\newcommand*\BGK{Bernstein--Greene--Kruskal}
\newcommand*\BibTeX{Bib\TeX}
\newcommand*\BL{Balescu--Lenard}
\newcommand*\BLe{\BL\ equation}
\newcommand*\BLE{\BLe}		
\newcommand*\BS{Bethe--Salpeter}
\newcommand*\BSE{\BS\ equation}
\newcommand*\Bhat{\Hat{B}}	
\newcommand*\Bo{B_0} 		
\newcommand*\Bp{B_p} 		
\newcommand*\Bphi{B_\phi} 	
\newcommand*\Br{B_r} 		
\newcommand*\Bstar{B\conj}	
\newcommand*\Bth{B_\theta} 	
\newcommand*\Bz{B_z} 		
\newcommand*\backreaction{backreaction} 
\newcommand*\bBar{{\Bar b}}	
\newcommand*\bbar{{\Bar\beta}}  
\newcommand*\bfit{\rmfamily\bfseries\itshape} 
\newcommand*\bdot{\boldsymbol{\cdot}}	
\newcommand*\bhat{\unit{b}}	
\newcommand*\bohat{\bhat_0}	

\newcommand*\bigast{{\mathchoice{\asterisk}
	{\displaystyle\asterisk}
	{\textstyle\asterisk}
	{\scriptstyle\asterisk}
	}}

\newcommand*\bin[1]{\D_{#1}}	
\newcommand*\bo{b_0} 		
\newcommand*\boe{b_{0e}}	

\newcommand*\CLandau{C_{s,\sbar}\on{f} &= 2\pi\Fr{q^2,m}_s(\nbar
	q^2)_\sbar\ln\Lambda_{s,\sbar} 
		\Partial{}{\vv}\bdot\Int d\vvbar\,\mU(\vv-\vvbar)
\NN
\\
	&\qquad\bdot
		\(\fr{1,m_{\sbar}}\Partial{}{\vvbar}-\fr{1,m_s}\Partial{}{\vv} 
	\)\!f_s(\vv) f_\sbar(\vvbar)}

\newcommand*\Collhat{\Hat C_{s,\sbar}\,\chi &=
	-2\pi\Fr{nq^2,\nbar m}_s(n q^2)_\sbar\ln\Lambda
\NN\\
	&\quad{}\times \Partial{}{\vv}\bdot\[\fM(\vv)\Int
d\vvbar\,\fM(\vbar)\U(\vv-\vvbar)\bdot 
		\(\fr{1,m}\Partial{\chi}{\vv} - \fr{1,\Bar m}
		\Partial{\Bar\chi}{\vvbar}\)\]}

\newcommand*\CC{convective cell}
\newcommand*\CE{Chapman--Enskog}
\newcommand*\CK{Chapman--Kolmogorov}
\newcommand*\CKE{\CK\ equation}
\newcommand*\CM{center manifold}
\newcommand*\Ch{Chebyshev}
\newcommand*\Chat{{\Hat{C}}}	
\newcommand*\Chatk{\Chat_\vk}
\newcommand*\Circ{\uP{c}}
\newcommand*\Comp{^{\Rm C}}	
\newcommand*\Cbar{\Bar{C}}	
\newcommand*\Ck{C_\vk} 		
\newcommand*\Ctk{\Tilde{C}_\vk}	
\newcommand*\Cp{C_\vp} 		
\newcommand*\Cq{C_\vq} 		
\newcommand*\Cpp{C$++$} 	
\newcommand*\Ct{{\Tilde{C}}}
\newcommand*\cA{{\mathcal{A}}} 
\newcommand*\cB{{\mathcal{B}}}
\newcommand*\cC{{\mathcal{C}}}
\newcommand*\cD{{\mathcal{D}}}	
\newcommand*\cDt{\Tilde{\cD}}	
\newcommand*\cDhat{{\Hat{\cD}}}
\newcommand*\cDbar{{\Bar{\cD}}}
\newcommand*\cE{{\mathcal{E}}} 	
\newcommand*\cEstar{\cE_{\bigast}}	
\newcommand*\cEt{{\widetilde{\cE}}}	
\newcommand*\cEbar{{\Bar{\cE}}}		
\newcommand*\cEhat{{\skew6\widehat{\boldsymbol{\cE}}}} 
\newcommand*\cEk{\cE_\vk}	
\newcommand*\cF{{\mathcal{F}}}
\newcommand*\cFbar{{\Bar{\cF}}} 
\newcommand*\cFhat{{\skew6\Hat{\boldsymbol{\cF}}}} 
\newcommand*\cG{{\mathcal{G}}}	
\newcommand*\cH{{\mathcal{H}}} 	
\newcommand*\cHbar{\Bar{\cH}}
\newcommand*\cHm{\cH_{\Rm m}}	
\newcommand*\cI{{\mathcal{I}}}	
\newcommand*\cIhat{\Hat{\cI}}	
\newcommand*\cIt{\Tilde{\cI}}	
\newcommand*\cJ{{\mathcal{J}}}	
\newcommand*\cJk{\cJ_\vk}	
\newcommand*\ck{c_\vk}		
\newcommand*\cK{{\mathcal{K}}} 	
\newcommand*\cKt{{\Tilde\cK}}	
\newcommand*\cL{{\mathcal{L}}}	
\newcommand*\cLhat{\Hat{\cL}}	
\newcommand*\cLk{\cL_\vk}	
\newcommand*\cLt{{\Tilde\cL}}
\newcommand*\cM{{\mathcal{M}}}
\newcommand*\cN{{\mathcal{N}}}
\newcommand*\cNk{\cN_\vk}	
\newcommand*\cNkhat{\Hat{\cN}_\vk}
\newcommand*\cNp{\cN_\vp}
\newcommand*\cNphat{\Hat{\cN}_\vp}
\newcommand*\cNq{\cN_\vq}
\newcommand*\cNqhat{\Hat{\cN}_\vq}
\newcommand*\cO{{\mathcal{O}}}
\newcommand*\cOmega{\mathit{\Omega}}
\newcommand*\coworker{co-worker} 
\newcommand*\cP{{\mathcal{P}}}	
\newcommand*\cPt{\Tilde{\cP}}	
\newcommand*\cPbar{{\Bar{\cP}}}
\newcommand*\cp{\text{c.p.}}	
\newcommand*\cQ{{\mathcal{Q}}}
\newcommand*\cR{{\mathcal{R}}}
\newcommand*\cS{{\mathcal{S}}}
\newcommand*\cT{{\mathcal{T}}}	
\newcommand*\cTt{\Tilde{\cT}}	
\newcommand*\cU{{\mathcal{U}}}
\newcommand*\cUt{{\Tilde{\cU}}}	
\newcommand*\cUbar{{\Bar{\cU}}}
\newcommand*\cUhat{\Hat{\cU}}
\newcommand*\cV{{\mathcal{V}}}
\newcommand*\cW{{\mathcal{W}}}	
\newcommand*\cWt{{\Tilde{\cW}}}	
\newcommand*\cWbar{{\Bar{\cW}}}
\newcommand*\cWk{\cW_\vk}	
\newcommand*\cX{{\mathcal{X}}}
\newcommand*\cZ{{\mathcal{Z}}}
\newcommand*\cZk{\cZ_\vk}       
\newcommand*\cZt{{\Tilde{\cZ}}}	

\newcommand*\Cerenkov{Cerenkov} 
\newcommand*\cAl{c_A}		
\newcommand*\cbar{{\Bar{c}}}
\newcommand*\cf{\LatinAIP{cf.}}
\newcommand*\chat{\unit{c}}	
\newcommand*\chit{\Tilde{\chi}} 
\newcommand*\chie{\chi_e}	
\newcommand*\chii{\chi_i}	
\newcommand*\chik{\chi_\vk}	
\newcommand*\chip{\chi_\vp}
\newcommand*\chiq{\chi_\vq}
\newcommand*\chibar{{\Bar{\chi}}}
\newcommand*\chihat{\Hat{\chi}} 
\newcommand*\circa{\Latin{circa}}
\newcommand*\cs{c_{\rm s}} 	
\newcommand*\cl{_{\rm cl}}	
\newcommand*\clumps{^{\rm clumps}}
\newcommand*\coh{^{\rm coh}}
\newcommand*\conj{^\bigast}	
\newcommand*\const{\text{const}}
\newcommand*\cross{\boldsymbol{\times}} 
\newcommand*\closeup[1]{\mkern-#1mu} 
\newcommand*\cum[1]{\mathopen\langle\closeup2\langle#1\unkern
	\rangle\closeup2\mathclose\rangle} 
\newcommand*\curl{\vgrad\cross}


\newcommand*\DB{D_{\Rm B}}	
\newcommand*\DgB{D_{\textrm{gB}}} 
\newcommand*\Dbar{{\Bar{D}}}	
\newcommand*\Dcl{D\cl}		
\newcommand*\Df{\Delta f}       
\newcommand*\Dh{\Dbar_{\Rm h}}	
\newcommand*\Dpar{D_\parallel}	
\newcommand*\Dperp{D_\perp}	
\newcommand*\Dm{D_{\Rm m}} 	
\newcommand*\Dn{D_n}		
\newcommand*\Dq{\Dbar_{\Rm q}}	
\newcommand*\Dv{D_v}		

\newcommand*\Dele{\Partial{}{\e}}
\newcommand*\Delk{{\Delta_\vk}}	
\newcommand*\Delt{\Partial{}{t}}
\newcommand*\DelT[1]{\Partial{#1}{t}}
\newcommand*\Deltau{\Partial{}{\tau}}
\newcommand*\Delv{\Partial{}{v}}
\newcommand*\DelV[1]{\Partial{{#1}}{v}}
\newcommand*\Delvpar{\Partial{}{\vpar}}
\newcommand*\DelvpaR[1]{\Partial{{#1}}{\vpar}}
\newcommand*\Delvv{\Partial{}{\vv}}
\newcommand*\DelVV[1]{\Partial{{#1}}{\vv}}
\newcommand*\Delr{\Partial{}{r}}
\newcommand*\DelR[1]{\Partial{{#1}}{r}}
\newcommand*\DelTau[1]{\Partial{{#1}}{\tau}}
\newcommand*\Delx{\Partial{}{x}}
\newcommand*\DelX[1]{\Partial{{#1}}{x}}
\newcommand*\Dely{\Partial{}{y}}
\newcommand*\DelY[1]{\Partial{{#1}}{y}}
\newcommand*\Delz{\Partial{}{z}}
\newcommand*\DelZ[1]{\Partial{{#1}}{z}}

\newcommand*\Da{\Delta a}
\newcommand*\DG{\Delta\Gamma}
\newcommand*\DNS{direct numerical simulation}
\newcommand*\DcS{\Delta\cS}
\newcommand*\DIA{direct-interaction approximation}
\newcommand*\Dim[1]{#1D}		
\newcommand*\Diss{{\mathcal{D}}}	
\newcommand*\Div{\Mathop{div}} 		
\newcommand*\DK{drift kinetic}
\newcommand*\DKE{\DK\ equation}
\newcommand*\Dk{\Delta k}		
\newcommand*\Dkpar{\Delta\kpar}
\newcommand*\Dmu{\Delta\mu}
\newcommand*\Do{D_0}
\renewcommand*\Dot{^{\displaystyle.}}	
\newcommand*\Dr{\Delta r}
\newcommand*\Dt{\Delta t}		
\newcommand*\Dth{\Delta\theta}
\newcommand*\DS{\Delta S}
\newcommand*\DT{\Delta T}		
\newcommand*\DTc{\DT_{\Rm c}}
\newcommand*\DV{\Delta v}		
\newcommand*\Dw{\Delta\omega}		

\newcommand*\dA{\delta \Ekern A}
\newcommand*\da{\delta a}
\newcommand*\dq{\delta\vq}
\newcommand*\dB{\delta B}               
\newcommand*\dE{\delta \Ekern E} \newcommand*\Ekern{\kern-0.12em}
\newcommand*\dF{\delta F}		
\newcommand*\dH{\delta H}
\newcommand*\dS{\delta S}
\newcommand*\dT{\delta T}		
\newcommand*\dVt{\delta\Vt}
\newcommand*\dVto{\dVt_0}
\newcommand*\db{\delta b}
\newcommand*\dbar{{\Bar{\delta}}}
\newcommand*\dc{d_{\Rm c}}		
\newcommand*\down[1]{_{(#1)}}		
\newcommand*\dperp{d_\perp}

\newcommand*\degreeso{^\circ}		
\renewcommand*\degrees{\ifmmode\degreeso\else$\degreeso$\fi} 

\newcommand*\del{\partial}		
\newcommand*\dele{\partial_\e}
\newcommand*\delk{{\partial_\vk}}
\newcommand*\delpar{\partial_\parallel}
\newcommand*\delphi{\partial_\phi}
\newcommand*\delt{\partial_t}
\newcommand*\deltau{\partial_\tau}
\newcommand*\delr{\partial_r}
\newcommand*\delrho{\partial_\rho}
\newcommand*\delv{\partial_v}
\newcommand*\delvk{\partial_\vk}	
\newcommand*\delvpar{\partial_{\vpar}}
\newcommand*\delvv{\partial_{\vv}}
\newcommand*\delw{\partial_\omega}	
\newcommand*\delx{\partial_x}
\newcommand*\delvx{\partial_\vx}
\newcommand*\dely{\partial_y}
\newcommand*\delz{\partial_z}
\newcommand*\df{\delta \fkern f} \newcommand*\fkern{\kern-0.125em}
\newcommand*\dfc{\df_{\Rm c}}		
\newcommand*\dft{\delta \fkern\Tilde f} 
\newcommand*\dfk{\df_\vk}		
\renewcommand*\dh{\delta h}		
\newcommand*\dk{\delta k}		
\newcommand*\dm{\delta m}

\newcommand*\diel{{\mathcal{D}}} 	
\newcommand*\dielG{\diel_{\Rm G}}	
\newcommand*\dielo{\diel_0} 		
\newcommand*\dielperp{\diel_\perp} 	

\renewcommand*\div{\vgrad\bdot} 	

\newcommand*\dn{\delta n}
\newcommand*\dne{\dn_e}
\newcommand*\dni{\dn_i}
\newcommand*\dN{\d N}                   
\newcommand*\dNi{\d N_i}		

\newcommand*\dhat{\Hat{\delta}}		
\newcommand*\dphk{\delta\phk}		
\renewcommand*\dq{\delta q}
\newcommand*\dvq{\?d\vq}		
\newcommand*\Du{\Delta u}		
\newcommand*\du{\delta u}
\newcommand*\dv{\delta v}
\newcommand*\DW{\Delta\Omega}		
\newcommand*\dw{\delta\omega}
\newcommand*\dx{\delta x}
\newcommand*\dxt{\delta\xt}
\newcommand*\dy{\delta y}
\newcommand*\dV{\delta V}               
\newcommand*\dVEx{\delta\VEx}		
\newcommand*\Dirac[1]{\delta(#1)}
\newcommand*\DiraC[1]{\delta\vlp#1\vrp}
\newcommand*\dchi{\delta\chi}		
\newcommand*\dph{\delta\ph}		
\newcommand*\dphbar{\delta\Bar{\ph}}
\newcommand*\dpht{\delta\Tilde\ph}
\newcommand*\dpsi{\delta\psi}	
\newcommand*\dpsik{\dpsi_\vk}	
\newcommand*\dRt{\delta\Rt}	
\newcommand*\dX{\delta X}
\newcommand*\dwt{\delta\wtld}
\newcommand*\Dx{\Delta x}	
\newcommand*\Dy{\Delta y}		

\newcommand*\dotafter[1]{\setbox0=\hbox{$\displaystyle#1$}
#1\dot{\vphantom{\box0}}}

\newcommand*\EB{\vE\cross\vB}
\newcommand*\EDQNM{eddy-damped quasinormal Markovian}
\newcommand*\EDQNMA{\EDQNM\ approximation}
\newcommand*\Ebar{{\Bar{E}}}            
\newcommand*\Ehat{\Hat{E}}
\newcommand*\Ek{E_\vk}			
\newcommand*\Epar{E_\parallel}		
\newcommand*\Eparbar{{\Bar{E}}_\parallel} 
\newcommand*\Eparhat{\Hat{E}_\parallel}
\newcommand*\Epart{\Tilde{E}_\parallel}
\newcommand*\Eperp{E_\perp}		
\newcommand*\Ec{E_c}			
\newcommand*\Er{E_r}
\newcommand*\Et{{\Tilde E}}
\newcommand*\Ext{\updown{\textrm{ext}}}	
\newcommand*\Ez{E_z}			
\newcommand*\ec{\fr{e,c}}
\newcommand*\ed{\epsilon_\delta}
\newcommand*\eddy{_{\textrm{eddy}}}	
\newcommand*\eff{_{\textrm{eff}}}	
\newcommand*\ehalf{^{1/2}}		
\newcommand*\eg{\LatinAIP{e.g.}}
\newcommand*\ehat{\unit{e}}		
\newcommand*\ek{\v{\epsilon}_\vk}
\newcommand*\Em{\fR{e,m}}		
\newcommand*\edash{--}			
\newcommand*\ep{\epsilon_{\Rm p}}
\newcommand*\epsd{\eps_{\Rm d}}
\newcommand*\epst{\Tilde{\eps}}		
\newcommand*\eperp{\epsilon_\perp}	
\newcommand*\erf{\Mathop{erf}}          
\newcommand*\erfc{\Mathop{erfc}}	
\newcommand*\ew{\epsilon_\omega}

\newcommand*\ETAL{\Latin{et~al.}}	
\newcommand*\etal{\ETAL\kill@period}

\newcommand*\etabar{{\Bar{\eta}}}	
\newcommand*\etahat{{\widehat{\eta}}}
\newcommand*\etahatk{\etahat_\vk}
\newcommand*\etahatp{\etahat_\vp}
\newcommand*\etahatq{\etahat_\vq}
\newcommand*\etanlm{\etanl_-}		

\newcommand*\etanl{\etahat}	
\newcommand*\etaknl{\etanl_\vk}
\newcommand*\etanlk{\etanl_\vk}
\newcommand*\etanlp{\etanl_\vp}
\newcommand*\etanlq{\etanl_\vq}
\newcommand*\etaknld{\etak^{\textrm{nl(d)}}}	

\newcommand*\etam{\eta_-}		
\newcommand*\etapar{\eta_\parallel}	
\newcommand*\etaperp{\eta_\perp}	
\newcommand*\etai{\eta_i}		
\newcommand*\etak{\eta_\vk}		
\newcommand*\etap{\eta_\vp}
\newcommand*\etaq{\eta_\vq}
\newcommand*\etar{\eta_{\Rm r}}		
\newcommand*\etaim{\eta_{\Rm i}}	

\newcommand*\ETC{\Latin{etc.}}		
\newcommand*\etc{\ETC\kill@period}

\newcommand*\exP[1]{\exp[#1]}	

\newcommand*\Fbar{{\Bar{F}}}		
\newcommand*\FD{fluctuation--dissipation}
\newcommand*\FDT{\FD\ theorem}
\newcommand*\FLR{finite-Larmor-radius}
\newcommand*\Fhat{\Hat{F}}		
\newcommand*\Fhatk{\Fhat_\vk}		
\newcommand*\Fk{F_\vk}			

\newcommand*\Fnl{\Fhat}			
\newcommand*\Fknl{\Fnl_\vk}

\newcommand*\fM{f_{\Rm M}}		
\newcommand*\flM{f_{\textrm{lM}}}	
\newcommand*\FM{F_{\Rm M}}		
\newcommand*\FP{Fokker--Planck}
\newcommand*\FPE{\FP\ equation}
\newcommand*\Fo{F_0}			
\newcommand*\Fortran{\textsc{Fortran}}	
\newcommand*\Ft{\Tilde F}		
\newcommand*\FWEB{\texttt{FWEB}}	
\newcommand*\fbar{{\skew3\Bar{f}}}	
\newcommand*\fcum[1]{\mathopen[\closeup1[#1\unkern
	]\closeup1\mathclose]}	
\newcommand*\fhat{{\Hat{f}}}		
\newcommand*\finiteb{\mbox{finite-$\beta$}}
\newcommand*\finiteB{\mbox{finite~$\beta$}}
\newcommand*\fk{f_\vk}			
\newcommand*\fkt{\Tilde f_\vk}		
\newcommand*\Fluid{\updown{\textrm{fluid}}}
\newcommand*\fluid{_{\textrm{fluid}}}
\newcommand*\fm{f_{\Rm m}}		
\newcommand*\fnr{f\nres}		
\newcommand*\fo{f_0}			
\newcommand*\fourth{\case14}		
\newcommand*\free{_{\textrm{free}}}

\newcommand*\from[3]{\left.#1\right\verticalbar_{#2}^{#3}}
\newcommand*\FROM[3]{\left.#1\right._{#2}^{#3}} 

\newcommand*\ft{{\widetilde f}}		
\newcommand*\ftk{\ft_\vk}

\newcommand*\Gal{^{\Rm G}}		
\newcommand*\Gammat{{\Tilde\Gamma}}	
\newcommand*\Gbar{{\Bar{\Gamma}}}
\newcommand*\Ghat{{\Hat{\Gamma}}}
\newcommand*\GC{guiding center}
\newcommand*\gc{guiding-center}
\newcommand*\Gcl{\Gamma\cl}
\newcommand*\GF{gyrofluid}
\newcommand*\GFE{\GF\ equation}
\newcommand*\GK{gyrokinetic}
\newcommand*\GKE{\GK\ equation}
\newcommand*\GL{Ginzburg--Landau}
\newcommand*\GLE{\GL\ equation}
\newcommand*\GT{\Gamma_T}		
\newcommand*\Go{G^{(0)}}		
\newcommand*\Ge{\Gamma_e}		
\newcommand*\Gi{\Gamma_i}		
\newcommand*\gspace{\mbox{$\Gamma$-space}}
\newcommand*\Gspace{$\Gamma$~space}
\newcommand*\go{g_0}			
\newcommand*\Gt{{\Tilde G}}		
\newcommand*\Gtot{\Gamma\tot}		
\newcommand*\Gyro{^{\Rm G}}		
\newcommand*\gB{gyro-Bohm}
\newcommand*\gbar{{\Bar{\gamma}}}	
\newcommand*\gd{\gamma_{\Rm d}}		
\newcommand*\gh{\gbar_{\Rm h}}		
\newcommand*\ghat{{\Hat{g}}}		
\newcommand*\gk{\gamma_\vk}		
\newcommand*\gkone{\gamma_\vk\up{1}}	
\newcommand*\gtkone{\Tilde\gamma_\vk\up{1}} 
\newcommand*\gq{\gamma_\vq}		
\newcommand*\gql{\gbar_{\Rm q}}		
\newcommand*\grad{\nabla}		
\newcommand*\gradpar{\grad_\parallel}	
\newcommand*\gradperp{\grad_{\fkern\perp}}	
\newcommand*\gstar{\gamma_{\bigast}}	
\newcommand*\gt{\Tilde{g}}              


\hyphenation{Brown-ian}
\hyphenation{Krom-mes}
\hyphenation{Lo-rentz-ian}
\hyphenation{Max-well-ian}
\hyphenation{pa-ram-e-trized}

\renewcommand*\Hat[1]{{\widehat{#1}}}	
\newcommand*\Hbar{\Bar{H}}              
\newcommand*\HM{Hasegawa--Mima}
\newcommand*\HMe{\HM\ equation}
\newcommand*\HME{\HMe}
\newcommand*\HW{Hasegawa--Wakatani}
\newcommand*\HWe{\HW\ equation}
\newcommand*\HWE{\HWe}
\newcommand*\Ho{H_0}
\newcommand*\half{\case12}		

\newcommand*\Io{I_0}
\newcommand*\Idv{\Int d\vv}		
\newcommand*\Idvbar{\Int d\vvbar}
\newcommand*\Ito{It\=o}			


\newcommand*\Int{\int\!\?}		
\newcommand*\I[2]{\int_{#1}^{#2}\!\?}	
\newcommand*\InT{\I0\infty}		
\newcommand*\INT{\I{-\infty}\infty}	
\newcommand*\ID{\I\Delta{}}		
\newcommand*\IDprime{\I\Delta{'}}	

\newcommand*\Ibar{{\Bar{I}}}
\newcommand*\Ik{I_\vk}			
\renewcommand*\Im{\Mathop{Im}}		
\newcommand*\Intern{\updown{\textrm{int}}}	

\newcommand*\Ind{\updown{\textrm{ind}}}
\newcommand*\Ip{I_\vp}			
\newcommand*\Iq{I_\vq}			
\newcommand*\It[1]{\emph{#1}}		
\newcommand*\ibar{{\Bar{\imath}}}
\newcommand*\ie{\LatinAIP{i.e.}}	
\newcommand*\inc{^{\textrm{inc}}}		
\newcommand*\inWax[1]{reprinted in \textsl{Selected Papers on Noise and Stochastic
Processes}, edited by N.~Wax (Dover, New York, 1954), p.~#1} 

\newcommand*\Jbar{\Bar{J}}              
\newcommand*\Jij{J^{ij}}
\newcommand*\Jo{J_0}			
\newcommand*\jo{j_0}			
\newcommand*\jbar{\Bar{\jmath}}
\newcommand*\jpar{j_\parallel} 		
\newcommand*\jpare{j_{\parallel e}}
\newcommand*\jpari{j_{\parallel i}}
\newcommand*\jperp{j_\perp} 		
\newcommand*\jz{j_z}			

\newcommand*\K{\cK}			
\newcommand*\KAM{Kolmogorov--\Arnold--Moser}
\newcommand*\KAP{Kubo--Anderson process}
\newcommand*\Keps{$K$--$\eps$}
\newcommand*\KH{Kelvin--Helmholtz}
\newcommand*\Ko{K_0}			
\newcommand*\Kron[1]{\delta_{#1}} 	
\newcommand*\KS{Kuramoto--Sivashinsky}
\newcommand*\KSe{\KS\ equation}
\newcommand*\KSE{\KSe}

\newcommand*\Kspace{$k$~space}
\newcommand*\kspace{\mbox{$k$-space}}

\newcommand*\kD{k_{\Rm D}}		
\newcommand*\kDi{k_{{\Rm D}i}}		
\newcommand*\kDe{k_{{\Rm D}e}}		
\newcommand*\kDs{k_{{\Rm D}s}} 		

\newcommand*\kTbar{\kbar_{\Rm T}}
\newcommand*\ka{k_{\Rm a}} \newcommand*\kb{k_{\Rm b}} 
\newcommand*\kbar{{\Bar{k}}}		
\newcommand*\kc{\kappa_{\Rm c}}		
\newcommand*\kd{k_{\Rm d}}		
\newcommand*\kdbar{\kbar_{\Rm d}}	
\newcommand*\khat{\unit{k}}		

\newcommand*\kinetic{_{\textrm{kin}}}

\newcommand*\kill@period{\futurelet\nextchar\no@period}
\newcommand*\no@period{\ifx\nextchar.\skip@period\fi}	

\newcommand*\kf{k_{\Rm f}}		
\newcommand*\km{k_{\Rm m}}		
\newcommand*\kmin{k\min}		
\newcommand*\kmax{k\max}		
\newcommand*\kn{\kappa_n}		
\newcommand*\ko{\vk_0}			
\newcommand*\kT{\kappa_T}		
\newcommand*\kpar{k_\parallel}		
\newcommand*\kperp{k_\perp}		
\newcommand*\kperpbar{{\Bar{k}}_\perp} 

\newcommand*\kpq[4]{#1_{\v{#2}\v{#3}\v{#4}}}
\newcommand*\kpqi[4]{#1_{{#2}{#3}{#4}}}	
\newcommand*\kpqbar{\kbar,\pbar,\qbar}
\newcommand*\kr{k_r}
\newcommand*\ksh{\kappa_{\Rm s}}
\newcommand*\kstar{k_\star}
\newcommand*\kv{\vk\bdot\vv}		
\newcommand*\kvw{\fr{\kpar\vpar,\w}}
\newcommand*\kw{{k,\omega}}		
\newcommand*\kt{k_\theta}
\newcommand*\ktb{\kappa_{\Rm t}}
\newcommand*\kx{k_x}			
\newcommand*\ky{k_y}			
\newcommand*\kz{k_z}			

\newcommand*\Lac{L_{\textrm{ac}}}	
\newcommand*\Latin{\textit} 
\newcommand*\LatinAIP{\textrm}	
\newcommand*\Lbar{\Bar{L}}	
\newcommand*\Lhat{\Hat{L}}	
\newcommand*\Lie[1]{L_{#1}}	
\newcommand*\LB{L_B}		
\newcommand*\LH{\hbox{L--H}}	
\newcommand*\LIF{laser-induced fluorescence}
\newcommand*\Lc{L_{\Rm c}}	
\newcommand*\Ld{L_{\Rm d}}
\newcommand*\Levy{L\'evy}
\newcommand*\Lg{\Lie{g}}
\newcommand*\LK{L_{\Rm K}}	
\newcommand*\Lk{\cL_\vk}	
\newcommand*\Lperp{L_\perp}	
\newcommand*\Ls{L_{\Rm s}}	
\newcommand*\Lx{L_x}		
\newcommand*\Ly{L_y}		
\newcommand*\Lz{L_z}		

\newcommand*\LP{L_P}		
\newcommand*\Ln{L_n}		
\newcommand*\LT{L_T}		
\newcommand*\LTcrit{L_{T,{\textrm{crit}}}}

\newcommand*\Lu{\Mathop{Lu}}	

\newcommand*\lD{\lambda_{\Rm D}}	
\newcommand*\lDe{\lambda_{{\Rm D}e}}	
\newcommand*\lDi{\lambda_{{\Rm D}i}}	
\newcommand*\lDs{\lambda_{{\Rm D}s}}	

\newcommand*\lE{\lambda_E}		
\newcommand*\lT{\lambda_{\Rm T}}	
\newcommand*\lab{^{\rm lab}}		
\newcommand*\lc{\lambda_{\Rm c}}	
\newcommand*\ld{\lambda_{\Rm d}}	
\newcommand*\lmfp{\lambda_{\textrm{mfp}}}	
\newcommand*\lmfpo{\lambda_{{\textrm{mfp}},0}} 
\newcommand*\lpar{\lambda_\parallel}
\newcommand*\lhat{\Hat{\lambda}}
\newcommand*\lhs{left-hand side}
\newcommand*\like{like}			
\newcommand*\Like{-like}		
\newcommand*\lin{^{\textrm{lin}}}		
\newcommand*\lino{^{(0)}}		

\newcommand*\Mach{{\mathcal{M}}} 	
\newcommand*\MATH{\texttt{MATHEMATICA}}
\newcommand*\mathbox[1]{{\setlength\fboxrule{3pt}\framebox{$\displaystyle #1$}}}
\newcommand*\Mathop[1]{\mathop{\hbox{\rm #1}}\nolimits}
\newcommand*\MC{Monte Carlo}
\newcommand*\Mhat{{\widehat{M}}}
\newcommand*\Mhatl{{\widehat{M}}}
\newcommand*\MHD{magnetohydrodynamic}
\newcommand*\More{\emph{More\dots}}	
\newcommand*\Mbar{{\Bar{M}}}
\newcommand*\Max{\mathop{\operator@font max}}	
\renewcommand*\max{_{\textrm{max}}}
\newcommand*\Min{\mathop{\operator@font min}}	
\newcommand*\MRT{$\hbox{M}(\hbox{RT})^2$}	
\newcommand*\MSR{Martin, Siggia, and Rose}
\newcommand*\mCk{\mC_\vk}		
\renewcommand*\min{_{\textrm{min}}}

\renewcommand*\tensor{\textsf}		

\newcommand*\mA{\tensor{A}}
\newcommand*\ma{\tensor{a}}
\newcommand*\mB{\tensor{B}}
\newcommand*\mC{\tensor{C}}
\newcommand*\mD{\tensor{D}}
\newcommand*\md{\tensor{d}}
\newcommand*\mdiel{\tensor{D}}		
\newcommand*\mF{\tensor{F}}
\newcommand*\mG{\tensor{G}}
\newcommand*\mI{\tensor{I}}
\newcommand*\mone{\mI}			
\newcommand*\mJ{\tensor{J}}
\newcommand*\mK{\tensor{K}}		
\newcommand*\mL{\tensor{L}}
\newcommand*\mLambda{\boldsymbol{\Lambda}}
\newcommand*\mM{\tensor{M}}
\newcommand*\mOmega{\boldsymbol{\Omega}}
\newcommand*\mP{\tensor{P}}
\newcommand*\mPi{\boldsymbol{\Pi}}
\newcommand*\mpi{\boldsymbol{\pi}}
\newcommand*\mQ{\tensor{Q}}
\newcommand*\mR{\tensor{R}}
\newcommand*\mS{\tensor{S}}
\newcommand*\mT{\tensor{T}}
\newcommand*\mU{\tensor{U}}
\newcommand*\mW{\tensor{W}}
\newcommand*\mX{\tensor{X}}
\newcommand*\mepsilon{\boldsymbol{\epsilon}}
\newcommand*\mg{\tensor{g}}		
\newcommand*\mchi{\boldsymbol{\chi}}
\newcommand*\msigma{\boldsymbol{\sigma}}

\newcommand*\mtau{\boldsymbol{\tau}}
\newcommand*\maybebreak{\discretionary{}{}{}} 
\newcommand*\mhalf{\m{1/2}}
\newcommand*\mix{^{\textrm{ml}}}
\newcommand*\mk{m_\vk}

\newcommand*\ml{^{\textrm{ml}}}		
\newcommand*\m[1]{^{-#1}}		
\newcommand*\mc{^{\textrm{mc}}}

\newcommand*\me{m_e}
\newcommand*\mi{m_i}
\newcommand*\ms{m_s}

\newcommand*\muL{\mu_L}
\newcommand*\mubar{{\Bar{\mu}}}
\newcommand*\muc{\mu_{c}}		
\newcommand*\muo{\mu}			
\newcommand*\mucl{\muo\cl}		
\newcommand*\muk{\mu_\vk}
\newcommand*\mup{\mu_\vp}
\newcommand*\muq{\mu_\vq}
\newcommand*\mum{\muo_{\Rm m}}		
\newcommand*\mumcl{\muo_{\Rm m,\textrm{cl}}}	
\newcommand*\muhat{\Hat{\mu}}
\newcommand*\muhatk{\muhat_\vk}
\newcommand*\muhatp{\muhat_\vp}
\newcommand*\muhatq{\muhat_\vq}

\newcommand*\Nbar{\Bar{N}}		
\newcommand*\Nhat{{\widehat N}}		
\newcommand*\NH{\Nose--Hoover}
\newcommand*\NS{Navier--Stokes}
\newcommand*\NSe{\NS\ equation}
\newcommand*\NSE{\NSe}			
\newcommand*\Nk{N_\vk}
\newcommand*\Nm{{N_m}}			
\newcommand*\Nt{N_T}
\newcommand*\nd{\nu_{\Rm d}}
\newcommand*\Nose{Nos\'e}

\newcommand*\nbar{{\Bar{n}}}		
\newcommand*\nbarm{\nbar_{\Rm m}}	
\renewcommand*\ne{n_e}			
\newcommand*\nebar{\nbar_e} 		
\newcommand*\neG{\ne\Gyro} 		
\renewcommand*\ni{n_i}			
\newcommand*\Ni{N_i}			
\newcommand*\nibar{\nbar_i} 		
\newcommand*\niG{\ni\Gyro} 		
\newcommand*\nipol{\ni\pol} 		
\newcommand*\no{n_0}			
\newcommand*\np{n_{\Rm p}}		
\newcommand*\nsbar{\nbar_s}		

\newcommand*\nhat{\unit{n}}		
\newcommand*\nk{n_\vk}
\newcommand*\nl{^{\textrm{nl}}}		
\newcommand*\nlin{^{\textrm{nlin}}}		
\newcommand*\noise{^{\textrm{noise}}}
\newcommand*\norm[1]{\mathopen{\parallel}#1\mathclose{\parallel}}
\newcommand*\nq[1]{(\nbar q)_{#1}}
\newcommand*\nr{_{\textrm{nres}}}		
\newcommand*\nres{_{\textrm{nres}}}		
\newcommand*\nth{$n$th}

\newcommand*\nue{\nu_e}		
\newcommand*\nuee{\nu_{ee}}	
\newcommand*\nui{\nu_i}		
\newcommand*\nuii{\nu_{ii}}	
\newcommand*\nuei{\nu_{ei}}	
\newcommand*\nuie{\nu_{ie}}	

\newcommand*\nud{\nu_{\Rm d}}	
\newcommand*\nuk{\nu_\vk}	
\newcommand*\nuhat{\Hat{\nu}}
\newcommand*\nustar{\nu_\bigast} 

\newcommand*\OC{oscillation center}
\newcommand*\Od{{\Omega_d}}		
\newcommand*\Ohat{{\widehat O}}
\newcommand*\Omegahat{\Hat{\Omega}}	
\newcommand*\Omegak{\Omega_\vk}		
\newcommand*\Order[1]{O(#1)} 		
\newcommand*\OrdeR[1]{O\vlp#1\vrp} 	
\newcommand*\on[1]{[#1]}			
\newcommand*\order[1]{o(#1)}		

\newcommand*\PDF{probability density function}
\newcommand*\Pade{Pad\'e}
\newcommand*\Pb[1]{\set{#1}}		
\newcommand*\Pbar{\Bar{P}}
\newcommand*\Pd{P_{\Rm d}}		
\newcommand*\Phat{{\Hat{\Phi}}}		
\newcommand*\Poincare{Poincar\'e}
\newcommand*\Pos{\cP}			
\newcommand*\PS{Pfirsch--Schl\"uter}
\newcommand*\PV{\Mathop{P}}  		
\newcommand*\Pvalue{\Pr}		
\renewcommand*\Pr{\Mathop{Pr}} 		
\newcommand*\Prandtl{\Pr}		
\newcommand*\Pt{{\Tilde P}}
\newcommand*\pMatrix[1]{\begin{pmatrix}#1\end{pmatrix}} 
\newcommand*\pbar{{\Bar{p}}}
\newcommand*\perse{\Latin{per~se}}
\newcommand*\phbar{{\Bar{\ph}}}		
\newcommand*\phat{{\widehat\psi}}
\newcommand*\phihat{\Hat{\boldsymbol{\phi}}}   
\newcommand*\phibar{{\Bar{\phi}}}       
\newcommand*\phk{\ph_\vk}		
\newcommand*\php{\ph_\vp}		
\newcommand*\phq{\ph_\vq}		
\newcommand*\pht{\Tilde\ph}		
\newcommand*\phtt{\<\d\pht^2>}		
\newcommand*\pol{^{\textrm{pol}}}		
\newcommand*\prima{\Latin{prima facie}}

\newcommand*\psik{\psi_\vk}
\newcommand*\psip{\psi_\vp}
\newcommand*\psiq{\psi_\vq}
\newcommand*\psibar{{\Bar{\psi}}}
\newcommand*\psihat{{\widehat\psi}}
\newcommand*\psit{\Tilde{\psi}}         
\newcommand*\px{p_x}
\newcommand*\py{p_y}

\newcommand*\Qhat{\Hat{Q}}		
\newcommand*\Qhattilde{{\skew6\Tilde{\Qhat}}}
\newcommand*\QL{quasilinear}
\newcommand*\QLT{\QL\ theory}
\newcommand*\Qzero{\text{``}0\text{''}}	
\newcommand*\qzero{\text{``}0\text{''}} 
\newcommand*\qbar{{\Bar{q}}}
\newcommand*\qc{\fr{q,c}}
\newcommand*\qhat{\Hat{q}}		
\newcommand*\qo{q_0}
\newcommand*\ql{^{\textrm{QL}}}		
\newcommand*\qm{\(\frac{q}{m}\)}		
\newcommand*\qpar{q_\parallel}		
\newcommand*\qs{q_s}			
\newcommand*\qx{q_x}			
\newcommand*\qy{q_y}			
\newcommand*\qz{q_z}			

\newcommand*\Ratfor{\textsc{Ratfor}}	
\newcommand*\RB{resonance-broadening}
\newcommand*\RBT{\RB\ theory}
\newcommand*\RCM{random-coupling model}
\newcommand*\RMC{realizable Markovian closure}
\newcommand*\REVTeX{REV\TeX}
\newcommand*\RGI{random Galilean invariance}
\newcommand*\RPA{random-phase approximation}
\newcommand*\Rc{\R_c}			
\newcommand*\Ren{\R_e}			
\newcommand*\Rhat{\Hat{R}}		
\newcommand*\Rtt{R_{\Rm t}}		
\newcommand*\Rll{R_{\Rm l}}		
\newcommand*\Rlin{\R_l}			

\newcommand*\Rm[1]{#1}			

\renewcommand*\Re{\Mathop{Re}}		
\newcommand*\Reynolds{\Re}		

\newcommand*\Ro{R^{(0)}}		
\newcommand*\Rk{R_\vk}
\newcommand*\Rp{R_\vp}
\newcommand*\Rq{R_\vq}
\newcommand*\Rt{\widetilde R}		

\newcommand*\rbar{{\Bar{r}}}
\newcommand*\rd{r_{\Rm d}} 		

\newcommand*\re{\rho_e}
\newcommand*\ri{\rho_i}
\newcommand*\rhos{\rho_s}		
\newcommand*\rs{\rho_{\textrm{s}}}	

\newcommand*\res{_{\textrm{res}}}
\newcommand*\ro{r_0}
\newcommand*\rhat{\unit{r}} 		
\newcommand*\rhobar{{\bar\rho}}
\newcommand*\rhom{\rho_m}              
\newcommand*\rhs{right-hand side}
\newcommand*\rhomass{\rho_{\Rm m}}	
\newcommand*\rms{_{\textrm{rms}}}

\newcommand*\Sbar{{\Bar{S}}}
\newcommand*\Sc{S_c}
\newcommand*\Schr{Schr\"odinger}
\newcommand*\Sigmabar{\Bar{\Sigma}}	
\newcommand*\Sigmad{\Sigma\uP{d}}
\newcommand*\Sigmahat{\Hat{\Sigma}}
\newcommand*\Sigmahatg{\Sigmahat_g}
\newcommand*\Sigmak{\Sigma_\vk}
\newcommand*\Sigmahatk{\Sigmahat_\vk}

\newcommand*\Sigmanl{\Sigmahat}		
\newcommand*\Sigmaknl{\Sigmanl_\vk}

\newcommand*\Sigmanlg{\Sigmanl_g}
\newcommand*\Sigmaknlg{\Sigmanl_{g,\vk}}
\newcommand*\SO{stochastic oscillator}
\newcommand*\SOC{self-organized criticality}
\newcommand*\Sin{\Sum(i=1,n)}
\newcommand*\Sk{\sum_\vk} 		
\newcommand*\Sq{\sum_\vq}		
\newcommand*\Skt{{\textstyle\sum_\vk}} 		
\newcommand*\Skpq{\sum_{k+p+q=0}}
\newcommand*\Sol{^{\Rm S}}		
\newcommand*\STb{strong-turbulence}
\newcommand*\St{{\Tilde S}}		
\newcommand*\Stoss{Stosszahlansatz}
\newcommand*\sD{\sum_\Delta}		
\newcommand*\sans{\Latin{sans}}
\newcommand*\sbar{{\Bar{s}}}
\newcommand*\sd{\sigma_d}
\newcommand*\shat{{\Hat{s}}}		
\newcommand*\sigmak{\sigma_\vk}		
\newcommand*\sk{\sum_\vk} 		
\newcommand*\set[1]{{\let|\mid \{#1\}}}	
\newcommand*\sgn{\Mathop{sgn}}
\newcommand*\sink{_{\textrm{out}}}
\newcommand*\src{\eta}			
\newcommand*\srcHat{\Hat{\eta}}		
\newcommand*\srchat{\hat{\eta}}		
\newcommand*\strain{{\tensor s}} \newcommand*\antistrain{{\tensor a}}
\newcommand*\Sstate{steady state}
\newcommand*\sstate{steady-state}
\newcommand*\submax{_{\textrm{max}}}
\newcommand*\submin{_{\textrm{min}}}
\newcommand*\svk{s_\vk}			
\newcommand*\svp{s_\vp}
\newcommand*\svq{s_\vq}
\newcommand*\sx{\sigma_x}		
\newcommand*\sy{\sigma_y}		

\renewcommand*\term[2]{\mathord{\mathop{#2}^{(#1)}}}
\renewcommand*\TH{Terry--Horton}
\newcommand*\THE{\TH\ equation}
\newcommand*\Tbar{{\Bar{T}}}
\newcommand*\That{\Hat{T}}
\newcommand*\Tr{^{\Rm T}} 		
\newcommand*\Trace{\Mathop{Tr}}		
\newcommand*\Trap{\uP{t}}

\newcommand*\Te{T_e} 		
\newcommand*\Ti{T_i} 		
\newcommand*\Ts{T_s} 		
\newcommand*\Tpar{T_\parallel} 	
\newcommand*\Tperp{T_\perp} 	
\newcommand*\Tt{{\Tilde T}} 	
\newcommand*\taue{\tau_e}       
\newcommand*\taui{\tau_i}       

\renewcommand*\Tilde{\widetilde}	
\newcommand*\Tot{\updown{\textrm{tot}}}
\newcommand*\tK{\tau_{\Rm K}}	

\newcommand*\tac{\tau_{\textrm{ac}}}	
\newcommand*\tacE{\tac^E}               
\newcommand*\tacL{\tac^L}               
\newcommand*\taco{\tau_{\textrm{ac}}\up0{}} 
\newcommand*\tacbar{{\Bar\tau}_{\textrm{ac}}}
\newcommand*\tacWTT{\tac^{\textrm{WTT}}} 

\newcommand*\tbar{{\overline t}}
\newcommand*\taubar{{\overline\tau}}
\newcommand*\tauhat{\Hat{\tau}}
\newcommand*\taut{\tau_{\Rm t}}
\newcommand*\tc{\tau_{\Rm c}}		
\newcommand*\tcl{\tau_{\textrm{cl}}} 	
\newcommand*\tD{\tau_D}			
\newcommand*\td{\tau_{\Rm d}}		
\newcommand*\tdq{\tau_{{\Rm d},q}}	
\newcommand*\tdperp{\tau_{{\Rm d}\perp}} 
\newcommand*\tdperpk{\tau_{{\Rm d}\perp,\vk}} 
\newcommand*\tdperpq{\tau_{{\Rm d}\perp,q}} 
\newcommand*\tdpar{\tau_{{\Rm d}\parallel}} 
\newcommand*\teddy{\tau\eddy} 		
\newcommand*\Test{\updown{\textrm{test}}}
\newcommand*\thbar{\Bar{\theta}}	
\newcommand*\thydro{\tau_{\Rm h}}	
\newcommand*\tL{\tau_L}
\newcommand*\thhat{\unit{\theta}}
\newcommand*\that{\Hat{t}}		
\newcommand*\timescale{timescale}	
\newcommand*\ttheta{\tau_\theta}
\newcommand*\third{\case13}
\newcommand*\tot{_{\textrm{tot}}} 		
\newcommand*\tR{\tau_{\Rm R}}
\newcommand*\tr{\tau_{\Rm r}} 		
\newcommand*\transit{_{\textrm{tr}}}
\newcommand*\true{^{\textrm{true}}}
\newcommand*\ttilde{{\Tilde t}}
\newcommand*\tw{{\widetilde w}}

\newcommand*\uE{u_E}            
\newcommand*\Uk{U_\vk}
\newcommand*\Unsym{^{\Rm U}}	
\newcommand*\Unsymconj{^{\Rm U\bigast}}	
\newcommand*\Ut{\widetilde U}	
\newcommand*\Ubar{{\Bar{U}}}
\newcommand*\Uhat{{\Hat{U}}}	
\newcommand*\Upar{U_\parallel}
\newcommand*\ubar{{\Bar{u}}}	
\newcommand*\uhat{\Hat{u}}	
\newcommand*\uk{u_k}
\newcommand*\unit[1]{{\widehat{\v{#1}}}}	
\newcommand*\uo{u_0}
\newcommand*\up[1]{^{(#1)}}
\newcommand*\uP[1]{^{\textrm{(#1)}}}
\newcommand*\upar{u_\parallel}		
\newcommand*\upare{u_{\parallel e}}     
\newcommand*\upari{u_{\parallel i}}     
\newcommand*\updown{_} 
\newcommand*\uspace{\mbox{$\muo$-space}}  
\newcommand\muspace{\uspace} 	
\newcommand*\Uspace{$\muo$~space} 
\newcommand*\Mspace{\Uspace} 	
\newcommand*\Muspace{\Uspace} 	
\newcommand*\ut{\widetilde u}

\newcommand*\Vhat{\Hat{V}}
\newcommand*\Vk{V_\vk}		
\newcommand*\Vkern{\kern-0.1em} 

\renewcommand*\v[1]{{\bm{#1}}} 	
\newcommand*\vVt{{\bm{\Vt}}}	
\newcommand*\vTh{\boldsymbol{\Theta}}
\newcommand*\vth{\boldsymbol{\theta}}
\newcommand*\Vth{V_\theta}
\newcommand*\Vto{\Vt_0}
\newcommand*\vA{\v{A}} 		
\newcommand*\vAl{v_A} 		
\newcommand*\vAperp{\vA_\perp}  
\newcommand*\va{\v{a}}		
\newcommand*\vahat{{\Hat{\va}}}	
\newcommand*\vat{{\Tilde\va}}	
\newcommand*\valpha{\boldsymbol{\alpha}} 
\newcommand*\vB{\v{B}} 		
\newcommand*\vBt{{\Tilde{\vB}}} 
\newcommand*\vb{\v{b}}          
\newcommand*\vbeta{\v{\beta}}
\newcommand*\vcE{\v{\cE}}
\newcommand*\vcP{\v{\cP}}
\newcommand*\vC{\v{C}}
\newcommand*\vD{\v{D}} 		
\newcommand*\vd{\vv_d}          
\newcommand*\vE{\v{E}} 		
\newcommand*\vEt{{\Tilde{\vE}}}	
\newcommand*\vEhat{{\Hat{\vE}}}	
\newcommand*\vEperp{\vE_\perp}
\newcommand*\vF{\v{F}}
\newcommand*\vG{\v{\Gamma}} 
\newcommand*\vgamma{\boldsymbol{\gamma}} 
\newcommand*\vK{\v{K}}
\newcommand*\vJ{\v{J}}
\newcommand*\vJJ{\v{\cJ}}
\newcommand*\vN{\v{N}}
\newcommand*\vNhat{{\bm{\Nhat}}}
\newcommand*\vGamma{\boldsymbol{\Gamma}}  
\newcommand*\vP{\v{P}}		
\newcommand*\vPhat{{\Hat{\boldsymbol{\Phi}}}}	
\newcommand*\vQ{\v{Q}}
\newcommand*\vR{\v{R}}		
\newcommand*\vRbar{\Bar{\vR}}	
\newcommand*\vS{\v{S}}
\newcommand*\vSc{\vS_{\Rm c}}	
\newcommand*\vSchat{\Hat{\vS}_{\Rm c}}
\newcommand*\vSw{\vS_{\Rm w}}	
\newcommand*\vSstar{\vS_{\bigast}}
\newcommand*\vsigma{\boldsymbol{\sigma}} 
\newcommand*\vbar{{\Bar{v}}}
\newcommand*\Vd{V_{\Rm d}}	
\newcommand*\Vdhat{\Hat{V}_{\Rm d}} 
\newcommand*\Vshat{\Hat{V}_{\bigast}} 
\newcommand*\vVd{\v{V}_{\Vkern\Rm d}} 
\newcommand*\vVs{\v{V}_{\bigast}} 
\newcommand*\vVp{\v{V}_{\Vkern\Rm p}}	
\newcommand*\vdel{\boldsymbol{\del}}
\newcommand*\vek{\vepsilon_\vk}
\newcommand*\vell{\v{\ell}}
\newcommand*\vepsilon{\v{\epsilon}}
\newcommand*\veps{\vepsilon} 	
\newcommand*\veta{\v{\eta}}		
\newcommand*\vetahat{\v{\etahat}}	
\newcommand*\vf{\v{f}}
\newcommand*\vgrad{\v{\nabla}}
\newcommand*\vgradperp{\vgrad_{\fkern\perp}}
\newcommand*\vh{\v{h}}
\newcommand*\vhat{\unit{v}}
\newcommand*\vj{\v{j}}
\newcommand*\vjpar{\v{j}_\parallel}	
\newcommand*\vjperp{\v{j}_\perp}	
\newcommand*\vk{\v{k}}
\newcommand*\vkappa{{\boldsymbol{\kappa}}} 
\newcommand*\vkhat{\Hat{\vk}}		
\newcommand*\vkstar{\vk_{\bigast}} 	
\newcommand*\vkperp{\v{k}_\perp}	
\newcommand*\vko{\vk_0}
\newcommand*\vKspace{$\vk$~space}
\newcommand*\vkspace{\mbox{$\vk$-space}}
\newcommand*\vkw{{\vk,\omega}}		
\newcommand*\vkx{\vk\bdot\vx}
\newcommand*\vkbar{{\bm{\kbar}}}
\newcommand*\vl{\v{l}}
\newcommand*\vlambda{{\boldsymbol{\lambda}}}
\newcommand*\vonKarman{von K\'arm\'an}
\newcommand*\VonKarman{Von K\'arm\'an}
\newcommand*\vor{\varpi}		
\newcommand*\vvor{\boldsymbol{\varpi}}  
\newcommand*\vp{\v{p}}
\newcommand*\vph{{\boldsymbol{\ph}}}	
\newcommand*\vphat{\Hat{\vp}}
\newcommand*\vpstar{\vp_{\bigast}}
\newcommand*\vpbar{{\Bar{\vp}}}
\newcommand*\vgroup{v_{\textrm{gr}}} 	
\newcommand*\vvgroup{\vv_{\textrm{gr}}} 	
\newcommand*\vphase{v_{\textrm{ph}}} 	
\newcommand*\vvphase{\vv_{\textrm{ph}}} 
\newcommand*\vpar{v_\parallel}
\newcommand*\vparbar{\Bar{v}_\parallel}
\newcommand*\vpare{v_{\parallel e}}
\newcommand*\vperp{v_\perp}
\newcommand*\vperpbar{\Bar{v}_\perp}	
\newcommand*\vpsi{\v{\psi}}		
\newcommand*\vpsihat{{\bm{\psihat}}}
\newcommand*\vq{\v{q}}
\newcommand*\vqhat{\Hat{\vq}}		
\newcommand*\vqstar{\vq_{\bigast}}
\newcommand*\vqbar{{\Bar{\vq}}}
\newcommand*\vlb{\mathopen{\boldsymbol[}} 
\newcommand*\vlp{\mathopen{\boldsymbol(}} 
\newcommand*\vm{\v{m}}
\newcommand*\vn{\v{n}}
\newcommand*\vo{v_0}
\newcommand*\vOmega{\v{\Omega}}		
\newcommand*\vomega{\v{\omega}}		
\renewcommand*\vr{\v{r}}
\newcommand*\vrho{\v{\rho}}
\newcommand*\vrb{\mathclose{\boldsymbol]}} 
\newcommand*\vrp{\mathclose{\boldsymbol)}} 

\newcommand*\vt{v_{\Rm t}}	
\newcommand*\vT{v_T}		
\newcommand*\vte{v_{{\Rm t}e}}	
\newcommand*\vti{v_{{\Rm t}i}}	
\newcommand*\vts{v_{{\Rm t}s}}	
\newcommand*\vtu{v_{{\Rm t}\mu}} 

\newcommand*\vtr{v_{\textrm{tr}}}	
\newcommand*\Vtr{V_{\textrm{tr}}}	

\newcommand*\vtheta{\v{\theta}}
\newcommand*\vI{\v{I}}
\newcommand*\vU{\v{U}}		

\newcommand*\vu{\v{u}}		
\newcommand*\vuk{\vu_\vk}	
\newcommand*\vuperp{\vu_\perp}	
\newcommand*\vuE{\vu_\vE}	
\newcommand*\vuEbar{\Bar{\vu}_\vE}
\newcommand*\vue{\vu_e}		
\newcommand*\vui{\vu_i}		

\newcommand*\vud{\vu_{\Rm d}}	
\newcommand*\vuds{\vu_{\Rm d, s}}
\newcommand*\vude{\vu_{\Rm d, e}}
\newcommand*\vudi{\vu_{\Rm d, i}}
\newcommand*\vup{\vu_{\Rm p}}	
\newcommand*\vV{\v{V}}		
\newcommand*\vVbar{\Bar{\vV}}	
\newcommand*\vVE{\vV_{\Vkern E}}	
\newcommand*\vVEbar{\Bar{\vV}_{\Vkern E}} 
\newcommand*\vVEhat{\Hat{\vV}_{\Vkern E}}
\newcommand*\vVEperp{\vV_{\vE,\perp}}
\newcommand*\vVEt{\Tilde{\vV}_{\Vkern E}}
\newcommand*\vW{\v{W}}
\newcommand*\vX{\v{X}}		
\newcommand*\vY{\v{Y}}		
\newcommand*\vZ{\v{Z}}		
\newcommand*\vZbar{\Bar{\vZ}}   
\newcommand*\vXbar{\Bar{\vX}}	
\newcommand*\vv{\v{v}}
\newcommand*\vvperp{\vv_\perp}
\newcommand*\vvbar{{\Bar{\vv}}}
\newcommand*\vvperpbar{\vvbar_\perp}
\newcommand*\vw{\v{w}}
\newcommand*\vx{\v{x}}		
\newcommand*\vxbar{\Bar{\vx}}	
\newcommand*\vxperp{\vx_\perp}	
\newcommand*\vxt{\vx,t}		
\newcommand*\vy{\v{y}}		
\newcommand*\vz{v_z}		
\newcommand*\vzbar{\Bar{v}_z}
\newcommand*\vzero{\boldsymbol{0}} 

\newcommand*\Vbar{{\Bar{V}}}	
\newcommand*\VE{V_E}		
\newcommand*\VEx{V_{E,x}}	
\newcommand*\VEy{V_{E,y}}	
\newcommand*\Vo{V_0}
\newcommand*\Vstar{V_{\displaystyle \star}}
\newcommand*\Vt{{\Tilde V}}	

\newcommand*\vxspace{\mbox{$\vx$-space}}
\newcommand*\vXspace{$\vx$~space}
\newcommand*\vvspace{\mbox{$\vv$-space}}
\newcommand*\vVspace{$\vv$~space}
\newcommand*\vspce{\mbox{$v$-space}}
\newcommand*\Vspace{$v$~space}

\newcommand*\Vs{V_\bigast}	
\newcommand*\Vse{V_{\bigast e}} 
\newcommand*\Vsi{V_{\bigast i}} 
\newcommand*\Vsn{\Vs^n}
\newcommand*\VsT{\Vs^T}
\newcommand*\Vss{\Vs{}_s}
\newcommand*\vus{\vu_\bigast}	
\newcommand*\vusi{\vu_{\bigast i}}
\newcommand*\vuss{\vu_{\bigast s}} 

\newcommand*\vs{\LatinAIP{vs}}
\newcommand*\via{\Latin{via}}
\newcommand*\viz{\LatinAIP{viz.}}
\newcommand*\versa{\Latin{vice versa}}


\newcommand*\Wbar{\Bar{\Omega}}
\newcommand*\WKE{wave kinetic equation}
\newcommand*\WTb{weak-turbulence}
\newcommand*\WTT{\WTb\ theory}

\newcommand*\Wk{\Omega_\vk}		
\newcommand*\What{\Hat{\Omega}}		
\newcommand*\Whatk{\Hat\Omega_\vk}	
\newcommand*\Wkr{\Omega_{\vk,{\Rm r}}}
\newcommand*\Wp{\Omega_\vp}
\newcommand*\Wq{\Omega_\vq}
\newcommand*\Wt{{\Tilde\Omega}}

\newcommand*\wb{\omega_{\Rm b}}
\newcommand*\wbe{\omega_{{\Rm b}e}}
\newcommand*\wbi{\omega_{{\Rm b}i}}
\newcommand*\wbs{\omega_{{\Rm b}s}}

\newcommand*\wbar{{\Bar{\omega}}}	
\newcommand*\wz{\omega_z}		

\newcommand*\wc{\omega_{\Rm c}}
\newcommand*\wce{\omega_{{\Rm c}e}}
\newcommand*\wci{\omega_{{\Rm c}i}}
\newcommand*\wcs{\omega_{{\Rm c}s}}

\newcommand*\wD{\omega_{\Rm d}} 
\newcommand*\wR{\omega_R} 

\newcommand*\what{{\Hat\omega}}
\newcommand*\wk{\omega_\vk} 	
\newcommand*\wq{\omega_\vq} 	
\newcommand*\wij{\Hat\omega_{ij}}
\newcommand*\wmn{\omega_{\mu\nu}}
\newcommand*\wo{\omega_0}	
\newcommand*\Wo{\Omega_0}	

\renewcommand*\wp{\omega_{\Rm p}}
\newcommand*\wpe{\omega_{{\Rm p}e}}
\newcommand*\wpi{\omega_{{\Rm p}i}}
\newcommand*\wps{\omega_{{\Rm p}s}}

\newcommand*\whitenoise{{\widetilde{w}}} 	
\renewcommand*\wr{\omega_{\Rm r}}
\newcommand*\wrt{with respect to}

\newcommand*\ws{\omega_\bigast}
\newcommand*\wsk{\omega_{\bigast\vk}}
\newcommand*\wsbar{\Bar{\omega}_\bigast}	
\newcommand*\wss{\ws{}_s}
\newcommand*\wsi{\ws{}_i}
\newcommand*\wse{\ws{}_e}
\newcommand*\wsn{\ws^n}
\newcommand*\wsT{\ws^T}
\newcommand*\wsTi{\omega^T_{\bigast i}}

\newcommand*\wsh{\omega_{\Rm s}} 	

\newcommand*\Wspace{$\omega$~space}
\newcommand*\wspace{\mbox{$\omega$-space}}

\newcommand*\wt{\omega_{\Rm t}}
\newcommand*\wte{\omega_{{\Rm t}e}}
\newcommand*\wti{\omega_{{\Rm t}i}}
\newcommand*\wts{\omega_{{\Rm t}s}}

\newcommand*\wtld{\widetilde\omega}

\newcommand*\wtr{\omega_{\textrm{tr}}} 
\newcommand*\Wtr{\Omega_{\textrm{tr}}} 

\newcommand*\WorldPhysics[2]{Reprinted in J. H. Weaver, \textsl{The World of
Physics} (Simon and Schuster, New York, 1987), vol.~#1, p.~#2}

\newcommand*\Wiki{\texttt{Wikipedia}}

\newcommand*\Xk{X_\vk}			
\newcommand*\xbar{{\Bar{x}}}
\newcommand*\xdot{\dot x}
\newcommand*\xE{x_e}
\newcommand*\xI{x_i}
\newcommand*\xit{\Tilde{\xi}}		
\newcommand*\xp{\xit_\vp}
\newcommand*\xq{\xit_\vq}
\newcommand*\xspace{\mbox{$x$-space}}
\newcommand*\Xspace{$x$~space}

\newcommand*\xt{\widetilde x}
\newcommand*\xtilde{\Tilde{x}}
\newcommand*\xT{x_{\Rm T}}

\newcommand*\ybar{{\Bar{y}}}
\newcommand*\yt{\widetilde y}
\newcommand*\yw{y_{\Rm w}}		

\newcommand*\xhat{\unit{x}}
\newcommand*\yhat{\unit{y}}
\newcommand*\zhat{\unit{z}}
\newcommand*\Zhat{\unit{Z}}

\newcommand*\zbar{{\Bar{z}}}
\newcommand*\Zaslavskii{Zaslavski\hacek{\i}}
\newcommand*\ZF{zonal flow}	
\newcommand*\Zb{Z_{\Rm b}}	
\newcommand*\Zf{Z_{\Rm f}}	
\newcommand*\Zbar{{\Bar{Z}}}    
\newcommand*\ZDIA{Z_{\textrm{DIA}}}
\newcommand*\ZDIAck[1]{Z^{{\Rm c}#1}_{{\textrm{DIA}},\vk}}
\newcommand*\ZG{Z^{\Rm G}} 	
\newcommand*\Zc{Z^{\Rm c}} 	
\newcommand*\Zk{Z_\vk}		
\newcommand*\Zt{\Tilde{Z}}	
\newcommand*\zs{\z_{\Rm s}}


\newcommand*\Deriv[3]{\Derivo{#1}{#2}#3\Derivo}


\newcommand*\Partial[1]{\Deriv\partial\fr{#1}}
\newcommand*\Partialt[1]{{\textstyle\Partial{#1}}}
\newcommand*\PartiaL[1]{\Deriv\partial\fR{#1}}

\newcommand*\Func[1]{\Deriv\delta\fr{#1}}
\newcommand*\Funct[1]{{\textstyle\Func{#1}}}
\newcommand*\FunC[1]{\Deriv\delta\fR{#1}}

\newcommand*\Total[1]{\Deriv{\?d}\fr{#1}}
\newcommand*\Totalt[1]{{\textstyle\Total{#1}}}
\newcommand*\TotaL[1]{\Deriv{\?D}\fr{#1}}


\newcommand*\mathselect[4]{\mathchoice{#1{#3}{#4}}
	{#2{#3}{#4}}{#2{#3}{#4}}{#2{#3}{#4}}}

\newcommand*\frD{\frac} 
\newcommand*\frT[2]{#1/#2} 

\newcommand*\fr[1]{\fro#1\fro} 
\newcommand*\frt[1]{{\textstyle\fr{#1}}} 

\newcommand*\fR[1]{\left(\fr{#1}\right)}	
\newcommand*\fRt[1]{{\textstyle(\fr{#1})}} 

\newcommand*\bR[1]{\left[\fr{#1}\right]}
\newcommand*\bRt[1]{{\textstyle[\fr{#1}]}}

\newcommand*\choice[1]{\choiceo#1\choiceo}

\newcommand*\choicE[1]{\left(\choice{#1}\right)}

\newcommand*\Casefr[2]{\frac{#1}{#2}}	
\newcommand*\Case{\Casefr}



\newcommand*\Half{\Casefr12}	\newcommand*\HALF{\frac{1}{2}}

\newcommand*\Third{\Casefr13}	\newcommand*\THIRD{\frac{1}{3}}

\newcommand*\Fourth{\Casefr14}	\newcommand*\FOURTH{\frac{1}{4}}

\newcommand*\sixth{\casefr16}	
\newcommand*\Sixth{\Casefr16}	\newcommand*\SIXTH{\frac{1}{6}}

\newcommand*\eighth{\casefr18}  
\newcommand*\Eighth{\Casefr18}	\newcommand*\EIGHTH{\frac{1}{8}}


\newcommand*\ordspacing{%
	\normalbar
	\mathcode`|="226A
	\mathcode`+="002B
	\mathcode`-="0200
	\mathcode`*="0203
	\mathcode`=="003D
	}

\newcommand*\of[1]{({\ordspacing #1})}

\newcommand*\oF[1]{\vlp{\ordspacing #1}\vrp}

\newcommand*\Of[1]{(\mkern1.5mu{\ordspacing #1}\mkern1.5mu)}

\newcommand*\OF[1]{\vlp{\mkern1.5mu%
	{\ordspacing #1}\mkern1.5mu}\vrp}



%
	\renewcommand*\BE{\begin{equation}}
	\renewcommand*\EE{\end{equation}}
	\renewcommand*\BEA{\begin{eqnarray}}
	\renewcommand*\BAL[1][]{\BM[#1]\BA}
	\newcommand*\BALams[1][]{\BM[#1]\BAams}
\def\BALams{\@ifnextchar[\BALams@{\BALams@[]}}
\def\BALams@[#1]#2\EALams{\BM[#1]\BAams#2\EAams\EM}
	\renewcommand*\EM{\end{subequations}}
	\newcommand*\WT{\begin{widetext}}
	\newcommand*\NT{\end{widetext}}
	\renewcommand*\etal{\ETAL}	
	\renewcommand*\etc{\ETC}	
	\renewcommand*\Of[1]{{(\,#1\,)}} 
	\renewcommand*\fkern{\!}	
	\renewcommand*\Vkern{}		
	\renewcommand*\Unskip{}		


\newcommand*\journalno{}




\newcommand*\avg[1]{\langle#1\rangle}
\renewcommand*\Partial[2]{\frac{\partial#1}{\partial#2}}
\renewcommand*\PartiaL[2]{\left(\frac{\partial#1}{\partial#2}\right)}
\renewcommand*\Total[2]{\frac{d#1}{d#2}}
\renewcommand*\TotaL[2]{\frac{D#1}{D#2}}
\renewcommand*\Func[2]{\frac{\d#1}{\d#2}}
\renewcommand*\fr{\frac}
\renewcommand*\fR[2]{\left(\frac{#1}{#2}\right)}
\renewcommand*\bR[2]{\left[\frac{#1}{#2}\right]}
\let\asterisk*
\def\.{\cdot}

\def\Ext{^{\rm ext}}
\def\Intern{^{\rm int}}
\def\Tot{^{\rm tot}}

\def\crossout#1{\setbox0=\hbox{$\displaystyle #1$}
\dimen0=\ht0
\advance\dimen0 by \dp0 
\dimen1=0.5\wd0
\advance\dimen1 by -\dimen0
\kern\dimen1
\dimen2=\dp0
\advance\dimen2 by0.25\dimen0
\lower\dimen2\rlap{\hbox{%
\setlength{\unitlength}{1.5\dimen0}
\begin{picture}(1,1)
\vector(1,1){1}
\end{picture}
}}
\kern-\dimen1
{#1}%
}

\renewcommand*\Lk{L_\vk}
\def\bk{\beta_\vk}
\def\less{_q}
\def\great{_k}
\def\psiQ{\psi\less}
\def\dpsiQ{\d\psiQ}
\def\psiK{\psi\great}
\def\dpsiK{\d\psiK}
\def\cEQ{\cE\less}
\def\cEK{\cE\great}
\def\SD{\sum_\Delta}
\def\sk{\sigma_\vk}
\def\ak{\alpha_\vk}
\long\def\comment#1\endcomment{}
\def\<#1>{\langle#1\rangle}
\def\set#1{\{#1\}}
\def\kmin{k_{\rm min}}
\def\Ct{\Tilde C}
\def\pk{\phi_\vk}
\def\pp{\phi_\vp}
\def\pq{\phi_\vq}
\def\Zk{\Zbar_\vk}
\def\Zq{\Zbar_\vq}
\def\Zbar{\cN}
\def\Zqdot{\dot \Zbar_\vq}
\def\noise{^{\rm noise}}
\def\SUMo(#1,#2){\sum_{#1}}
\def\thetaE(#1,#2,#3){\theta^{(E)}_{#1,#2,#3}}
\def\q{\Bar{q}}
\def\C#1{C_{#1}}
\let\V\v
\def\vxt{\Tilde{\vx}}
\def\Phat{\Hat{P}}
\def\Dhat{\Hat{D}}
\let\Ref\citet
\def\Rk{R_\vk}
\def\Rq{R_\vq}
\def\cite{\ \citep}
\newcommand*\casefr[2]{\mathchoice{{\textstyle\frac{#1}{#2}}}%
	{{\textstyle\frac{#1}{#2}}}
	{{\scriptstyle\frac{#1}{#2}}}%
	{{\scriptscriptstyle\frac{#1}{#2}}}}
\let\case\casefr

\def\Nt{\Tilde{N}}
\def\cZbar{\Zbar}
\def\ks{k_\bigast}
\def\half{\casefr12}
\def\wt{\Tilde{\omega}}
\def\Sd{\sum_\Delta}
\def\Tt{\Tilde{T}}
\def\M(#1,#2,#3){M_{\v{#1}\v{#2}\v{#3}}}
\def\Placeholder{\emph{Placeholder:  In part, this article serves as an
    introduction to the remainder of the material in Chaps.~5 and~6.
    Although I have already included pointers to some of the articles in
    the body of the present article, I propose to rewrite this Discussion
    section after I see the first drafts of the other articles so that I
    have a better understanding of their content.}}
\def\T(#1,#2,#3){\theta_{\v{#1}\v{#2}\v{#3}}}
\renewcommand*\mg{\tensor{g}}		
\newcommand*\mghat{\Hat{\mg}}
\newcommand*\PBB[1]{\[#1\]}
\newcommand*\PB[1]{[#1]}
\def\pages#1#2{#1}
\def\delT{\del_T}
\def\Pb#1{\Pbo#1\relax}
\def\Pbo#1,#2\relax{\set{#1,\,#2}}

\def\fnlabel#1{\edef\@currentlabel{\the\value{footnote}}%
\label{#1}}

\def\fnref#1{footnote~\ref{#1}%
\edef\temp{\the\value{page}\null}%
\edef\tempref{\pageref{#1}}%
\ifx\temp\tempref\else
\ on page~\pageref{#1}\fi}

\def\Paragraph#1{\paragraph{#1 ---}}

\def\FIGURE#1#2#3{\begin{figure}
\figurebox{#3\columnwidth}{}{#1.eps}
\caption{#2}
\label{Fg.#1}
\end{figure}
}

\def\sq{\sigma_\vq}
\def\qhat{\unit{q}}
\def\Ko{K_0}
\def\Ky{K_y}
\def\Kx{K_x}
\def\vK{\v{K}}
\def\vQ{\v{Q}}
\def\MI{modulational instability}
\def\Vhat{\Hat{V}}
\def\uo{u_0}
\def\Kx{\vK\.\vx}
\def\vKhat{\unit{K}}
\def\Kbar{\Bar{K}}
\def\vKbar{\Bar{\vK}}
\def\GT{^>}
\def\LT{^<}
\def\mDk{\mD_\vk}
\def\cEq{\cE_\vq}
\def\Kobar{\Bar{K}_0}
\def\atld{\Tilde{a}}
\def\ptld{\Tilde{\phi}}
\def\bt{\Tilde{b}}
\def\Rt{\Tilde{R}}
\def\Kt{\Tilde{K}}
\def\vGk{\v{\G}_\vk}
\def\LD{L_d}
\def\kD{k_d}

\catcode`\@=12

\newcommand{\defineas}{=}


We found in \Sec{jp:sec:isotropic}, using the CE2 approximation, that for the zonostrophic instability the behavior of the effective forcing on the zonal flows depended on whether the deformation radius~$\LD$ was infinite or finite. For $\LD = \infty$, wave numbers $k > q$ of an isotropic spectrum produce no net forcing, whereas there is net forcing for finite~$\LD$. We also saw in \Sec{jp:sec:parametricinst} that the standard equations for generalized \MI\ are a special case of those for zonostrophic instability. In this appendix we will discuss some of the connections between these various results. 

Various proposed mechanisms for the formation of zonal jets have been summarized by \Ref{bakasioannou2013}.  They listed ``turbulent cascades, modulational instability, mixing of potential vorticity, and statistical theories''; their work focused on the implications of the S3T closure. As they pointed out, one of the key points to be reckoned with is that ``previous studies have shown that shearing of isotropic eddies on an infinite domain and in the absence of dissipation and~$\b$ does not produce any net momentum fluxes\cite{shepherd1985,farrell1987,holloway2010}.'' Note that none of Bakas and Ioannou, Shepherd, or Farrell cited the closely related, detailed, and compelling discussion given by \Ref[Sec.~5]{kraichnan1976} of the physical mechanisms that underlie long-wavelength flow generation for 2D \NS\ turbulence in both coherent and stochastic situations. The implications of that work also do not seem to be appreciated by many workers on the modulational instability. \Ref{holloway2010} did cite it, discussed why the works of Kraichnan and Shepherd seem to have had limited impact, and went on to provide valuable new insights about some of the apparent contradictions that arise in various descriptions of eddy shearing. Our discussion below adds additional perspectives.

Although Holloway provided some description of Kraichnan's calculations, we find it necessary to discuss them here as well. (Essential background can be found in the article by Krommes and Parker in \Sec{KP} of this book, which we will abbreviate as~KP.)  Kraichnan's original analysis was for 2D homogeneous \NS\ turbulence (for which $\LD$ is infinite).  The analysis, which is generalized here to finite~$\LD$, turns out to be useful not only for understanding long-wavelength flow generation in homogeneous turbulence (see KP), but also for gaining an intuitive understanding of the physics of zonostrophic instability and bifurcation to inhomogeneous turbulence.  We find a connection to various limits reported in \Sec{jp:sec:isotropic}.  In order to provide necessary background, we will first review Kraichnan's original analysis; we will also discuss how it is related to conventional calculations of \MI, thereby making a connection to our observation in \Sec{jp:sec:parametricinst} that \MI\ is a special case of the zonostrophic instability. Then we will generalize the basic ideas to situations with finite~$\LD$. For those cases, we will show that some of Kraichnan's conclusions are nontrivially modified in a way that is consistent with the results described in \Sec{jp:sec:isotropic}, and we will provide some heuristic understanding.

\subsubsection{Review of Kraichnan's discussion of negative eddy viscosity}

Kraichnan framed his analysis as a calculation of a statistical eddy viscosity $\mu(q \mid \kmin)$ felt by resolved scales (wave number $< \kmin$) due to the interactions with unresolved sub-grid scales; see the discussion in Sec.~5.1.4.2 of KP. In the asymptotic limit $q \ll \kmin$, he found that $\mu(q \mid \kmin) < 0$ in 2D; this is the famous negative eddy viscosity. \Ref{krommeskim2000} discussed an important connection between that result and a certain formula for the rate of zonal flow generation, and aspects of that analysis will be useful here as well. Kraichnan also pointed out that $\mu(q \mid \kmin)$ actually vanishes for situations in which the interactions are dominated by long-wavelength straining of turbulent excitations confined to $k \ge \kmin$. We will generalize that result to models with finite~$\LD$. 

Kraichnan described the transfer of energy from short to long wavelengths in 2D turbulence as resulting from the generation of a `secondary flow,' a concept closely related to the mechanism of `secondary instability' considered by various authors\cite{rogersdorlandetal2000,plunk2007,pueschelgorleretal2013}. He began with a blob of short-wavelength vorticity (having central wave vector~$\vK = K\yhat$) initially localized within a circular domain of radius~$D$ ($K D \gg 1$) and possessing the stream function\footnote{Kraichnan used $\Ko = 1$ and $\uo = 1$, but we prefer to leave them general so that the dimensions of various quantities are correct. We have changed some of his notation as well. For example, we have used uppercase~$\vK$ and~$\vQ$ for the specific wave vectors of the turbulence and the straining field, respectively.}
\BE
\psi(\vx,t) = \fR{\Ko\uo}{K^2(t)}f(\vx)\cos(\vK\.\vx),
\eq{psi}
\EE
where $\Ko \defineas K(0)$ and 
\BE
f(\vx) = \exp\(-\Half\fr{\r^2}{D^2}\),
\quad
\hbox{where\ }
\r^2 \defineas x^2 + y^2.
\EE
The resulting velocity is 
\BALams
\vu &= \zhat\cross\vgrad\psi
\\
&= -\fR{\Ko}{K}\uo f(\vx)[\zhat\cross\vKhat\sin(\Kx) 
\NN\\
&\qquad\qquad +
  (KD)\m1\zhat\cross(\vx/D)\cos(\Kx)]
\eq{u}
\EALams
(\Fig{u}),
 and the vorticity is
\BALams
\w &=
\grad^2\psi
\\
&= -\Ko\uo f(\vx)\big(\{[1 + (KD)\m2[2 - (\r/D)^2]\}\cos(\Kx)
\NN\\
&\qquad - 2(KD)\m1\vKhat\.(\vx/D)\sin(\Kx)\big)
\eq{omega}
\EALams
(\Fig{omega1}).

\FIGURE{u}{The velocity field corresponding to Eq.~(\ref{psi}).  Lengths
  are normalized to~$D$; $\Ky = 2\pi$.}{0.9}

\FIGURE{omega1}{The vorticity field corresponding to
  Eq.~(\ref{psi}). 
}{0.9}

One way of understanding the role of the shaping function~$f$ is by inquiring about the spectral content of~$\psi$.  One has 
\BALams
&\psi(\vk,t) = \fR{\Ko\uo}{K^2}\Int d\vx\,
  e^{-i\vk\.\vx}f(\vx)\cos(\vK\.\vx)
\\
&= \fR{\Ko\uo}{K^2} \pi D^2 \Bigl(e^{-\half\abso{\vk-\vK}^2 D^2} +
  e^{-\half\abso{\vk+\vK}^2 D^2} \Bigr). 
\EALams
With $Q \defineas D\m1$ (underlying vector~$\vQ$'s will be introduced later), this describes a spectrum containing the primary mode~$\vK$ and all sidebands having magnitudes up to $P_\pm \defineas \abso{\mp\vK - \vQ}$. The role of~$f$ is thus to introduce sidebands that are necessary in order that triad interactions can occur between the primary, the sidebands, and a long-wavelength disturbance with characteristic wave vector~$\vQ$; compare the minimal system of four wave vectors $\vK$, $\vP_\pm$, and~$\vQ$ [KP, 
Fig.~5.4 (right)] used in modulational-instability calculations.

Kraichnan now introduces a long-wavelength straining field having potential
\BE
V(\vx) = -a x y,
\EE
where $a$~is an unspecified amplitude (having the dimensions of frequency or vorticity).  The straining velocity is
\BE
\vv(\vx) = \zhat\cross\vgrad V = a(x\,\xhat - y\,\yhat)
\EE
and is visualized in \Fig{straining}.  This is a flow with pure rate of strain, \ie, it is irrotational: $\zhat\.\curl\vv = \grad^2 V = 0$.  To make contact with the calculations of \MI, consider its spectral content, which is 
\BE
V(\vq) = (2\pi)^2a\d'(q_x)\d'(q_y) \equiv (2\pi)^2 a \d'(\vq).
\EE
This is a somewhat unusual and pathological function.  However, it can be regularized by replacing the derivatives of the delta functions with finite-difference representations, \eg, $\d'(q) \approx [\Dirac{q + Q} - \Dirac{q - Q}]/2Q$ for small~$Q$.  One is led naturally to this approximation by noting that since $\vv(\vx)$ will be interacting with the shaped blob of short-wavelength vorticity, which localizes distances to $Qx < 1$, it makes little qualitative difference if one replaces~$V(\vx)$ by 
\BE
\Vhat(\vx) \defineas -aD^2\sin(Qx)\sin(Qy) = \Vhat_+(\vx) - \Vhat_-(\vx),
\EE
where
\BE
\Vhat_\pm \defineas \Half a D^2 \cos(\vQ_\pm\.\vx)
\EE
with
\BE
\vQ_\pm \defineas Q(\xhat \pm \yhat).
\EE
One has
\BE
\Vhat_\pm(\vq) = \pi^2 a D^2[\Dirac{\vq - \vQ_\pm} + \Dirac{\vq +
    \vQ_\pm}].
\EE
Thus the original irrotational straining field is the difference of two fields, each possessing both strain and vorticity,\footnote{In plasma physics and possibly elsewhere, it is ubiquitous to illustrate physics related to eddy `shearing' with velocity fields like the one shown in \protect\Fig{vplus}, which possess vorticity as well as strain. Usually the rotational part of the interaction is not remarked upon. While that often does not matter for simple heuristics, some arguments and illustrations would be clearer if Kraichnan's example were followed and a field with pure rate of strain were used.} whose wave vectors~$\vQ_\pm$ are oriented along the $\pm 45\degrees$ diagonals, as illustrated in \Figs{vplus} and \FIG{vminus}.

\FIGURE{straining}{The long-wavelength straining field.}
{0.9}

\FIGURE{vplus}{The velocity field corresponding to~$\Vhat_+$. It is built from $Q_+ \defineas (1,1)\Tr$ and has both strain and vorticity.}{0.9}

\FIGURE{vminus}{The velocity field corresponding to~$\Vhat_-$, built from   $Q_- \defineas (1,-1)\Tr$.}{0.9}

Conventional \MI\ calculations\cite[and references therein]{nazarenkoconnaughtonetal2014} begin with a single~$\vQ$ and its negative.  The initial state of the instability thus possesses both vorticity and strain.  We will see shortly how such an instability is related to Kraichnan's procedure.

We continue to review his calculations. The next step is to find an expression for the time rate of change of short-scale energy due to the straining. The vorticity equation $\delt\w + \vv\.\vgrad\w$ becomes, in a Lagrangian representation, $d\w/dt = 0$ with the characteristic equations $d\vx/dt = \vv(\vx) = a(x,-y)\Tr$. As Kraichnan observed, it follows that an initially circular blob is stretched into an ellipse with major axis in the $x$~direction, while the central wave vector is stretched according to $K_x(t) = K_x(0)e^{-at}$, $\Ky(t) = \Ky(0)e^{at}$. Direct calculation of the time rate of change of the spatially-integrated energy\footnote{There is a crucial misprint in Kraichnan's formula for the initial kinetic energy in the second line after his Eq.~(5.7); a factor of~$k\m2$ is omitted.} ~$\Bar{\cE} \defineas \half\Bar{u^2} \propto K\m2 + \OrdeR{(KD)\m2}$ then leads, for $\vK(0) = \Ky\yhat$, to the initial energy loss rate $\dot{\Bar{\cE}} = -2a\Bar{\cE}$ to lowest order. By considering the secondary flow that is generated by~$\vu$ (\ie, by evaluating $\delt\D\w = - \vu\.\vgrad\w$) at $t = 0$), Kraichnan demonstrated that the lost energy shows up as energy of interaction between the secondary flow and the straining flow. \Figure{vdif} illustrates that secondary flow, which is such as to reinforce the original straining flow near the origin (for positive~$a$). 

\FIGURE{vdif}{The secondary flow, containing four vortices, that arises by self-interaction of the small-scale motion.}{0.9}

This nonrandom mechanism, with energy transfer mediated by the amplitude~$a$, is closely related to conventional calculations of \MI. Those describe an eigenvalue problem in which the unstable eigenvector possesses both strain and vorticity and grows coherently. In Kraichnan's calculation, the original straining field is reinforced by the vorticity of the secondary flow. If that reinforced field were taken as a new initial condition and the process were repeated, the evolving long-wavelength flow would contain vorticity as well as strain, as in the modulational-instability calculations. To understand the time scale for the reinforcement, consider the secondary-flow equation
\BE
\delt\D\w = -\vu\.\vgrad\w.
\eq{Dw_dot}
\EE
It is straightforward to use the results \EQ{u} and \EQ{omega} to show that the secular part of the \rhs\ of \Eq{Dw_dot} is at $t = 0$ 
\BALams
&-(\vu\.\vgrad\w)_{\rm secular}
\NN\\
&\qquad = -2\uo^2f^2
D\m4(\zhat\.\vKhat\cross\vx)(\vKhat\.\vx) 
\\
&\qquad\propto \fR{x}{D}\fR{y}{D}f^2(Q\uo)^2 \sim (Q\uo)^2.
\EALams
The frequency $Q\uo$ is the circulation rate or vorticity of one of the vortices shown in \Fig{vdif}. That should also be the characteristic rate of the reinforcement, so one concludes that the characteristic rate for the growth of the long-wavelength flow is $\l \sim Q\uo$. This agrees with the result of a modulational-instability calculation in which wave effects are neglected and the asymptotic limit of small~$Q/K$ is taken\cite{krommes2006}; it is also consistent with the implications of \Eq{jp:disprelation_connaughton}. 
\comment
This is consistent with the result for the change of interaction energy~$I$, namely $d\Bar{I}/dt = 2a\Bar{\cE}$, if
\endcomment

We now turn to the implications of this analysis of coherent interactions for statistical scenarios. Kraichnan addressed this in the context of the 2D \NSE; we are interested in the generalization of his analysis for cases with finite deformation radius. The basic calculation makes the straining amplitude~$a$ a random function $\atld(t)$ and also assumes that the wave vector~$\Tilde{\vK}$ of the small-scale motion is oriented randomly, having angle~$\ptld$ \wrt\ the $y$~axis. With the assumption of passive advection of the small scales by the straining field, it is straightforward to find that 
\BE
\Kt^2(t)/\Ko^2 = \cosh[2\bt(t)] + \sinh[2\bt(t)]\cos(2\ptld),
\EE
where
\BE
\bt(t) \defineas \I0t\,d\tbar\,\atld(\tbar).
\eq{beta_def}
\EE
Kraichnan noted that $\<\Kt^2>_\p$ (averaged over~$\p$ but not~$a$) typically grows in mean square, consistent with general results of \Ref{cocke1969}. However, according to \Eq{u} the 2D \NS\ energy $\Bar{\cE} \defineas \half\Bar{u^2}$ is proportional to $K\m2$, and Kraichnan found that 
\BE
\<\Kt\m2(t)>_\p = \Ko\m2,
\eq{<Km2>}
\EE
\ie, $\Bar{\cE}$~is independent of time in spite of the random stretching and squeezing of~$\Tilde{\vK}(t)$. It is worth quoting Kraichnan's interpretation of this in his own words, since we will shortly give a more general discussion. He was concerned with the physics of the isotropic 2D eddy viscosity, which we repeat here for convenience\footnote{Following the conventions used by~KP, we indicate discrete Fourier transforms by subscripts (\eg, $\cNk$) and integral transforms by arguments [\eg, $\cN(\vk)$]. Two-point spectra are normalized such that the fluctuation intensity is $\cN = \sum_\vk\cNk = (2\pi)\m{d}\Int d\vk\,\cN(\vk)$, where $d$~is the dimensionality of space ($ = 2$ for the present discussion). The velocity spectrum is~$\cU_\vk$. Energy and enstrophy spectra are defined with a factor of~$\half$ relative to~$\cU$. We will not have occasion to use omnidirectional spectra such as the common~$E(k)$, which incorporate the wave-number volume element.}: 

\BE
\mu(q \mid \kmin) = \fr{\pi}{4}\I{\kmin}\infty
dk\,\theta_{qkk}\Partial{[k^2\cU(k)]}{k}. 
\eq{mu_NS}
\EE
Regarding \Eq{mu_NS}, he observed\footnote{For consistency with our   notation and that of Kraichnan's model, we have interchanged~$k$ and~$q$   from Kraichnan's original usage in his Sec.~4.  We also write~$\kmin$   instead of~$k_m$ and~$\cU$ instead of~$U$.}
\begin{extract}
``The integrand is a total derivative except for the $k$~dependence of~$\theta_{kkq}$.  This means that any addition to the spectrum~$\cU(k)$ for $k > \kmin$ which vanishes at $k = \kmin$ would add nothing to $\mu(q \mid \kmin)$ were it not for the $k$ dependence of $\theta_{kkq}$.''
\end{extract}
\noindent
He then interpreted the results of his model calculation as follows:
\begin{extract}
``If $\theta_{kkq}$ is dominated by low-wavenumber straining, in correspondence to our present discussion, it is independent of~$k$ and the integrand of [\Eq{mu_NS}] is a total derivative.  Thus any excitation, described by~$\cU(k)$, which is totally confined to\footnote{The published text contains the typographical error $k < \kmin$ instead of the correct $k > \kmin$.} $k > \kmin$, gives zero contribution to the effective eddy viscosity exerted on $q \ll k$.  This is a direct consequence of [\Eq{<Km2>}] which says that low-wavenumber straining of the small scales gives a diffusion process in wavenumber with \emph{no} average loss of kinetic energy.  By conservation, there is then no net gain of kinetic energy by the straining scales. On the other hand, if $\kmin$~falls within the small-scale excitation, the diffusion of the excitation to smaller~$k$ occurs at wavenumbers $ < \kmin$ and is not counted in [\Eq{mu_NS}] which then includes only the outward diffusion.  The latter \emph{does} involve a net loss of kinetic energy by the small scales and thus gives rise to a negative contribution to the eddy viscosity.''
\end{extract}

Kraichnan's insights here are deep and important, but two points require further discussion. First, he attributes the nonvanishing of~$\muq$ to the $k$~dependence of $\theta_{qkk}$, but he does not give a satisfactory explanation of why that quantity should be fundamental. Second, he uses the phrase ``diffusion process in wavenumber'' without clearly specifying exactly what quantity is diffusing. Given the immediate context, the reader would be forgiven for pondering whether it is energy diffusion that is meant, but further thought and reference to the discussion of nonlinear invariants in KP, 
Sec.~1.1.4.1, lead one to conclude that it is actually enstrophy that diffuses (total enstrophy being the nonlinear conserved quantity). We will see that a proper understanding of this latter point will also clarify the role of~$\theta_{qkk}$; it is the autocorrelation time associated with the wave-number diffusion coefficient~$D_\vk$ of enstrophy, which is more fundamental than $\theta$~itself.

\subsubsection{The effects of finite deformation radius}

It is instructive to consider these issues for cases involving finite deformation radius~$L_d$, specifically the Charney--\HME\ (CHME) and the modified \HME\ (mHME).

\Paragraph{General formulas for energy gain and loss}

The relevant nonlinear invariant is (see the background material in KP, Sec.~1.1.4.2) $\cNk = \sk^2\cEk$, where $\sk^2 = k^2$ for the CHME and $\sk^2 = \kbar^2$ for the mHME\@. Here $\kbar^2 \defineas \ak + k^2$, where $\ak = 0$ for zonal modes and $\ak = \kD^2$ otherwise ($\kD \defineas \LD\m1$); also, $\cEk = \half\kbar^2\<\abso{\d\p_\vk}^2>$. (The 2D \NS\ case is recovered for $\ak = 0$.) We assume a homogeneous ensemble with random long-wavelength flows. \Ref{krommeskim2000} showed that, upon expansion in $\e \defineas q/k \ll 1$ of an anisotropic extension of Kraichnan's test field model, a diffusion equation ensues for the short-wavelength spectrum:
\BE
\Partial{\cNk\GT}{t} = \Partial{}{\vk}\.\mDk\.\Partial{\cNk\GT}{\vk},
\EE
where 
\BE
\mDk \defineas 2k^2\fR{\sk^2}{\kbar^2}^2\sum_\vq(\qhat\,\qhat)
\abso{\khat\cross\qhat}^2\fR{q^2}{\qbar^2}\fR{q^2}{\sq^2}\theta_{\vk,-\vk,\vq}\cNq\LT.
\eq{mDk}
\EE
(Krommes and Kim also gave a heuristic random-walk derivation of~$\mDk$.) Short-wavelength energy~$\cEk\GT$ evolves according to the nonconservative equation
\BE
\Partial{\cEk\GT}{t} =
\fr{1}{\sk^2}\Partial{}{\vk}\.\mDk\.\Partial{(\sk^2\cEk\GT)}{\vk}.
\EE
Upon writing this as much as possible in conservative form, one finds
\BAams
\Partial{\cEk\GT}{t} &= \Partial{}{\vk}\.\(\mDk\.\Partial{\cEk\GT}{\vk}\)
- \Partial{}{\vk}\.\(2\mDk\.\Partial{\ln\sk\m2}{\vk}\cEk\GT\) 
\NN\\
&\qquad
+ \Partial{}{\vk}\.\(\mDk\.\Partial{\sk\m2}{\vk}\)\cNk\GT.
\eq{E>_dot}
\EAams
Thus, while wave-number diffusion (first term) does act on the short-scale energy, $\cEk\GT$~also experiences drag (second term) as well as an intrinsic loss mechanism (last term). The loss term describes the second-order, statistically averaged effect of random refraction of the ray trajectories of the small-scale wave packets; it is built from the first-order refraction effect discussed by KP, 
Eqs.~(5.82) and~(5.83). Mathematically, it arises because the scale factor~$\sk^2$ that relates~$\cNk$ and~$\cEk$ does not commute with the Poisson bracket, involving large scale~$\vX$ and large wave number~$\vk$, that generates weakly inhomogeneous wave kinetics.

To verify that energy lost from the short scales shows up in the large scales, consider the equation for long-wavelength energy evolution~$\cEq\LT$, which from \Ref{krommeskim2000} is
\BE
\delt\cEq\LT = 2\gq\cEq\LT
\EE
with
\BE
\gq \defineas -2 q^2\fR{q^2}{\qbar^2}\sum_\vk
\fr{1}{k}\fR{k^2}{\kbar^2}^2\abso{\khat\cross\qhat}^2\khat\.\qhat\,\theta_{\vq,-\vk,\vk}\qhat\.\Partial{\cNk\GT}{\vk}.
\label{jp:krommeskimgammaq}
\EE
It is then straightforward to verify the energy conservation law
\BE
\delt\cE\LT = \sum_\vq\fr{1}{\sq^2}(2\gq\cNq\LT)
= -\delt\cE\GT =
-\sum_\vk\fr{1}{\sk^2}\Partial{}{\vk}\.\mDk\.\Partial{\cNk\GT}{\vk}
\EE
by integrating the last expression by parts.  We ignore surface terms, meaning that we consider excitations entirely localized within the domain of integration.

The form of \Eq{E>_dot} can be used to give further insight to Kraichnan's observation that in the isotropic 2D \NS\ case the energy transfer would vanish for localized excitations were it not for the $k$~dependence of $\theta_{qkk}$.  Clearly the first two terms contribute nothing; they merely rearrange short-scale energy locally in \vKspace. The last term of \Eq{E>_dot} can be written as
\BE
(\del_\vk\.\vGk)\cNk\GT,
\eq{last_term}
\EE
where the `flux of inverse scale factor' is
\BE
\vGk \defineas -\mDk\.\del_\vk\sk\m2.
\EE
It is this term, the statistical manifestation of the ray equation $\dot\vk = -\vgrad\Wk$, where $\Wk$~is the nonlinear advection frequency (see the discussion of the first-order distension rate $\gk\up1$ by KP, Sec.~5.1.4.2), that has the potential to transfer energy. Because of the factor of~$\cNk\GT$ in \Eq{last_term}, the term is not conservative. Rather than describing a rate of redistribution of energy among the small scales, $\vGk$~gives the rate of transfer to the secondary flow and thus to the large scales. But if the divergence of that flux vanishes, no net energy transfer ensues (no secular contributions to secondary flow are generated). There are two contributions to that divergence, namely the $\vk$~dependencies of~$\mDk$ [$\propto k^2(\sk^2/\kbar^2)^2\theta_{\vq,\vk,-\vk}$] and of~$\del_\vk\sk\m2 = -\sk\m4\delk\sk^2 = -2\sk\m4\vk$. Note that the dependence on~$\sk$ cancels out between this term and~$\mDk$. For the isotropic 2D \NS\ case ($\ak = 0$), one finds $\vGk \propto \khat\,(k^2\theta_{qkk})\times(k\m3) = \khat\, k\m1\theta_{qkk}$; thus $\del_\vk\.\vGk$ vanishes to the extent that $\theta_{qkk}$ is independent of~$k$. For the cases with finite deformation radius, the result is instead $\vGk \propto \khat\, k\m1(k^4/\kbar^4)\theta_{qkk}$, which has nontrivial divergence even if $\theta_{qkk}$ is independent of~$k$. One sees that Kraichnan's result that the energy transfer is controlled by the $k$~dependence of~$\theta_{qkk}$ is a special case; of more fundamental relevance is the $\vk$~dependence of~$\vGk$, which stems from the underlying physics of random ray refraction.

Upon summing \Eq{E>_dot} over~$\vk$, one finds that the explicit result for a localized isotropic spectrum is
\BAams
\PartiaL{\cE\GT}{t}_{\rm iso} &=
-\fr{\pi}{2}\I0{q_{\rm max}} q\,dq\,\fR{q^2}{\qbar^2}\fR{q^2}{\sq^2}\cN\LT(q)
\NN\\
&\qquad\times
\I\kmin\infty
dk\,\Partial{}{k}\[\fR{k^4}{\sk^4}\theta_{kkq}\]\cN\GT(k).
\eq{Edot_iso}
\EAams
By virtue of energy conservation, this reduces to $-2\sum_\vq \gq\cEq$ where $\gq \defineas -q^2\muq$ and $\muq$, which generalizes the 2D \NS\ result \EQ{mu_NS}, is
\BE
\mu(q \mid \kmin) = -\fr{\pi}{4}\fR{q^2}{\qbar^2}\I{\kmin}\infty dk\,
\Partial{}{k}\[\fR{k^4}{\kbar^4}\theta_{kkq}\]\cN\GT(k).
\eq{mu_gen}
\EE
For $\ak = 0$ ($\qbar = q$ and $\kbar = k$), this reduces correctly to Kraichnan's result \EQ{mu_NS}.

The interpretation of the ratio $\Rk \defineas k^2/\sk^2$ is that it is a measure of the portion of the physics devoted to perpendicular advection. To be specific, we discuss the plasma case. The \HME\ for the magnetized plasma, Eq.~(V.1.49), embodies the two quite different physical processes of (i)~perpendicular advection of vorticity (the $\gradperp^2$ term), and (ii)~parallel electron response, which is rapid and adjusts essentially instantaneously to changes in the electrostatic potential (the $\ak$~term). The total energy is the sum of (i)~the kinetic energy associated with the perpendicular flow, and (ii)~the compressional energy associated with the parallel motion. $\Rk$~is the fraction of total energy associated with the perpendicular processes. (It approaches~1 for a mode whose wavelength is much smaller than~$\LD$.) It is only that fraction that is relevant for the random ray refraction. More directly, the presence of~$\Rk$ in \Eq{mDk} for~$\mDk$ arises from the fact that the effective frequency for advection of the short scales is reduced for the CHME by a factor of~$\Rk$ from the nominal $\vk\.\vV_\vq$ of the mHME; it appears squared because the random nature of the refraction leads to wave-number diffusion, a second-order effect.

\Paragraph{Generalization of Kraichnan's model to include finite
  deformation radius}

We now show that these results are consistent with a generalization of Kraichnan's model. For definiteness, we consider the modified \HME. In order to construct a stream function that corresponds to a short-scale blob of generalized vorticity, and in view of the forms of~$\cEk$ and~$\cNk$, one must replace~$K^2$ in \Eq{psi} by~$\Kbar^2$ (but not~$\vK$ by~$\vKbar$). Because $\cEk = \kbar\m2\cNk$ and $\cN$~is conserved under the disparate-scale interaction, it is useful to consider 
\BALams
\Rt(t) &\defineas \Ko^2\<\Kbar\m2(t)>_\p \\
&= \fr{1}{2\pi}\I0{2\pi}d\p\,\fr{\Ko^2}{\a_{\vK} + [\cosh(2\bt) +
    \sinh(2\bt)\cos(2\p)]\Ko^2}
\\
&= \{1 + 2\abar \cosh[2\bt(t)] + \abar^2\}\m{1/2},
\eq{R(t)}
\EALams
where $\abar \defineas \a/\Ko^2$.  ($\a_{\vK} = \kD\m2$ for short-scale modes.)  This correctly reduces to Kraichnan's result \EQ{<Km2>} for $\abar = 0$, but depends on the random straining otherwise.

At $t = 0$, one finds
\BE
R(0) = \fr{\Ko^2}{\a + \Ko^2} = \fr{\Ko^2}{\Kobar^2}.
\EE
This is trivial (straining has not yet acted at $t = 0$); it should not be confused with \Eq{<Km2>}, which holds for all times, and merely confirms that an average over an isotropic wave-number distribution does not change the isotropic quantity~$\Kbar\m2$. The results in the presence of the random straining are more interesting. We now show that an appropriate average of \Eq{R(t)} over random~$\bt$ gives a result for short-scale energy loss in accord with \Eq{Edot_iso}. Upon recalling the definition of~$\bt$ [\Eq{beta_def}], and noting that the formula \EQ{R(t)} is even in~$\bt$, one sees that $\Rt(t) = R(0) + \Order{t^2}$; thus $\dot\cE\GT \propto \dot R(t)$ vanishes at $t = 0$. This is not in conflict with formulas like \EQ{Edot_iso}, however, because those follow from a Markovian closure; one must therefore consider times greater than the autocorrelation time~$\tac$ of the straining and evaluate the coarse-grained derivative $\lim_{t \to \hbox{`0'}}\delt\cE$, where `0'~implies the restriction $t \gg \tac$. A useful general formula for~$\<R(t)>$ for arbitrary statistics of~$\atld$ (assumed to be stationary) seems difficult to obtain; however, one may extract the short-time result by expanding 
\BE
\Rt(t) = (1 + \abar)\m1 - \fr{2\abar}{(1+\abar)^3}\bt^2(t) + \Order{\bt^4}.
\EE
One has $\<\bt^2(t)> = \I0t d\tbar\I0t d\tbar'\<\atld(\tbar)\atld(\tbar')> \approx 2\<a^2>\tac t$ for $t \gg \tac$, which is a standard diffusion law.  Thus the coarse-grained time derivative is
\BE
\left. \Total{\<R>}{t}\right\rvert_{t = 0} \approx -4\fR{\Ko^4}{\Kobar^6}\<a^2>\tac.
\EE
From $\cE\GT = \Kobar\m2\cN\GT$ and using the fact that $\cN\GT$ is conserved, one finds
\BE
\Total{\cE\GT}{t} = \Ko\m2\Total{\<R>}{t}\cN\GT =
-4\fR{\Ko^2}{\Kobar^6}\<a^2>\tac\cN\GT.
\eq{cE_dot_model}
\EE
To compare this result with \Eq{Edot_iso}, we observe that in the present model we are assuming that long-wavelength straining dominates, so we should assume that $\theta_{kkq}$ is independent of~$k$.  Also, the derivative that is required in \Eq{Edot_iso} is explicitly
\BE
\Total{}{k}\fR{k^4}{\kbar^4} = \fr{4k^3}{\kbar^6}.
\EE
Since the model contains a single~$\vK$, we take the isotropic spectrum $\cNk\GT = (2\pi)^2k\m1\Dirac{k - \Ko}\cN\GT$.  One then obtains exact agreement between \Eqs{cE_dot_model} and \EQ{Edot_iso} by replacing~$\tac$ by~$\theta_q$ and choosing
\BE
\<a^2> = 2\fR{q^4}{\qbar^4}\cNq\LT.
\EE
This is nothing but the mean-square strain $q^2\cU(q)$; the factors of $q^2/\qbar^2$ correct~$\cNq$ by removing compressional energy:  $\<q^2 \abso{u_\vq}^2> = 2q^2(q^2/\qbar^2)\cEq\LT = 2(q^2/\qbar^2)^2\cNq\LT$.

Kraichnan's model and its generalization assume that long-wavelength straining dominates. In general, that is not necessarily the case. If short-wavelength decorrelation dominates~$\theta_{\vk,-\vk,\vq}$, one must ask whether the factor of~$\Rk^2$ under the $k$~derivative in \Eq{Edot_iso} is all or partly canceled by the $k$~dependence of~$\theta_{kkq}$. For the Galilean-invariant $\etak^S$ at large~$\vk$, one can show that in the absence of linear frequencies $\etak^S \propto \Rk^2 (\etak^S)\m1$, or $\etak^S \propto \Rk$. In the presence of linear frequencies, a dependence on~$\Rk$ remains as well, although the general case is somewhat complicated. In any case, the fact that $\theta_{qkk} \propto (\etak^S)\m1$ at large~$k$ means that the result $\vGk \propto \khat\,k\m1(\Rk^2\theta_{qkk})$ depends less strongly on~$\Rk$ than~$\Rk^2$ but is not independent of~$\Rk$. Clearly the basic conclusion that the energy transfer to the large scales is controlled by the $\vk$~dependence of $\Rk^2\theta_{qkk}$ still holds.

\subsubsection{Relation to zonostrophic instability}
Let us consider the relation between these results and zonostrophic instability. For the general anistropic case, if in the CE2 zonostrophic instability the $\l$ in the denominator of \Eq{jp:lambda_pm} were replaced by an inverse triad interaction time, then \Eq{jp:dispersionrelation} in the small $q$ limit agrees with \Eq{jp:krommeskimgammaq}. $\theta$~does not appear naturally in \Eq{jp:dispersionrelation} because the CE2 closure omits eddy damping~$\etak$; a more sophisticated closure should retain it. A consequence is that the zonostrophic dispersion relation derived from~CE2 is not correct in all details. Nevertheless, we expect that many of its qualitative predictions are robust.  The close connection between zonostrophic instability and the results derived in this appendix show the relevance of the physical mechanism discussed here.  In addition to this physical picture, our discussion has elucidated the reason behind the appearance of the factor of~$\Rk^2$ that controls the mathematical behavior of the asymptotic results. 


\endgroup 

%% file: jp_appendices.tex
\begingroup


\renewcommand{\a}{\alpha}
\renewcommand{\b}{\beta}
\newcommand{\de}{\delta}
\newcommand{\D}{\Delta}
\newcommand{\e}{\epsilon}
\newcommand{\ve}{\varepsilon}
\newcommand{\g}{\gamma}
\newcommand{\G}{\Gamma}
\renewcommand{\k}{\kappa}
\renewcommand{\l}{\lambda}
\renewcommand{\L}{\Lambda}
\newcommand{\m}{\mu}
\newcommand{\n}{\nu}
\newcommand{\p}{\phi}
\newcommand{\vp}{\varphi}
\renewcommand{\P}{\Phi}
\renewcommand{\r}{\rho}
\newcommand{\s}{\sigma}
\renewcommand{\t}{\tau}
\renewcommand{\th}{\theta}
\newcommand{\w}{\omega}
\newcommand{\W}{\Omega}
\newcommand{\z}{\zeta}

\newcommand{\la}{\langle}
\newcommand{\ra}{\rangle}

\renewcommand{\Re}{\operatorname{Re}}
\renewcommand{\Im}{\operatorname{Im}}
\newcommand{\sign}{\operatorname{sign}}

\renewcommand{\d}[2]{\frac{d #1}{d #2}}						
\newcommand{\dd}[2]{\frac{d^2 #1}{d #2^2}}					
\renewcommand{\v}[1]{\mathbf{#1}}				
\newcommand{\unit}[1]{{\v{\hat{#1}}}}			
\newcommand{\pd}[2]{\frac{\partial #1}{\partial #2}}		
\newcommand{\pdd}[2]{\frac{\partial^2 #1}{\partial #2^2}}	
\newcommand{\pddm}[3]{\frac{\partial^2 #1}{\partial #2 \partial #3}}	
\newcommand{\pddd}[2]{\frac{\partial^3 #1}{\partial #2^3}}	
\newcommand{\fd}[2]{\frac{\delta #1}{\delta #2}}					
\newcommand{\avg}[1]{\langle #1 \rangle}							
\newcommand{\bavg}[1]{\left\langle #1 \right\rangle}			

\newcommand{\eref}[1]{Eq.~\eqref{#1}}
\newcommand{\eqnref}[1]{Eq.~\eqref{eqn:#1}}					

\newcommand{\comments}[1]{}									

\renewcommand{\O}{O}	
\newcommand{\defineas}{\equiv}
\newcommand{\wh}[1]{\widehat{#1}}

\providecommand{\ol}{}		
\renewcommand{\ol}[1]{\overline{#1}}

\newcommand{\vk}{\v{k}}
\newcommand{\ti}[1]{\widetilde{#1}}		

\newcommand{\azf}{\hat{\a}_{ZF}}
\newcommand{\LD}{L_d^{-2}}
\newcommand{\nablabarsq}{\ol{\nabla}^2}
\newcommand{\kbsq}{\ol{k}^2}
\newcommand{\xbar}{{\ol{x}}}
\newcommand{\ybar}{{\ol{y}}}
\newcommand{\qbsq}{\ol{q}^2}
\newcommand{\pbsq}{\ol{p}^2}
\newcommand{\hbpsq}{\ol{h}_+^2}
\newcommand{\hbmsq}{\ol{h}_-^2}

\newcommand{\RB}{Rayleigh--B\'{e}nard\ }
\newcommand{\todo}[1]{\textbf{\emph{TODO:}#1}}
\renewcommand{\cite}{\citep}

\subsection{Correlation function corresponding to a wave}
\label{jp:app:wavecorr}
We consider in this section the one-time, two-point correlation function corresponding to a wave.  First we consider the general case of a superposition of waves.  Let
	\begin{equation}
		\psi'(x,y,t) = 2 \sum_\v{k} c_\v{k} \cos(k_x x + k_y y - \w_\v{k} t + \p_\v{k}).
	\end{equation}
Then, letting $\psi'_1 = \psi'(x_1,y_1,t)$ and $\psi'_2 = \psi'(x_2,y_2,t)$, we have
	\begin{align}
		\psi'_1 \psi'_2 = &\sum_\v{k} \sum_{\v{k}'} 2 c_\v{k} c_{\v{k}'} \big\{ \cos \big[ \tfrac12 (k_x + k_x')x + (k_x - k_x')\xbar \notag \\
			& + \tfrac12 (k_y + k_y') y + (k_y - k_y') \ybar - z_{\v{k} \v{k}'}^-\big] \notag \\
			& + \cos \big[ \tfrac12 (k_x - k_x')x + (k_x + k_x')\xbar \notag \\
			& + \tfrac12 (k_y - k_y') y + (k_y + k_y') \ybar - z_{\v{k} \v{k}'}^+ \big] \big\},
	\end{align}
where $x = x_1-x_2$, $\xbar = \frac12 (x_1 + x_2)$, and $z_{\v{k} \v{k}'}^\pm = (\w_\v{k} \pm \w_{\v{k}'})t - (\p_\v{k} \pm \p_{\v{k}'})$.  Using a zonal average, the correlation function is obtained by integrating over $\xbar$ with $x$ held fixed:
	\begin{equation}
		\Psi(x,y \mid \ybar) = \frac{1}{L_x} \int_0^{L_x} d\ol{x}|_x \psi'_1 \psi'_2,
	\end{equation}
The first cosine vanishes unless $k_x' = k_x$, while the second cosine vanishes unless $k_x' = -k_x$.  For simplicity assume all the $k_x, k_x' > 0$.  Then we are left with
	\begin{align}
		\Psi(x,y \mid \ybar) = & \sum_\v{k} \sum_{k_y'} 2 c_\v{k} c_{\v{k}'} \cos[ k_x x + \tfrac12 (k_y + k_y')y \notag \\
			& + (k_y - k_y') \ybar - (\w_\v{k} - \w_{\v{k}'})t + \p_\v{k} - \p_{\v{k}'} ].
	\end{align}
If we separate out in the sum the terms for which $k_y' = k_y$, then we have
	\begin{align}
		\Psi(x,y \mid \ybar) &= \sum_\v{k} 2 c_\v{k}^2 \cos( k_x x + k_y y) \notag \\
			&+ \sum_\v{k} \sum_{k_y' \neq k_y} 2 c_\v{k} c_{\v{k}'} \cos[k_x x + \tfrac12(k_y + k_y') y \notag \\
			& \qquad + (k_y - k_y')\ybar - (\w_\v{k} - \w_{\v{k}'})t + \p_\v{k} - \p_{\v{k}'} ]. \label{jp:W_manywaves}
	\end{align}
It can be verified by substitution that this is a solution to the unforced, undamped CE2 equations without zonal flow, $\partial_t W = 2 \b \partial_\ybar \partial_y \partial_x \Psi$ (and using $\w_\v{k} = -k_x \b / \kbsq$).  The first term of \eref{jp:W_manywaves}, which corresponds to the covariance of individual waves, is unchanging in time and homogeneous in space.  But in the second term, waves with different $k_y$ give rise to a correlation function that oscillates in time and has $\ybar$ dependence.  This is a manifestation of the coherent beating between waves.

One can imagine using another averaging procedure instead of the zonal average.  With the zonal average, the only coherent structures allowed are zonally symmetric.  One might also want to investigate zonally asymmetric structures, which precludes the use of a zonal average \cite{bakasioannou2013b}.  To study these more general coherent structures, the correlation function can be defined using a coarse graining in time or space (this approach typically requires the mean field and fluctuations to obey a scale-separation assumption) or an ensemble average.
	
To illustrate an alternate derivation for a single wave, let
	\begin{equation}
		\psi'(\v{x}) = \psi_0 \left( e^{i\v{p}\cdot\v{x} - i\w t} + e^{-i \v{p}\cdot\v{x} + i\w t}\right).
	\end{equation}
Then
	\begin{align}
	 	\psi'_1 \psi'_2 &= \psi_0^2\left( e^{2i\v{p} \cdot \ol{\v{x}}} e^{-2i\w t} + e^{i\v{p}\cdot\v{x}} + e^{-i\v{p}\cdot\v{x}} + e^{-2i\v{p} \cdot \ol{\v{x}}} e^{2i\w t} \right).
	\end{align}
At this point, a coarse graining in time over an intermediate time between $\w^{-1}$ and the timescale of the coherent structure eliminates the oscillating terms.  Equivalently, one could perform a coarse graining in space over an intermediate scale between $p^{-1}$ and the size of the coherent structure.  Then, one obtains
	\begin{equation}
		\Psi = \psi_0^2 \left( e^{i\v{p}\cdot\v{x}} + e^{-i\v{p}\cdot\v{x}} \right).
	\end{equation}
This $\Psi$ is homogeneous (independent of $\ol{\v{x}}$).  Its Fourier transform is
	\begin{equation}
		\Psi_H(k_x,k_y) = (2\pi)^2 \psi_0^2 \left[ \de(\v{k} - \v{p}) + \de(\v{k} + \v{p}) \right].
	\end{equation}
The inclusion of the mode at $-\v{p}$ as well as the mode at $\v{p}$ is essential and arises from the reality condition.

\subsection{Dispersion Relation for Arbitrary Primary and Arbitrary Secondary Wave}
\label{jp:app:arbwave}
We show here that for an arbitrary primary wave and arbitrary secondary wave, exact agreement is obtained between the dispersion relations from CE2 and from the 4-wave modulational instability.  This generalizes Section \ref{jp:sec:parametricinst}, which shows agreement in the special case where the primary wave has $p_y=0$ and the secondary wave has $q_x=0$.

The 4-wave modulational instability has a dispersion relation \cite{connaughtonnadigaetal2010}\footnote{This formula corrects a typographical error in Eq.~(5.1) of \citet{connaughtonnadigaetal2010}.}
	\begin{align}
		(q^2& +\LD) \l  - i\b q_x = \psi_0^2 |\v{p} \times \v{q}|^2 (p^2 - q^2)  \notag \\
			& \times \left( \frac{p_+^2 - p^2}{(p_+^2 + \LD)(\l - i \w) - i \b (p_x + q_x)} \right. \notag \\
			& \quad \left. + \frac{p_-^2 - p^2}{(p_-^2 + \LD)(\l + i \w) + i \b (p_x - q_x)} \right),
			\label{jp:connaughtongeneral}
	\end{align}
where $\v{p}_\pm = \v{p} \pm \v{q}$ and $\w = - \b p_x / (p^2 + \LD)$.

To allow for an arbitrary secondary wave within the CE2 formalism, we use the recent formulation of \citet{bakasioannou2013b,bakasioannou2013c}.  That formulation allows for coherent structures of arbitrary spatial dependence rather than restricting to zonally symmetric $q_x=0$ structures.  Their formulation also assumed infinite deformation radius, though that could be modified.  The dispersion relation in the small forcing and small dissipation limit is \cite{bakasioannou2013c}\footnote{There is a seeming factor of $2\pi$ different from the formula in \citet{bakasioannou2013c} but that is merely due to the choice of Fourier transform convention.}
	\begin{equation}
		\l q^2 - i \b q_x = \int \frac{dk_x\, dk_y}{(2\pi)^2} \frac{ N }{D} \left(1 - \frac{q^2}{k^2} \right) W_H(k_x,k_y),
		\label{jp:bakasdispersion}
	\end{equation}
where
	\begin{align}
		N &= 2 (k_x  q_y- k_y q_x) \bigg\{ q_x q_y \left[ \left(k_x + \frac{q_x}{2}\right)^2 - \left(k_y + \frac{q_y}{2} \right)^2 \right] \notag \\
			& \qquad \qquad + (q_y^2 - q_x^2)\left(k_x + \frac{q_x}{2} \right) \left(k_y + \frac{q_y}{2} \right)\bigg\}, \\
		D &= \l k^2 k_+^2 - \frac12 i q_x \b \left[k^2 + k_+^2 \right] \notag \\ 
			& \quad +  2 i \b \left(k_x + \frac{q_x}{2}\right) \left[ \left(k_x + \frac{q_x}{2} \right) q_x + \left(k_y + \frac{q_y}{2}\right) q_y \right],
	\end{align}
and $\v{k}_+ = \v{k} + \v{q}$.  As in Section \ref{jp:sec:parametricinst}, the appropriate background spectrum to correspond with that of \eref{jp:connaughtongeneral} is $W_H = (2\pi)^2 \psi_0^2 p^4  \left[ \de(\v{k} - \v{p}) + \de(\v{k} + \v{p}) \right]$.  With sufficient algebra, it is possible to show that \eref{jp:bakasdispersion} reduces exactly to the $\LD=0$ limit of \eref{jp:connaughtongeneral}.  The key is in recognizing that
	\begin{gather}
		N = (k_x q_y - k_y q_x)^2 \left(k_+^2 - k^2\right), \\
		D = k^2 \left[ \left( \l + \frac{i \b k_x}{k^2} \right) k_+^2 - i\b (k_x + q_x) \right].
	\end{gather}
	
\endgroup